\documentclass[amsmath,amssymb,aps,10pt,prd,letterpaper,balancelastpage,notitlepage,twocolumn,floatfix,superscriptaddress,nofootinbib,preprintnumbers]{revtex4-2}

\usepackage{graphicx}
\usepackage{epsfig}
\usepackage{dcolumn}
\usepackage{dsfont}
\usepackage[utf8]{inputenc}
\usepackage{ragged2e}
\usepackage{adjustbox}
\usepackage{xcolor}

\usepackage[sort&compress]{natbib}	 
	\setcitestyle{square,numbers,comma}
\usepackage[colorlinks=true,urlcolor=blue,linkcolor=black,citecolor=blue]{hyperref}
\newcommand{\secref}[2][]{Sec{#1}.~\ref{#2}}		
\newcommand{\appref}[2][x]{Appendi{#1}~\ref{#2}}		
\newcommand{\tabref}[2][]{Table{#1}~\ref{#2}}		
\newcommand{\figref}[2][]{Fig{#1}.~\ref{#2}}		
\renewcommand{\eqref}[2][]{Eq{#1}.~(\ref{#2})}		
\newcommand{\eqrefRange}[2]{Eqs.~(\ref{#1})--(\ref{#2})}		
\newcommand{\citeR}[2][]{Ref{#1}.~\cite{#2}}		

\usepackage{bm}
\newcommand{\lb}{\ensuremath{\left}}				
\newcommand{\rb}{\ensuremath{\right}}				
\newcommand{\nl}{\nonumber\\&\qquad}

\newcommand{\keV}{\mathinner{\mathrm{keV}}}
\newcommand{\MeV}{\mathinner{\mathrm{MeV}}}
\newcommand{\GeV}{\mathinner{\mathrm{GeV}}}
\newcommand{\TeV}{\mathinner{\mathrm{TeV}}}

\newcommand{\antiHe}[1]{{}^{#1}\overline{\text{He}}}
\newcommand{\antiD}{{\mkern-2mu\bar{\mkern2mu\text{D}}}}
\newcommand{\antiT}{{\bar{\text{T}}}}

\usepackage{fontawesome5}
\definecolor{orcidlogocol}{rgb}{0.65, 0.807, 0.223}
\newcommand{\orcid}[1]{\,\href{https://orcid.org/#1}{\textcolor{orcidlogocol}{\footnotesize\faOrcid}}}

\begin{document}

\title{Fireball anti-nucleosynthesis}

\author{Michael A.~Fedderke\orcid{0000-0002-1319-1622}}
\email{mfedderke@perimeterinstitute.ca}
\affiliation{Perimeter Institute for Theoretical Physics, Waterloo, Ontario, N2L 2Y5, Canada}
\affiliation{The William H.~Miller III Department of Physics and Astronomy, The Johns Hopkins University, Baltimore, Maryland, 21218, USA}
\author{David E.~Kaplan\orcid{0000-0001-8175-4506}}
\email{david.kaplan@jhu.edu}
\affiliation{The William H.~Miller III Department of Physics and Astronomy, The Johns Hopkins University, Baltimore, Maryland, 21218, USA}
\author{Anubhav Mathur\orcid{0000-0003-0973-1793}}
\email{a.mathur@jhu.edu}
\affiliation{The William H.~Miller III Department of Physics and Astronomy, The Johns Hopkins University, Baltimore, Maryland, 21218, USA}
\author{Surjeet Rajendran\orcid{0000-0001-9915-3573}}
\email{srajend4@jhu.edu}
\affiliation{The William H.~Miller III Department of Physics and Astronomy, The Johns Hopkins University, Baltimore, Maryland, 21218, USA}
\author{Erwin H.~Tanin\orcid{0000-0002-3204-0969}\,}
\email{ehtanin@stanford.edu}
\affiliation{Stanford Institute for Theoretical Physics, Stanford University, Stanford, California, 94305, USA}
\affiliation{The William H.~Miller III Department of Physics and Astronomy, The Johns Hopkins University, Baltimore, Maryland, 21218, USA}

\begin{abstract}
    The tentative identification of approximately ten relativistic anti-helium ($\antiHe{}$) cosmic-ray events at AMS-02 would, if confirmed, challenge our understanding of the astrophysical synthesis of heavy anti-nuclei.
    We propose a novel scenario for the enhanced production of such anti-nuclei that is triggered by isolated, catastrophic injections of large quantities of energetic Standard Model (SM) anti-quarks in our galaxy by physics beyond the Standard Model (BSM).
    We demonstrate that SM anti-nucleosynthetic processes that occur in the resulting rapidly expanding, thermalized fireballs of SM plasma can, for a reasonable range of parameters, produce the reported tentative $\sim 2:1$ ratio of $\antiHe{3}$ to $\antiHe{4}$ events at AMS-02, as well as their relativistic boosts.
    Moreover, we show that this can be achieved without violating anti-deuterium or anti-proton flux constraints for the appropriate anti-helium fluxes.
    A plausible BSM paradigm for the catastrophic injections is the collision of macroscopic composite dark-matter objects carrying large net anti-baryon number.
    Such a scenario would require these objects to be cosmologically stable, but to destabilize upon collision, promptly releasing a fraction of their mass energy into SM anti-particles within a tiny volume.
    We show that, in principle, the injection rate needed to attain the necessary anti-helium fluxes and the energetic conditions required to seed the fireballs appear possible to obtain in such a paradigm.
    We leave open the question of constructing a BSM particle physics model to realize this, but we suggest two concrete scenarios as promising targets for further investigation.
\end{abstract}
\maketitle
\tableofcontents

\section{Introduction}
\label{sec:Intro}

The AMS-02 Collaboration%
\footnote{\label{ftnt:AMS02descr}%
    The Alpha Magnetic Spectrometer (AMS-02) is a particle-physics detector located on the International Space Station~\cite{AMS:2021nhj}.} %
has unofficially reported~\cite{TingCERNslides2016,TingCERNslides2018,SchaelCOSPARslides,OlivaSlides2019,ZucconMIAPbPslides,TingCERNslides2023} $\mathcal{O}(10)$ highly relativistic cosmic-ray events detected in $\sim 10$ years of data that are consistent with tentative identification as anti-helium~\cite{ZucconMIAPbPslides,TingCERNslides2023}.%
\footnote{\label{ftnt:noDetection}%
    These unofficial reports~\cite{TingCERNslides2016,TingCERNslides2018,SchaelCOSPARslides,OlivaSlides2019,ZucconMIAPbPslides,TingCERNslides2023} have taken the form of public oral presentations on behalf of the Collaboration in the context of scientific conferences or major colloquia, as well as the associated publicly available presentation slides.
    We stress however that these data have not to date been published, have never been officially claimed by the AMS-02 Collaboration to present a formal detection of anti-helium cosmic rays, and have always been accompanied by disclaimers and caveats that the origin of these candidate events requires more study. 
    Additionally, only partial data is available for these events.} %
Although publicly available mass determinations are uncertain~\cite{ZucconMIAPbPslides}, the data are reported to be consistent with tentative identification of both $\antiHe{3}$ and $\antiHe{4}$ candidate events, with an event ratio of roughly $N_{\antiHe{3}}:N_{\antiHe{4}} \sim 2:1$ (albeit with small statistics and large uncertainties)~\cite{ZucconMIAPbPslides,TingCERNslides2023}.
Additionally, tentative identification of 7 anti-deuterium candidate events has recently been reported~\cite{TingCERNslides2023}.

While the tentative identifications of these candidate events require more work to confirm~\cite{TingCERNslides2023}, the anti-helium candidates in particular have been the subject of extensive recent interest in the literature~\cite{Carlson:2014ssa,Cirelli:2014qia,Korsmeier:2017xzj,Coogan:2017pwt,Poulin:2018wzu,Heeck:2019ego,Cholis:2020twh,Winkler:2020ltd,Arbey:2020ccg,Winkler:2022zdu,Bykov:2023nnr} because, taken at face-value, they are surprising: the formation of complex anti-nuclei in astrophysical environments is challenging and the rates of production implied by these candidate events are hard to reconcile with known Standard Model (SM) physics.

Within the SM, a known source of anti-nucleus cosmic rays is spallation induced by primary cosmic rays (hydrogen or helium) in the interstellar medium (ISM)~\cite{Poulin:2018wzu}.
Spallation is, however, inefficient at producing anti-nuclei; due to kinematics, this is particularly true for those anti-nuclei with higher atomic mass number $A$, such as $\antiHe{4}$.
Coalescence of two anti-nucleons or anti-nuclei (or of an anti-nucleon and an anti-nucleus) produced in spallation into a higher-$A$ anti-nucleus is probable only if their relative kinetic energy is comparable to or below the difference in the nuclear binding energies $E_B$ of the initial and final states; roughly, $E_B \sim A \times \mathcal{O}(\MeV)$. 
On the other hand, the threshold energy for production of nuclear anti-particles via spallation of a primary cosmic ray against the ISM is $E_{\text{th}} \sim m_{\bar{N}} \sim A \mu_a \sim A \times \mathcal{O}(\GeV)$.
Because primary cosmic-ray fluxes tend to be power laws as a function of energy (see, e.g., \citeR{AMS:2021nhj}), it follows that the kinetic energy of nuclear anti-particles produced by spallation of sufficiently energetic primary cosmic rays is also of typically $\mathcal{O}(\text{GeV})$, except very near threshold. 
The formation rates of heavier anti-nuclei via coalescence of such products therefore tend to suffer significant phase-space suppressions, leading to strong hierarchies between the expected numbers $N$ of anti-nucleus events with subsequently higher $A$ that would be observed at AMS-02: i.e., $N_{\bar{p}}\gg N_{\antiD}\gg N_{\antiHe{3}}\gg N_{\antiHe{4}}$. 

These conventional astrophysics predictions for anti-nucleus cosmic ray fluxes from spallation were quantified recently in \citeR{Poulin:2018wzu}, where the expected numbers of anti-nucleus events at AMS-02 were found to scale roughly as $N_{\bar{p}}\sim 10^{4} N_{\antiD}\sim 10^{8} N_{\antiHe{3}}\sim 10^{12} N_{\antiHe{4}}$, in line with the phase-space argument above.
For model parameters that reproduce the anti-proton flux observed by AMS-02, \citeR{Poulin:2018wzu} thus found that the predicted anti-helium fluxes are always orders of magnitude below AMS-02 sensitivity. 
Conversely, in order to reproduce, e.g., the tentatively identified $\antiHe{4}$ flux, one would have to overproduce, e.g., the observed anti-proton flux by many orders of magnitude. 

Similar challenges in producing anti-nuclei also occur in decaying or annihilating particle dark-matter (DM) models commonly studied in the context of indirect searches~\cite{Poulin:2018wzu,Korsmeier:2017xzj,Coogan:2017pwt,Carlson:2014ssa,Cirelli:2014qia}.
The anti-nucleus production rates in these models suffer from similar phase-space suppressions as for spallation of primary cosmic rays.%
\footnote{%
    The phase-space suppression argument made above, and the resulting hierarchy of anti-nucleus fluxes with higher $A$, should apply for any anti-nucleus production scenario that starts with high-energy ($E \gtrsim \GeV$) processes.} %
One should keep in mind however that there are considerable variations in the predicted anti-nucleus fluxes stemming from uncertainties in the parameters of the nuclear-coalescence model~\cite{Carlson:2014ssa,Cirelli:2014qia} and different choices of the cosmic-ray propagation model~\cite{Cholis:2020twh}. 
With optimistic assumptions~\cite{Cholis:2020twh,Coogan:2017pwt}, and possibly 
with enhancements from $\overline{\Lambda_b}$ decays~\cite{Winkler:2020ltd}, it has been found that it might be possible for annihilating particle dark matter to be the origin of the tentatively identified $\antiHe{3}$ flux.
However, even a single confirmed $\antiHe{4}$ event at AMS-02 would be challenging to explain.
See also \citeR{DeLaTorreLuque:2024htu} for a recent detailed investigation of these points.

If the candidate events are confirmed, explaining the presence of both the $\antiHe{3}$ and $\antiHe{4}$ events at AMS-02, with their comparable rates, seems to require the absence of the severe phase-space suppressions that inexorably lead to a strong hierarchical relation of the anti-nucleus fluxes. 
These suppressions can be ameliorated if the anti-nucleons that combined to form the anti-nuclei have low relative momenta, $\Delta p\ll \GeV$.
Various beyond-the-Standard-Model (BSM) scenarios have been proposed to achieve this.

\citeR[s]{Poulin:2018wzu,Arbey:2020ccg,Bykov:2023nnr} considered anti-nucleus production occurring in anti-matter--dominated regions of primordial origin that have cooled down significantly by the time of the anti-nucleus production. 
These scenarios require a BSM mechanism in the early Universe to cause the requisite matter--anti-matter segregation. 

Anti-nucleus production scenarios via the decay of a new particle carrying anti-baryon number (which may or may not be the dark matter) have been also considered. 
In \citeR{Heeck:2019ego}, the mass of the decaying particle was tuned to be very close to the mass of the desired anti-nucleus in order to restrict the final-state phase space, such that the produced particles are non-relativistic. 
Such a scenario however requires several new decaying particles, each with its own mass tuning to separately enhance the production of $\antiHe{4}$, $\antiHe{3}$, and perhaps $\antiD$. 
In \citeR{Winkler:2022zdu}, a strongly coupled dark sector was considered, where dark hadron showers triggered by the decay of a new particle simultaneously increase the multiplicity of the decay products and decrease their relative momenta $\Delta p$. 
The final decay products (i.e., the lightest dark bound states) then decay to SM anti-quarks which subsequently form anti-nuclei. 
A challenging aspect of this scenario is the need to model strong-coupling phenomena such as dark hadron showers.

Another challenging aspect of the tentatively identified events at AMS-02 is their relativistic nature (i.e., large Lorentz boosts).
Overcoming phase-space suppressions of nuclear coalescence rates by considering scenarios where the colliding particles have low relative momenta usually also results in anti-helium products that are non-relativistic.
There are however ways around this: the anti-helium products in scenarios that start with the decay of a particle, such as those considered in \citeR[s]{Heeck:2019ego,Winkler:2022zdu}, can be made relativistic if the decaying particle is already boosted in the galactic rest frame. 
This can be achieved if, e.g., the decaying particle is produced in turn from the earlier decay of another heavier particle. 
Other works~\cite{Poulin:2018wzu,Arbey:2020ccg,Bykov:2023nnr} have either suggested acceleration mechanisms based on supernova shockwaves (similar to the Fermi mechanism), or did not address this question.

In this paper, we propose a scenario where the anti-helium nuclei observed at AMS-02 originate from sudden and localized ``injections'' (of BSM origin) of anti-baryon number in our Galaxy in the form of energetic SM anti-quarks. 
These particles subsequently thermalize into relativistically expanding, optically thick fireballs with a net anti-baryon number. 
Anti-nuclei are produced in each of these fireballs thermally through a nuclear reaction chain similar to that operating during Big Bang Nucleosynthesis (BBN), albeit with remarkable qualitative and quantitative differences due to the very different (anti-)baryon--to--entropy ratios, timescales, and expansion dynamics involved. 
The evolution of the fireball plasma after the initial injection of SM particles is purely dictated by SM physics and, moreover, since this process involves thermalization, depends only on certain bulk properties being achieved by the injection, and for the most part not on the exact details of the latter. 
However, owing to the inefficiency of weak interactions, there are regions of parameter space where the anti-neutron--to--anti-proton ratio in the plasma at the onset of the anti-nucleosynthetic processes may depend on the details of the initial injection of SM particles.
With that one exception, the predicted anti-particle cosmic ray fluxes produced in our proposed scenario are therefore both predictive and largely agnostic as to the microphysical origin of the injections that seed these fireballs. 

Assuming that the requisite injections can occur, we show that this scenario could explain not only the tentatively identified anti-nucleus fluxes at AMS-02, but also the relativistic Lorentz boosts of the detected particles.
The seemingly paradoxical requirements of a low-energy environment to foster production of higher-$A$ anti-nuclei and the high energies required for relativistic anti-nucleus products are reconciled naturally in our scenario by the expansion dynamics of the fireball plasma: as the plasma expands, its temperature falls adiabatically while its thermal energy is converted to bulk kinetic energy by the work of its internal pressure.
This allows a low-energy environment for anti-nucleosynthesis to proceed, while at the same time accelerating its products to relativistic speeds.

Of course, injections of the requisite amounts of SM anti-particles with the correct properties to generate these fireballs cannot occur spontaneously: a BSM mechanism is required.
Suggestively, we show that collisions of certain very heavy, macroscopic, composite dark objects (possibly a sub-fraction of all of the dark matter) that carry SM anti-baryon number could at least achieve a high-enough rate of injections with appropriate parameters to explain the tentatively identified anti-helium events at AMS-02, provided that a substantial fraction of the dark objects' mass energy can be converted into SM anti-quarks as a result of dark-sector dynamics triggered by the collision (in some parameter regions, there may be a requirement to inject also a comparably sized asymmetry of charged leptons).
While this is encouraging, we have not yet developed a detailed microphysical model that achieves the necessary dark-sector dynamics in a way that is amenable to robust and controlled understanding; however, we offer some speculative initial thoughts on certain model constructions that we believe are promising avenues to explore toward that goal.
For the purposes of this paper, we ultimately leave this question open; as it is crucial to providing a concrete realization of the scenario we advance, however, we both intend to return to it in our own future work and we encourage other work on it.

The remainder of this paper is structured as follows.
In \secref{sec:AMSEvents}, we review the candidate anti-nuclei events observed by AMS-02.
In \secref{sec:FireballAntiNucleo}, we discuss synthesis of anti-nuclei in an expanding relativistic fireball, discussing first questions of thermalization after energy injection [\secref{sec:thermalization}], then turning to the fireball expansion dynamics [\secref{sec:expansionDynamics}] and its termination at the point of photon decoupling [\secref{sec:photonDecoupling}], and the nuclear reactions at play during the expansion [\secref{sec:antinucleosynthesis}], before summarizing [\secref{sec:summary}].
In \secref{sec:prop-detection}, we then discuss how the anti-nuclei thus produced would propagate to AMS-02 and give our scenario's projections for the anti-nuclei spectra and event rates [\secref{sec:Galprop}]; we also discuss other potential observables [\secref{sec:bursts}].
In \secref{sec:DMseedModel}, we discuss a potential origin for the fireballs in collisions of composite dark-matter states, showing first that the rates could work [\secref{sec:BlobCollRates}], that the fireballs could be appropriately seeded if certain benchmarks can be met [\secref{sec:fireballMapping}], and then offering some speculative thoughts toward particle physics models that may be worth further investigation to see it they are able to achieve the necessary conditions [\secref{sec:DMmodelIdeas}].
We conclude in \secref{sec:DiscussionConclusion}.
A number of appendices add relevant detail.
\appref{app: fireballscalings} gives derivations of various scaling laws for fireball expansion that we rely on in the main text.
\appref{app:antinucleonFO} discusses the ratio of anti-nucleons from which the anti-nucleosynthesis is initiated.
\appref[ces]{app:reactionNetwork} and \ref{app:nuclearCrossSections}, respectively, give details of the nuclear reaction networks and cross-sections we have used.
\appref{app:dof} discusses whether dynamical changes in the number of degrees of freedom during fireball expansion are relevant.
\appref{app:BurstWind} discusses prompt versus slow injections.
Finally, \appref{app:AMSantiHeSensitivity} reviews an estimate of the AMS-02 rigidity-dependent sensitivity to anti-helium events.

\section{The AMS-02 candidate anti-helium events}
\label{sec:AMSEvents}

In this section, we summarize the data that is publicly available~\cite{TingCERNslides2016,TingCERNslides2018,SchaelCOSPARslides,OlivaSlides2019,ZucconMIAPbPslides,TingCERNslides2023} regarding the AMS-02 candidate anti-helium and anti-deuterium events.

\begin{table*}[!t]
\begin{ruledtabular}
    \begin{tabular}{lp{3.75cm}llllll}
    \#  & Event Date [mm/dd/yyyy] (Day of Year)   &   $p$ [GeV]   &   $m$ [GeV]   &   $Q/e$   &   $\Gamma$    &   $\mathcal{R}$ [GV]    &   Ref(s).    \\ \hline \hline
    1 & 09/26/2011 (269)&   $33.1(1.6)$    &   $2.93(36)$    &   $-1.97(5)$   &   $11.3(1.5)$    &   $-16.80(92)$   &   \cite{TingCERNslides2018,SchaelCOSPARslides} \\
    2 & 12/08/2016 (---)\footnote{No later than} & $40.3(2.9)$    &   $2.96(33)$    &   $-2$(---)      &   $13.6(1.3)$    &   $-20.2(1.5)$   &   
    \cite{TingCERNslides2016,TingCERNslides2018}\\ 
    3 & 06/22/2017 (173)& $32.6(2.5)$    &   $3.81(29)$    &   $-2.05(5)$   &   $8.61(92)$     &   $-15.9(1.3)$   &    \cite{TingCERNslides2018,OlivaSlides2019,TingCERNslides2023} \\ \hline
    4 & 09/20/2022 (265) & ---        &    $3.15(53)$  &   $-2$    &   --- &   --- &   \cite{TingCERNslides2023}
    \end{tabular}
    \end{ruledtabular}
    \caption{\label{tab:events}%
        Parameters for individual candidate anti-helium events displayed in the identified references: $p$ is momentum, $m$ is mass, $Q$ is charge ($e>0$ is the elementary charge), $\Gamma$ is the Lorentz factor, and $\mathcal{R}=p/Q$ is rigidity.
        For identification purposes across the references, the reported event date and corresponding day of year are given (the date of event 2 is not given in the references; we give the date of its first public presentation in \citeR{TingCERNslides2016}).
        Unavailable data are denoted by ``---''.
        We have set $\hbar=c=1$.
        Data for $p$, $m$, and $Q/e$ are given in the references [speed data, $v = 0.9973(5)$, is additionally given for event 2]; we derived $\Gamma = \sqrt{1+(p/m)^2}$ [or $\Gamma = 1/\sqrt{1-v^2}$ for event 2] and $\mathcal{R} = p/Q$ and propagated uncertainties na\"ively.
        For comparison, $m_{\antiHe{3}} = 2.81\,\GeV$ and $m_{\antiHe{4}} = 3.73\,\GeV$.
        Event~\#~3 is clearly tentatively identified in \citeR[s]{TingCERNslides2018,TingCERNslides2023} as a candidate $\antiHe{4}$ event.
        Event \# 4, which is similarly identified as an $\antiHe{4}$ candidate in the oral presentation of \citeR{TingCERNslides2023}, is not included in our analysis; see discussion in text.
    }
\end{table*}

To our knowledge, the most recent scientific presentations on these candidate events are \citeR[s]{ZucconMIAPbPslides,TingCERNslides2023}.
\citeR{ZucconMIAPbPslides} provides graphical mass and rigidity histogram data for 9 candidate events collected in $\sim 10 \ \text{years}$ of AMS-02 observations.
\citeR{TingCERNslides2023} also appears to show an additional anti-helium candidate event dated after \citeR{ZucconMIAPbPslides}, with a mass measurement consistent with either $\antiHe{3}$ or $\antiHe{4}$, but closer to the former.
\tabref{tab:events} shows the detailed basic parameter values for 3 of the 9 candidate events that have been shown publicly~\cite{TingCERNslides2016,TingCERNslides2018,SchaelCOSPARslides,OlivaSlides2019,TingCERNslides2023}, as well as the data available for the additional candidate event~\cite{TingCERNslides2023}; such detailed data has not been presented for the other candidate events.

We base our analysis on the data for the 9 candidate events presented in \citeR{ZucconMIAPbPslides}, which are consistent with the identification of 6 candidate $\antiHe{3}$ events, and 3 candidate $\antiHe{4}$ events in $T\sim 10\,\text{years}$ of AMS-02 data (see discussion about $T$ below).%
\footnote{\label{ftnt:EarlierData}%
    This identification is also consistent with earlier presentations based on a smaller dataset with a total of 8 candidate anti-helium events~\cite{TingCERNslides2018,OlivaSlides2019}, of which 2 were tentatively identified as $\antiHe{4}$ candidates~\cite{TingCERNslides2018,SchaelCOSPARslides}, giving a 3:1 ratio for $\antiHe{3}:\antiHe{4}$.}\textsuperscript{,}%
\footnote{\label{ftnt:newEvent}%
    We do not explicitly consider the ``additional'' event shown in \citeR{TingCERNslides2023} (event \# 4 in \tabref{tab:events}).
    This event is labeled in the slide deck for \citeR{TingCERNslides2023} to have occurred on September 20, 2022, which is after the date of presentation of the histograms in \citeR{ZucconMIAPbPslides} on February 28, 2022; it is thus highly likely to be a new event not previously discussed in past references before \citeR{TingCERNslides2023}.
    However, at the level of uncertainty regarding these events that we work in this paper, whether or not we include this event has almost no relevant impact on our discussion of the overall or relative event rates for the $\antiHe{3}$ and $\antiHe{4}$ events.
    Assuming either that $T\sim 8.5\,\text{years}$ or $T\sim 10\,\text{years}$ for the data in \citeR{ZucconMIAPbPslides} (see discussion in main text), it is also entirely consistent within statistical errors for one additional event to occur in the additional integration time of relevance for the $T\sim 11\,\text{years}$ of data that appears to be discussed in \citeR{TingCERNslides2023}.} %
The mass histogram shown in \citeR{ZucconMIAPbPslides} is however more than broad enough to support an inference that some of the candidate $\antiHe{3}$ events thus classified could in fact be $\antiHe{4}$ events, and vice versa (the additional event shown in \citeR{TingCERNslides2023} could also be identified as either isotope within the uncertainties).
Nevertheless, throughout this paper, we adopt a $\sim 2:1$ event ratio for $\antiHe{3}$ vs.~$\antiHe{4}$ as fiducial. 

Of the candidate events in \citeR{ZucconMIAPbPslides}, 3 have rigidities $\mathcal{R}=p/Q$ (where $p$ is the particle momentum and $Q=qe$ its charge) in the approximate range $-40\,\text{GV}\lesssim \mathcal{R} \lesssim -35\,\text{GV}$, while the other 6 candidate events have rigidities in the range $-25\,\text{GV} \lesssim \mathcal{R} \lesssim -15\,\text{GV}$.
The data in \tabref{tab:events} however make clear that care should be taken \emph{not} to associate the 3 candidate events with larger $|\mathcal{R}|$ shown in the histogram in \citeR{ZucconMIAPbPslides} with the 3 candidate $\antiHe{4}$ events and the remainder with the 6 candidate $\antiHe{3}$ events: indeed, event \# 3 in \tabref{tab:events} has a rigidity in the low range, but a mass most consistent with $\antiHe{4}$.
One robust inference however is that all of the candidate events for which the necessary data are available to make this determination are highly relativistic, with $\Gamma \sim 10$ being a typical fiducial value (an extreme range of possible Lorentz factors for the other candidate events based on the rigidity and mass data shown in \citeR{ZucconMIAPbPslides} is roughly $6 \lesssim \Gamma \lesssim 36$, assuming $q=-2$).

Given the $\sim 1.3\times 10^8$ confirmed helium events in the AMS-02 data as of \citeR{ZucconMIAPbPslides}, the ratio of the 9 candidate anti-helium events in that reference to the confirmed helium events is approximately $7 \times 10^{-8}$.
Likewise, \citeR{TingCERNslides2023} reports $\sim 1.45\times 10^8$ confirmed helium events; with 10 total anti-helium events, this yields the same ratio within uncertainties.

Finally, for the purposes of converting total candidate anti-helium event numbers into rate estimates, we assume that the relevant AMS-02 data-taking period over which all 9 events discussed in \citeR{ZucconMIAPbPslides} were observed is $T=10\,\text{years}$.
There is some uncertainty on this given available information: the data-taking period of relevance to the 9 events reported as of \citeR{ZucconMIAPbPslides} may actually be slightly shorter, $T\approx 8.5\,\text{years}$ (this point is not made unambiguously clear in \citeR{ZucconMIAPbPslides}).
There is thus an uncertainty on the required rates of $\mathcal{O}(15\%)$ arising from the incompleteness of the publicly available information on this point; this is however significantly smaller than the uncertainty on the rates owing to the small statistics of the relevant event samples.

Additionally, \citeR{TingCERNslides2023} makes a new report of 7 candidate anti-deuterium ($\antiD$) events in $T\sim 11\,\text{years}$ of AMS-02 data. 
However, only scant information is available regarding these events: a single histogram showing that the charge-sign--mass product for these events lies in the range $-2.1 \leq \text{sgn}[q]\times (m / \text{GeV}) \leq -1.8$.

\section{Fireball anti-nucleosynthesis}
\label{sec:FireballAntiNucleo}

In this section, we show how an abrupt localized injection in a region of space of a large amount of energy and anti-baryon number in the form of SM anti-quarks can lead to formation of a locally thermalized \textit{fireball} comprised of a plasma mixture of free anti-nucleons, pions, leptons, and photons.
As this fireball expands hydrodynamically~\cite{Piran:1999kx,Goodman:1986az,Paczynski:1986px,Shemi:1990rv,paczynski1990super,Meszaros:2006rc}, it cools adiabatically, eventually permitting Standard Model anti-nucleosynthetic processes to produce bound anti-nuclei, including anti-helium, in the hot and dense environment.
Owing to the dynamics of the expansion in the interesting region of parameter space, the radial bulk expansion velocity of the fireball constituents also becomes relativistic by the time of anti-nucleosynthesis, resulting in the anti-helium thus produced being released into the interstellar medium relativistically (assuming the fireball is located within our galaxy).
We also demonstrate that the fireball expansion shuts off anti-nucleosynthetic processes prior to the attainment of complete nuclear statistical equilibrium, thereby allowing the amount of $\antiHe{3}$ produced, after decays of unstable products, to be larger than the amount of $\antiHe{4}$.

The thermalized nature of the initial fireball caused by the requisite injection of energy and anti-baryon number conveniently erases most of the history of how such a thermal state comes to exist. 
As such, the conclusions we reach in this section as to the anti-nuclear outputs of the fireball expansion are largely independent of the model details of how such a fireball comes to exist; instead, they depend only on bulk physical properties of the fireball state, such as its temperature, anti-baryon--to--entropy ratio, and radius at certain critical points in its evolution. 
The exception to this is that, due to incomplete thermalization via inefficient weak interactions, the results we obtain can depend on the net charge on the hadronic sector that is injected primarily via the BSM process that seeds the fireball; this essentially becomes another parameter we must consider.

Of course, this history-erasure does not alleviate the requirement for a concrete BSM mechanism by which the requisite energy and anti-baryon number injection could occur; we comment on this aspect of the problem in \secref{sec:DMseedModel} but ultimately defer this to future work.

In the remainder of this section, we discuss: the thermalization following energy and anti-baryon number injection [\secref{sec:thermalization}], the dynamics of the fireball expansion [\secref{sec:expansionDynamics}], and the anti-nucleosynthetic and other outputs of the fireball expansion [\secref{sec:antinucleosynthesis}].
We summarize in \secref{sec:summary}.

\subsection{Thermalization}
\label{sec:thermalization}

We will be mainly concerned with SM fireballs whose initial size is of order $10^{-4}\,\text{m}\lesssim R_0\lesssim 1\,\text{m}$, and which double in size on characteristic timescales $\tau$ of order $10^{-13}\,\text{s} \lesssim \tau_0 \lesssim 10^{-9}\,\text{s}$. The injected particles will interact through various Standard Model processes which overall tend to bring themselves toward local thermal equilibrium. 

For simplicity, we limit ourselves to the parameter space where the injected SM energy density is such that the would-be temperature of the SM plasma is below the QCD scale,%
\footnote{\label{ftnt:higherTemp}%
    At higher energy densities, the plasma would lie above the QCD phase transition, a quark-gluon plasma (QGP) would form, and one would need more careful treatment of the evolution of the fireball back through the phase transition as it cools.
} %
$\sim 200\,\MeV$.
The path toward thermalization in this regime will involve a process of hadronization where the injected anti-quarks confine and fragment into anti-nuclei and copious pions. 
The rate of hadronization is set by the QCD scale $\Lambda_{\rm QCD}\sim 200\,\MeV\sim \left(3\times 10^{-24}\,\text{s}\right)^{-1}$ in the rest frame of the confining or fragmenting particles and will be time-dilated at some level in the center of mass frame of the injected gas of particles. Unless the initial Lorentz factors of the injected anti-quarks are extremely high, $\gamma_*\gtrsim 10^{11}$ (corresponding to energies $E\gtrsim 10^{10}\,\GeV$), a case we do not consider, hadronization occurs essentially instantaneously compared to the initial expansion timescale of the fireball, $\tau_0\gtrsim 10^{-13}\,\text{s}$.

The subsequent thermalization should proceed similarly to that in analogous setups involving hadrons found in the contexts of heavy-ion colliders~\cite{Letessier:2002ony,Noronha-Hostler:2009wof,Noronha-Hostler:2007fzh} and in the early Universe before~\cite{Jaikumar:2002iq,Fromerth:2002wb} and during~\cite{Pospelov:2010cw,Kawasaki:2004qu} BBN. 
A proper description of the thermalization process would require solving a complex set of Boltzmann equations dictating the time-evolution of the energy spectra of relevant particles and resonances. 
We instead provide rough estimates of the typical rates of the processes involved. 
The typical rates of strong interactions involving pions (e.g., $\bar{n}+\pi^-\rightarrow \bar{p}+\pi^0$), $\Gamma_{\rm strong}$; electromagnetic (EM) interactions for charged particles that are relativistic at a given temperature (e.g., $\gamma\gamma\leftrightarrow e^+ e^-$), $\Gamma_{\rm EM}$; and weak interactions for relativistic particles (e.g., $e^+e^-\leftrightarrow \bar{\nu}_e\nu_e$), $\Gamma_{\rm weak}$, are respectively given by
\begin{align}
    \Gamma_{\rm strong}&\sim \left(\frac{m_\pi T}{2\pi}\right)^{3/2}e^{-m_{\pi}/T}\left<\sigma_{\rm strong}v\right>\nonumber\\
    &\sim (1\times 10^{-20}\,\text{s})^{-1} \left(\frac{T}{100\,\MeV}\right)^{3/2}e^{-m_{\pi}/T} \label{eq:Gammastrong}\:, \\
    \Gamma_{\rm EM}&\sim \alpha_{\rm EM}^2 T\sim (1\times 10^{-19}\,\text{s})^{-1}\left(\frac{T}{100 \,\MeV}\right)\:, \\
    \Gamma_{\rm weak}&\sim G_{F}^2T^5\sim \left(5\times 10^{-10}\,\text{s}\right)^{-1} \left(\frac{T}{100\,\MeV}\right)^5  \:,  
\end{align}
with $m_\pi \approx 140\,\MeV$ and $\left<\sigma_{\text{strong}}v\right>\sim 1\,\text{mb}$~\cite{Kawasaki:2004qu,Boyarsky:2020dzc}. 
These rates suggest that within the timescale of expansion of the initial fireball $\tau_0\gtrsim 10^{-13}\,\text{s}$, the SM particles can interact efficiently through strong and electromagnetic processes, but generically not through weak processes, unless the temperature is significantly higher than $100\,\MeV$. 

In the cases that we consider, strong and electromagnetic processes are sufficient for thermally populating all SM particle species with masses below a given temperature.
For instance, anti-protons can be created from charged pions and anti-neutrons through strong interactions, photons can be produced via bremsstrahlung from charged pions and the decay of neutral pions, and leptons pairs can be produced in photon annihilations.

However, if the weak interactions are indeed always inefficient after SM particle injection, charge would need to be separately conserved in the leptonic and hadronic sectors, because reactions such as $\pi^+ + e^- \leftrightarrow \pi^0 + \nu_e$ (and similar crossed or charge conjugated reactions, as well as similar reactions with the anti-baryons) that would allow charge to be exchanged between those two sectors would not be efficient.
Therefore, were the initial injection of particles to be such that the the net charge is zero in both sectors (e.g., only net charge- and color-neutral combinations of anti-quarks are injected), the charged pions would have a chemical potential that fixes their population asymmetry to the number of anti-protons.
If the plasma temperature falls, this may have what we will see to be undesirable consequences for our scenario, such as a dramatic depletion of the anti-proton abundance when the symmetric, thermal pion abundance becomes Boltzmann suppressed while strong interactions between the asymmetric $\pi^+$ abundance and the $\bar{p}$ population remain efficient for some time.%
\footnote{%
    One possible injection-model-dependent solution to this issue is an initial injection of SM particles that is still net color- and charge-neutral, but which in addition to having a large net negative baryon number, also has a large \emph{charged lepton asymmetry} and therefore baked-in opposite-sign EM charge asymmetries in the hadronic and leptonic sectors that cannot be removed by the inefficient weak interactions.} %
We discuss this issue in more detail in \secref{ss:antineutronabundance} and \appref{app:antinucleonFO}; see also comments in \secref{sec:DMmodelIdeas}.

Immediately upon the completion of (partial) thermalization, the fireball energy is dominated by radiation comprised of a subset of photons and relativistic $e^{\pm}$, $\mu^{\pm}$, and $\pi^{0,\pm}$ (depending on the temperature); its mass is dominated by anti-baryons in the form of free $\bar{n}$ and $\bar{p}$. 
The thermal pressure of the trapped radiation drives an adiabatic expansion of the fireball, which can be described hydrodynamically owing to the short mean free path $\ell$ of the constituent particles, $\ell\sim v/\Gamma_{\rm \{strong,\, EM\}}<1/\Gamma_{\rm \{strong,\, EM\}}\ll R_0$.
This is the topic of the next subsection.

\subsection{Relativistic fireball expansion}
\label{sec:expansionDynamics}

We now turn to considering the dynamics of the expanding thermalized plasma.

We treat the plasma as a spherically symmetric perfect fluid and consider only its radial expansion.
In what follows, quantities defined in the comoving rest-frame of the radially moving fluid will be marked with a prime $'$, while those defined in the fireball center-of-mass (c.m.) frame are unprimed.

If $t$ is the CoM time co-ordinate and $r$ is the CoM-frame radial coordinate centered on the fireball, then the radial expansion of the fireball is described by two relativistic fluid equations that arise from the covariant conservation of the energy-momentum tensor for a perfect fluid assuming spherical symmetry: $\nabla_\mu T^{\mu\nu} = 0$, where $T^{\mu\nu} = (\rho'+p')v^\mu v^\nu - p' g^{\mu\nu}$ with $v^\mu = \gamma(1,v,0,0)$ for radial fluid flow ($\gamma \equiv 1/\sqrt{1-v^2}$) and $g_{\mu\nu} = \text{diag}[1,-1,-r^2,-r^2\sin^2\theta]$ is the metric for spherical co-ordinates $(t,r,\theta,\phi)$.
Respectively, the $\nu = 0$ equation encodes energy conservation and the $\nu = r$ equation encodes radial momentum conservation of the fireball fluid~\cite{Piran:1993jm, Bisnovatyi-Kogan:1995cyk}:
\begin{align}
    \partial_t\left[\gamma^2(\rho'+p')\right]+\frac{1}{r^2}\partial_r\left[r^2\gamma^2v\left(\rho'+p'\right)\right]&=\partial_t p' \:, \label{eq:FluidEqn1}\\
    \partial_t\left[\gamma^2v(\rho'+p')\right]+\frac{1}{r^2}\partial_r\left[r^2\gamma^2v^2\left(\rho'+p'\right)\right] &=-\partial_rp'\:, \label{eq:FluidEqn2}
\end{align}
where $\gamma(t,r) \equiv 1/ \sqrt{ 1 - [v(t,r)]^2}$ is the Lorentz factor associated with the CoM-frame bulk radial velocity $v(t,r)$ of the fluid at a given radius $r$, and $\rho'(t,r)$ and $p'(t,r)$ are the comoving energy density and pressure of the fluid.

While \eqref[s]{eq:FluidEqn1} and (\ref{eq:FluidEqn2}) are not analytically solvable, it is understood that the fireball will undergo an initial phase of rapid acceleration to relativistic speeds~\cite{Piran:1993jm,Kobayashi:1998faa}, whereupon analytically tractable evolution takes over. 
The acceleration is initially limited to a thin shell near the surface of the fireball where the pressure gradient is strong. 
This surface expansion then generates inward-traveling rarefaction waves which accelerate the bulk of the fireball. 
Numerical simulations~\cite{vitello1976hydrodynamic, Greiner:1996md,Yaresko:2010xe} (see also \citeR[s]{Fiorillo:2023cas,Fiorillo:2023ytr}) suggest that the radial layers comprising an initially static thermal fireball will accelerate to relativistic radial velocities within the time it takes for sound waves (whose speed in a radiation-dominated fluid is $c_s\approx 1/\sqrt{3}$) to cover the initial radius $R_0$ of the fireball.
Upon arrival at the center of the of the fireball, the waves get reflected outward, creating a strong underdensity at that point. 
This underdensity creates a hollow structure at the center of the fireball that becomes more pronounced over time, essentially turning the thermal plasma into a radially moving shell of density concentration with an initial thickness $\sim R_0$. 
The radial bulk velocities $v$ of all the radial layers of the plasma shell soon approach the speed of light $v\approx 1$ and the resulting nearly flat velocity profile keeps the shell thickness $\sim R_0$ constant until much later times.

Once the radial layers of the plasma shell are moving relativistically with Lorentz factors $\gamma\gtrsim\text{few}$, its subsequent evolution follows simple scaling laws, which can be parameterized by the initial temperature $T_0$ and the initial radius (and thickness) $R_0$ when $\gamma\sim\text{few}$, as well as the (conserved) anti-baryon--to--entropy ratio $\bar{\eta}$ of the plasma shell~\cite{Piran:1993jm,Kobayashi:1998faa}.%
\footnote{%
    Because the plasma temperature $T'$ will be related to the shell radius $R$ through simple scaling laws, the precise definitions of $T_0$ and $R_0$ beyond what we have described are not important.
    One can more generally define $T_0$ as the temperature when the radius of the shell is $R_0$ (or vice versa), where the reference $R_0$ (or $T_0$) is arbitrarily chosen. 
    So long as the reference point is sufficiently early in the shell evolution that the energy of the plasma is still dominated by radiation, the specific choice of reference point is unimportant. 
    For simplicity, in this paper we choose the reference point to be when $\gamma\sim\text{few}$.} %
Note that $T_0$ and $\bar{\eta}$ naturally have $\mathcal{O}(1)$ radial variation over the shell; however, thanks to the smoothing effect of sound waves, this variation is typically mild~\cite{Piran:1993jm}. 
For simplicity, our parametric estimates will assume a single (average) value of $T_0$ and $\bar{\eta}$ for the entire shell. 
We will however take into account the $\Delta \gamma/\gamma \sim 1$ radial variation of the Lorentz factor between the inner and outer radius of the shell because, as we will see, it has an important consequence at the late stages of the shell evolution; where relevant, we assume that $\gamma$ varies monotonically across the shell, being larger on the outer edge of the shell.

The expansion in the relativistic regime proceeds in three stages (see \appref{app: fireballscalings} for derivations):
\begin{enumerate}
    \item \textit{Acceleration.}~As long as the shell energy density is radiation-dominated, the internal pressure of the radiation continually accelerates the radial velocity of the shell, converting radiation energy into anti-baryon kinetic energy in the process. 
    In this phase, the shell thickness in the center-of-mass frame of the fireball $\Delta R$ remains approximately constant, $\Delta R\sim R_0$, and the bulk Lorentz factor scales with the radius of the shell $R$ as $\gamma\propto R$.
    See also \citeR[s]{Diamond:2023cto,Diamond:2023scc}.
    \item \textit{Coasting.}~Once the radiation energy drops below the anti-baryon kinetic energy, the radiation can no longer accelerate the shell appreciably and from then on the shell simply coasts at its terminal Lorentz factor
    \begin{align}
        \Gamma\sim \frac{T_0}{\bar{\eta} m_{p}} \:,
    \end{align}
    where $m_p$ is the proton mass.
    \item \textit{Spreading.}~The (assumed monotonic) variation of the Lorentz factor $\Delta \gamma/\gamma\sim 1$ between the inner and outer radii of the shell translates to velocity variation $\Delta v\sim \Delta\gamma/\gamma^3\sim 1/\gamma^2$. This leads to increasing shell thickness which becomes important when $R\gtrsim \Gamma^2R_0$ and is well captured by
    \begin{align}
        \Delta R\sim \left(1+\frac{1}{\gamma^2}\frac{R}{R_0}\right)R_0\:.  \label{eq:thicknessvsR}
    \end{align}
\end{enumerate}
To sum up, the Lorentz factor of the shell scales as
\begin{align}
    \gamma(R)\sim\begin{cases}
    R/R_0, &R\lesssim\Gamma R_0\:; \\
    \Gamma, &R\gtrsim\Gamma R_0 \label{eq:gammavsR}
    \end{cases} \:.
\end{align}
Neglecting changes in the number of degrees of freedom $g_*$ as various species (pions, muons, electrons, positrons) fall out of thermal equilibrium and get Boltzmann suppressed,%
\footnote{\label{ftnt:gStar}%
    This amounts to setting $g_*=2$ (from photons) throughout. 
    We discuss in \appref{app:dof} how accounting for changes in $g_*$ yields only mild, $\mathcal{O}(1)$ quantitative changes to the picture presented here, but leaves the qualitative evolution unchanged.
    } %
the average comoving temperature $T^\prime$ of the shell scales as 
\begin{align}
    T^\prime(R)\sim\begin{cases}
    T_0\left(R_0/R\right), &\phantom{i\Gamma R_0<}R\lesssim\Gamma R_0\:; \\
    (T_0/\Gamma)\left(\Gamma R_0/R\right)^{2/3}, &\Gamma R_0\lesssim R\lesssim\Gamma^2R_0\:; \\
    (T_0/\Gamma^{5/3})\left(\Gamma^2R_0/R\right), &\phantom{i\Gamma R_0<}r\gtrsim\Gamma^2R_0
    \end{cases}\:. \label{eq:Tprime}
\end{align}
These scaling laws are in agreement with the numerical simulations in \citeR[s]{Meszaros:1993cc, Piran:1993jm, Kobayashi:1998faa}.

\begin{figure}
    \centering
     \includegraphics[width=0.45\textwidth]{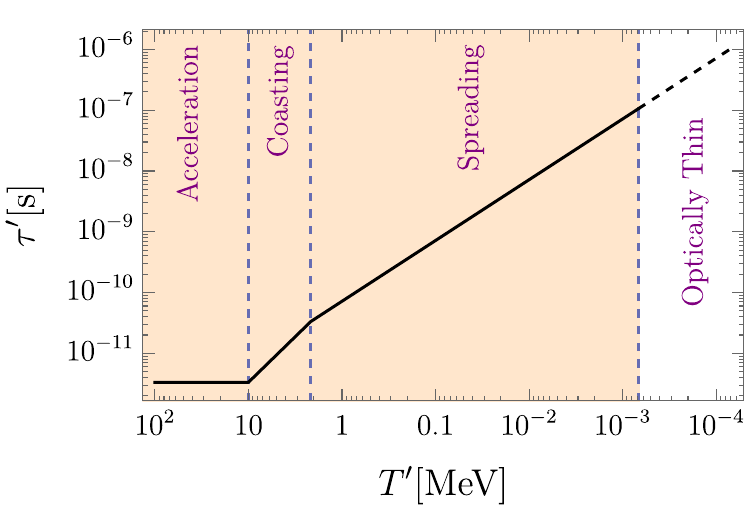}\\
    \caption{\label{fig:benchexp}%
        The comoving expansion timescale $\tau'$  (solid black line) as a function of the comoving temperature $T'$ [cf.~\eqref{eq:tauprime}] of a fireball with $T_0=100 \,\MeV$, $R_0=1\,\text{mm}$, and $\bar{\eta}=10^{-2}$. 
        The three vertical dashed lines mark, from left to right, the transition at $T'=T_0/\Gamma$ from the \textit{acceleration} phase to the \textit{coasting} phase, the transition at $T'=T_0/\Gamma^{5/3}$ from the \textit{coasting} phase to the \textit{spreading} phase, and the point at which photon decoupling occurs [corresponding to $R=R_{\rm thin}$; cf.~\eqref{eq:Rthin} with $\varepsilon n_{\bar{p}}\sim n_{\bar{B}}$].
        The orange shaded region is thus the region in which the plasma expands as a single tightly coupled fluid through its various expansion phases (as annotated), while the unshaded region where is the region where the plasma has decoupled; see \secref{sec:photonDecoupling}.
    }
\end{figure}

Note that the co-moving dynamical expansion timescale (i.e., $e$-folding timescale) for the fireball, as measured by the change in its comoving temperature,%
\footnote{%
    This turns out to be a convenient measure of fireball expansion for our later purposes; see also \appref{app:reactionNetwork} at \eqref{eq:epsilonnuc}.} %
can then be approximated as
\begin{align}
    \tau^\prime(T')&\equiv \frac{T'}{\gamma \left|dT'/dt\right|}\nonumber\\
    &\sim     \begin{cases}
    \displaystyle R_0, &\displaystyle  \phantom{i\frac{T_0}{\Gamma^{5/3}}\lesssim}T'\gtrsim \frac{T_0}{\Gamma}\:; \\[3ex]
    \displaystyle R_0\left(\frac{T_0/\Gamma}{T'}\right)^{3/2}, &\displaystyle  \frac{T_0}{\Gamma^{5/3}}\lesssim T'\lesssim \frac{T_0}{\Gamma}\:; \\[3ex]
    \displaystyle \Gamma R_0\left(\frac{T_0/\Gamma^{5/3}}{T'}\right), &\displaystyle  \phantom{i\frac{T_0}{\Gamma^{5/3}}\lesssim}T'\lesssim \frac{T_0}{\Gamma^{5/3}}\: ,
    \end{cases}  \label{eq:tauprime}
\end{align}
where the three cases here map to the three different expansion regimes for the fireballs discussed above.
We display the evolution of this timescale for a set of benchmark parameters that will be of interest (see \secref{sec:antinucleosynthesis}) in \figref{fig:benchexp}.

The evolution discussed here is valid so long as photons remain tightly coupled to the plasma; we turn to this topic in \secref{sec:photonDecoupling}.

\subsection{Photon decoupling}
\label{sec:photonDecoupling}

The fluid evolution described in \secref{sec:expansionDynamics} applies as long as the photons remain tightly coupled with the charged particles in the plasma.
Eventually, this assumption is violated and the photons decouple. 
In this section, we estimate when this occurs.

Photons in the plasma scatter dominantly with the electrons and positrons. 
In our parameter space of interest, we will show that it is the case that photon decoupling always occurs at a temperature well below the electron mass, $T' \ll m_e$, after any symmetric thermal population of charged leptons have annihilated away. 
The opacity of the plasma is then due to the remaining charged leptons that must exist to guarantee net neutrality of the plasma; while their identity depends on some details of the BSM injection process, residual positrons are most important to this estimate.%
\footnote{%
    As we discuss in the next section, typical expansion timescales are too short for any muon asymmetry that is present to be guaranteed to decay by the time of this decoupling. 
    Because of the $\sigma_T \propto m^{-2}$ scaling however, residual muons are not relevant to our estimates unless $\varepsilon \lesssim (m_e/m_\mu)^2 \sim 2\times 10^{-5} \ll 1$.} %
Let us therefore parametrize the population of residual positrons available for the photons to scatter from as having a CoM-frame density $n_{e^+} = \varepsilon n_{\bar{p}}$, where $\varepsilon$ can be a BSM-injection dependent parameter, but which we typically expect to be $\varepsilon\sim \mathcal{O}(1)$.
Note that our estimates will therefore conservatively underestimate the population of light charged particles available to scatter if they are na\"ively extended to $T' \gtrsim m_e$.

While all of the light charged particles are of course initially highly relativistic $(T' \gg m_e)$ upon thermalization in the comoving fireball frame (i.e., long before photon decoupling), because we will show that the fireball becomes optically thin only when $T' \ll m_e$, it will be appropriate in this section (when working in the comoving fireball frame) for us to self-consistently use the Thompson cross section $\sigma_T = (8\pi/3) ( \alpha^2/m_e^2)$ as the relevant scattering cross-section for photons from the residual positrons around the time of photon decoupling. 
In fact, we will also employ $\sigma_T$ as the relevant cross section even when na\"ively extending these results to $T' \gtrsim m_e$; the optical depth we derive should therefore not be understood to be accurate for $T' \gtrsim m_e$ except insofar as it indicates an optical depth $\kappa \gg 1$ (i.e., an opaque fireball), which qualitative conclusion we do not expect would be modified were we to instead use the full Klein--Nishina cross-section.

Now, consider a photon moving in the CoM of the fireball at an angle $\theta$ relative to the radial direction. 
If $T'\ll m_e$, its mean free path in the comoving frame of the fireball plasma is $\ell'=\left(\epsilon n_{\bar{p}}^\prime \sigma_T\right)^{-1}$.
This can be related via a Lorentz transformation to the mean-free path $\ell$ in the CoM frame as $\ell'=\gamma\ell\left(1-v\cos\theta \right)$. 
Solving for $\ell$ and using $n_{\bar{p}}=\gamma n_{\bar{p}}^\prime$, we find $\ell=\left[\epsilon n_{\bar{p}}\sigma_T(1-v\cos\theta)\right]^{-1}$. 
The optical depth for a photon emitted from a radius $r_i$ inside the shell and escaping to infinity is thus given by
\begin{align}
    \kappa(r_i)=\int \frac{ds}{\ell}=\int_{r_i}^\infty dr\frac{\epsilon n_{\bar{p}}(r)\sigma_T(1-v\cos\theta)}{\cos\theta}
\end{align}
where $ds=dr/\cos\theta$ is the displacement of the photon in the CoM frame. 
Note that $\cos\theta$ is in principle radius-dependent; however, we can assume $\cos\theta\approx 1$ for the following reason. 
Due to relativistic beaming, most of the thermal photons are concentrated within $\theta\lesssim \gamma^{-1}\ll 1$ in the CoM frame, yielding $1-\cos\theta\sim \gamma^{-2}\ll 1$ and $1-v\cos\theta\approx 1-v+v\theta^2/2\approx (\gamma^{-2}+\theta^2)/2\sim \gamma^{-2}\ll 1$.
Therefore, we can take $\cos\theta \approx 1$ and, up to a numerical factor that is not important for this parametric estimate, approximate $1-v\cos\theta \sim 1-v$.

The optical depth for a typical photon in the shell emitted from $r_i\sim R$ (when $T' \ll m_e$) is then~\cite{2013ApJ...772...11R}
\begin{align}
    \kappa(r_i\sim R)& \sim\int_{r_i\sim R}^{\infty} \varepsilon n_{\bar{p}}(r)\sigma_T (1-v)dr \:.  \label{eq:kappa1}
\end{align}
The antiproton density $n_{\bar{p}}(r)$ appearing in this expression needs to understood with some caution: it is the density experienced \emph{by the photon} when \emph{it} is located at CoM radial co-ordinate $r$, which differs from the density of the antiprotons as a function of the CoM radial co-ordinate $r$ when evaluated at a fixed instant of time; we denote that latter density here as $n_{\bar{p}}(r,R)$, using the fireball radius in the CoM frame $R$ as a proxy for time.
For the purposes of this estimate, suppose that $n_{\bar{p}}(r,R)$ is approximately constant over the interval $R\leq r \leq R+\Delta R(R)$, where $\Delta R(R)$ is the $R$-dependent thickness of the fireball shell when the inner edge of the fireball has radius $R$.

It will turn out (and we will show this \emph{a posteriori}) that the moment at which the photons decouple from the fireball plasma will be deep in the spreading phase of its expansion (consistent with $T'\ll m_e$). 
As a result, $v = \sqrt{1-1/\Gamma^2} \approx 1-1/(2\Gamma^2)$ (for $\Gamma \gg 1$) is approximately constant, and $\Delta R \sim R/\Gamma^2$.

Now, as the photon travels outward at speed $c=1$ from initial radius $r_i \sim R$, the fireball will also be expanding, so the anti-proton density it will experience is $n_{\bar{p}}(r) = n_{\bar{p}}(r,R_{\text{adv}}(r))$, where $R_{\text{adv}}(r) \equiv  R + v(r-R) = (1-v)R + vr$.
This density is non-zero on the interval from $R \leq r \leq r_f$ where $r_f$ is defined implicitly by $r_f = R_{\text{adv}}(r_f) +\Delta R(R_{\text{adv}}(r_f))$.
For the purposes of this estimate, let us take $\Delta R(R_{\text{adv}}(r_f))\approx \Delta R(R)$.%
\footnote{%
    Note that this estimate actually fails if we insert $\Delta R = R_{\text{adv}}(r_f)/\Gamma^2$; we assess that this is due to various $\mathcal{O}(1)$ factors that we have neglected here. 
    For instance, it would work for $\Delta R < R_{\text{adv}}(r_f)/(2\Gamma^2)$. 
    While we wish to be transparent about this issue, it is not a serious problem for this estimate, which is intended to be parametric only.} %
In the limit $\Gamma \gg 1$, this yields $r_f \approx R_{\text{adv}}(r_f) \approx 3R$.
Note that this yields a range of support for the integral in \eqref{eq:kappa1} over a range $\Delta r \sim R \sim \Gamma^2 \Delta R \gg \Delta R$.
This is really the important point that this parametric estimate leads to: the integral in \eqref{eq:kappa1} will run over a range $\Delta r \gtrsim R \gg \Delta R$. 
Physically, the relativistic expansion of the fireball at $v\sim 1$ implies that a photon spends much longer inside the fireball plasma than it would were the same (CoM-frame) thickness of plasma stationary (in the CoM frame).

The constant value that we assume that $n_{\bar{p}}(r,R)$ takes over its range of support is approximately $n_{\bar{p}}(R\leq r\leq R + \Delta R,R) \approx n_{\bar{p}} R_0^3 /(3R^2\Delta R)$, where $n_{\bar{p}}$ (without an argument) is the CoM fireball antiproton density when it is approximately spherical and has radius $R_0$ (this estimate assumes that anti-protons are not significantly depleted by nuclear reactions during expansion, which is true in the parameter region interest to us; see \secref{sec:antinucleosynthesis}).
Now, because $R^2\Delta R\propto R^3$ will increase by more than an order of magnitude as $R$ increases to $R_{\text{adv}}(r_f)\sim 3R$, it follows that $n_{\bar{p}}(r)$ will fall rapidly over the range of its support in the integral in \eqref{eq:kappa1}. 
As such, up to numerical factors that we neglect in this estimate, we will approximate the integral in \eqref{eq:kappa1} using the value of the integrand evaluated at $r=R$, multiplied by the range over which the integrand takes approximately that value, which will be $\Delta r \sim R$ according to the estimates above (i.e., the true integrand falls by $\mathcal{O}(1)$ over this characteristic change in the integration variable, and the upper limit of the integration is parametrically large enough as so to not cut it off before then).
That is, we will approximate \eqref{eq:kappa1} as
\begin{align}
    \kappa(R)& \sim \varepsilon \frac{n_{\bar{p}} R_0^3 }{3R^2\Delta R} \sigma_T (1-v) R 
        \sim \varepsilon \frac{n_{\bar{p}} R_0^3 }{6 R^2} \sigma_T  
        \sim \kappa_0 \frac{R_0^2}{R^2} \:,       
\end{align}
where we used $1-v\sim 1/(2\Gamma^2)$, $\Delta R \sim R/\Gamma^2$ (spreading phase), and we have defined $\kappa_0$ to be
\begin{align}
    \kappa_0 &\equiv \frac{1}{6} \varepsilon n_{\bar{p}} R_0 \sigma_T \label{eq:kappa2} \\
        &= \frac{1}{6} \frac{\varepsilon n_{\bar{p}}}{n_{\bar{B}}} n_{\bar{B}} R_0 \sigma_T \\
        &= \frac{g_*\zeta(3)}{6\pi^2} \frac{\varepsilon n_{\bar{p}}}{n_{\bar{B}}} \bar{\eta}T_0^3 R_0 \sigma_T \\ \displaybreak
        &\sim \frac{\varepsilon n_{\bar{p}}}{n_{\bar{B}}} \bar{\eta}T_0^3 R_0 \sigma_T\\
        &\sim 10^{11}\left(\frac{\bar{\eta}}{10^{-2}}\right)\left(\frac{T_0}{100\,\MeV}\right)^3\left(\frac{R_0}{1\,\text{mm}}\right)\left(\frac{\varepsilon n_{\bar{p}}}{n_{\bar{B}}}\right)\:,
\end{align}
where we inserted $n_{\bar{B}}$, which is understood here to be the baryon number density when the fireball is at radius $R_0$ and temperature $T_0$; applied the definition of the anti-baryon--to--photon ratio; and dropped a numerical constant $(g_*\zeta(3))/(6\pi^2) \sim 0.25$ for $g_* \sim 12$ (see \appref{app:dof}), because this estimate should be understood parametrically only.

If we define $t_{\rm thin}$ to be the approximate moment in time where the plasma becomes optically thin to photons emitted in its bulk, and the corresponding radius of the fireball at that time to be $R_{\text{thin}}$, we can write $\kappa(R_{\rm thin}) \sim 1$, leading to
\begin{align}
    R_{\rm thin}&\sim \kappa_0^{1/2}R_0 \\
        &\sim 300\,\text{m} \times \left(\frac{\bar{\eta}}{10^{-2}}\right)^{1/2} \left(\frac{T_0}{100\,\MeV}\right)^{3/2} \nl \times \left(\frac{R_0}{1\,\text{mm}}\right)^{3/2} \left(\frac{\varepsilon n_{\bar{p}}}{n_{\bar{B}}}\right)^{1/2}\: . \label{eq:Rthin}
\end{align}
At the fiducial parameter point, $\tau'(t_{\text{thin}}) \sim R_{\text{thin}}/\Gamma \sim 10^{-7} \,\text{s}$, which is less than the rest-frame muon lifetime, as noted above.
This is also deep in the spreading regime for these parameters and we also have $T'(t_{\text{thin}}) \ll m_e$~(cf.~\figref{fig:benchexp}), validating our assumptions above.

When the fireball radius hits $R_{\rm thin}$, the bulk of the plasma becomes optically thin and a burst of photons is released.
This is also the approximate moment at which the anti-baryons and other particles that were coupled to the fireball plasma (see \secref{sec:antinucleosynthesis}) are released into the interstellar medium (see \secref{sec:prop-detection}), assuming that the fireball was located in our Galaxy.

\subsection{Nuclear physics}
\label{sec:antinucleosynthesis}

The monotonically decreasing co-moving temperature $T'(R)$ of the fireball plasma given at \eqref{eq:Tprime} implies that it eventually becomes thermodynamically favorable for bound anti-nuclei to form.

Were thermodynamic equilibrium among the lightest few anti-nuclei species to be achieved (analogous to the situation in Big Bang Nucleosynthesis [BBN]), almost the entirety of the available anti-neutron abundance would be converted into $\antiHe{4}$, leading to a final configuration dominated by $\bar{p}$ and $\antiHe{4}$, with only trace amounts of other complex light anti-nuclei.
Given that AMS-02 has tentatively identified similar numbers of $\antiHe{3}$ and $\antiHe{4}$ candidates (up to a factor of a few; statistics are small), such an outcome would not be phenomenologically viable.

To understand how to avoid this outcome, consider a key feature of how the analogous process of ordinary BBN proceeds.
The most efficient nuclear-reaction pathway to ${}^4\text{He}$ has as an initial step free-neutron capture on hydrogen to form deuterium $\text{D}$: $n+p\rightarrow \text{D}+\gamma$~\cite{Wagoner:1972jh}, with the $\text{D}$ then being processed by further nuclear burning to other light elements.
However, the relatively low deuterium binding energy $B_{\text{D}}\approx 2\,\MeV$ and the low baryon-to-entropy ratio $\eta\approx  6\times 10^{-10}$ during BBN make deuterium prone to photodissociation back to free neutrons and protons: this is the famous ``deuterium bottleneck''~\cite{Cyburt:2004cq,Serpico:2004gx,Iocco:2008va,Cyburt:2015mya}.
Consequently, ${}^4\text{He}$ production in BBN was delayed until the temperature of the primordial plasma cooled down significantly below $B_{\text{D}}$, whereupon the abundance of photons capable of dissociating deuterium was considerably Boltzmann suppressed, enabling the deuterium abundance to rise.
Crucially, in the BBN realized in our Universe, the deuterium bottleneck was overcome while thermodynamic equilibrium was still being maintained among the light species: nuclear reaction rates were still sufficiently fast compared to Hubble expansion that, once deuterium was capable of being created without being photodissociated, it was rapidly burned to tritium and ${}^3\text{He}$, and then further to ${}^4\text{He}$.

But standard BBN successfully overcame the deuterium bottleneck while maintaining thermodynamic equilibrium only marginally. 
If the expansion rate of the Universe were to have been sufficiently larger, neutron capture on hydrogen would have decoupled (i.e., frozen out) before the bottleneck could have been overcome.
In that case, the output of the nucleosynthesis would not have been dictated by thermodynamic equilibrium among the light nuclei.
Instead, the immediate nucleosynthesis products in this scenario would have been mostly free protons and neutrons, with smaller abundances of deuterium, tritium, ${}^{3}\text{He}$, and ${}^{4}\text{He}$ being produced in amounts controlled by the relative rates of nuclear reactions that produce them, which can be comparable to one another.
Accounting for the fact that unstable neutrons later decay to protons and that tritium later decays to ${}^{3}\text{He}$, the final output in this counterfactual case could easily have been such that $n_p\gg n_{^3\text{He}}\sim n_{^4\text{He}}$.

In what follows, we show that parameter space exists for which the analogous anti-nucleosynthesis occurring in our expanding fireball remains in this ``stuck in the bottleneck" regime, yielding phenomenologically viable amounts of $\antiHe{3}$ as compared to $\antiHe{4}$.

\subsubsection{Preliminaries}

In order to separate changes in the number density of an element due to nuclear reactions from that due to the fireball expansion, in this section we describe the evolution of a nuclear species $i$ in terms of its fractional abundance $X_i$, defined as
\begin{align}
    X_i\equiv\frac{n_i^{\prime}}{n_{\bar{B}}^{\prime}} \:,
\end{align}
where $n_i^{\prime}$ is the number density of element $i$ and $n^{\prime}_{\bar{B}}$ is the anti-baryon number density. 
We consider only $i\in\{\bar{p},\bar{n},\antiD,\antiT, \antiHe{3},\antiHe{4} \}$. 
We obtain the evolution of the abundances of nuclear elements $X_i$ by numerically solving the Boltzmann equations detailed in \appref{app:reactionNetwork} describing the simplified network of nuclear reactions among these species,%
\footnote{%
        We solve a simplified, partial reaction network accounting only for light species, which is acceptably accurate for our purposes. 
        In principle, more accurate results could be obtained by using a modified version of BBN nucleosynthesis codes such as PRyMordial~\cite{Burns:2023sgx}, PRIMAT~\cite{Pitrou:2019nub}, or AlterBBN~\cite{Arbey:2011nf}.
        } %
using the nuclear cross sections shown in \appref{app:nuclearCrossSections}. 
The results of these numerical computations are summarized in 
\figref[s]{fig:Xvsx}--\ref{fig:outputspace}.

While those numerical results are of course more accurate, we also wish to gain an understanding of, and intuition for, the most important nuclear processes at work, and determine the dependencies of the final anti-helium isotope abundances on the fireball parameters $(T_0,R_0,\bar{\eta})$. 
In what follows, we therefore develop an analytical understanding that reproduces the gross features of the numerical results.

The main approximation we employ in our analytical arguments is as follows.
In general, nuclear species with higher mass numbers $A$ are produced from those with lower $A$ in a sequence of successive two-body nuclear reactions. 
Our numerical analysis shows that in the fireball parameter space that yields $X_{\antiHe{3}}\gtrsim X_{\antiHe4}$, a hierarchy is maintained between the nuclear abundances with successive mass numbers, namely $X_{\bar{n}}+X_{\bar{p}}\gtrsim X_{\antiD}\gtrsim X_{\antiT}+X_{\antiHe{3}}\gtrsim X_{\antiHe{4}}$, as depicted in \figref[s]{fig:Xvsx} and \ref{fig:Xvsr0}. 
Furthermore, that analysis shows that anti-nucleosynthesis occurs mainly at temperatures $T'\sim 100-200\,\keV$ low enough that all of the endothermic reverse nuclear reactions to the exothermic ones considered in this subsection are negligible, except for the photodissociation of anti-deuterium responsible for the bottleneck in the first place,%
\footnote{%
    This endothermic process is an exception because (a) it involves photons, which are highly abundant compared to anti-baryons due to the low anti-baryon--to--entropy ratio $\bar{\eta}\ll 1$ we consider; and (b) the binding energy of anti-deuterium is unusually small, $B_\antiD\approx 2.2\,\MeV$.} %
which we thus take into account in our analysis. 
These observations suggest that, instead of solving the whole nuclear reaction network at once, we can treat the nuclear reactions sequentially; that is, we can consider elements produced earlier in the chain of successive two-body reactions as fixed sources for reactions later in the chain of successive reactions, neglecting back-reaction on those sources arising from those later reactions.

In what follows, we first discuss relevant expansion timescales, and then proceed to discuss in turn heavier and heavier anti-nuclei synthesized in this approximate sequential paradigm.
Finally, we summarize and discuss other, non--anti-nucleosynthetic outputs.

\subsubsection{Expansion timescales}

Anti-nucleosynthesis in the fluid rest frame of the expanding fireball is qualitatively similar to BBN in that there are various nuclear reactions occurring in an adiabatically expanding background~\cite{Cyburt:2004cq,Serpico:2004gx,Iocco:2008va,Cyburt:2015mya}. 
However, it differs from the BBN in important ways. 
The plasma in our scenario is spatially finite and expands relativistically into vacuum.%
\footnote{\label{ftnt:otherAnalogousCases}%
    Our anti-nucleosynthesis process resembles in this respect that occurring in the context of gamma-ray burst~\cite{Pruet:2002hi,Beloborodov:2002af,Lemoine:2002vg,Inoue:2003pb} or heavy-ion collision~\cite{Chen:2018tnh}, but is otherwise very different.} %
This of course leads to a non-trivial dependence of the dynamical expansion timescale $\tau^\prime$ on the comoving temperature $T^\prime$, as shown at \eqref{eq:tauprime}.

The parameter space that is viable for our model is roughly $10^{-4}\,\text{m}\lesssim R_0\lesssim 1\,\text{m}$ (i.e., $10^{-13}\,\text{s}\lesssim \tau_0\lesssim 10^{-9}\,\text{s}$), $10\,\MeV\lesssim T_0\lesssim 200\,\MeV$,
and $\Gamma\sim T_0/\bar{\eta} m_p\sim 10$ (i.e., $10^{-3}\lesssim \bar{\eta}\lesssim 10^{-2}$). 
We displayed the scalings of $\tau'(T')$ in \figref{fig:benchexp} for benchmark parameters in these ranges. 
The results are quantitatively and qualitatively very different compared to the Hubble time as a function of temperature during BBN; as we will see, this leads to important differences between fireball anti-nucleosynthesis and BBN.
As we will show, fireball anti-nucleosynthesis in this parameter space commences at temperature $100\,\keV\lesssim T_{\antiD}^\prime\lesssim 200\,\keV$, which satisfies $T_{\antiD}^\prime\lesssim T_0/\Gamma^{5/3}$ and thereby always falls in the ``spreading'' phase of the fireball expansion.

\begin{figure*}[t]
    \centering
    \includegraphics[width=0.65\textwidth]{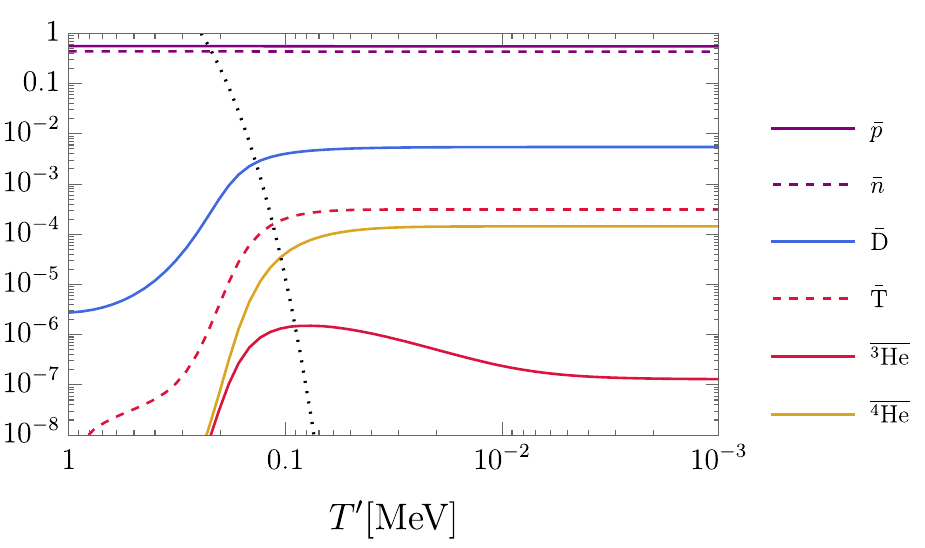}
    \caption{\label{fig:Xvsx}%
        The abundances of nuclear species  $X_i=n^{\prime}_i/n^{\prime}_{\bar{B}}$ (solid and dashed colored lines, as identified in the legend) as a function of the comoving fireball temperature $T'$, computed by numerically solving the Boltzmann equations for a simplified nuclear-reaction network detailed in \appref{app:reactionNetwork}, for a fireball with $T_0=100\,\MeV$, $R_0=1.5\,\text{mm}$, $\bar{\eta}=10^{-2}$ ($\Gamma=10$). 
        Also shown is $Y_{\antiD,\gamma}=n^{\prime}_{\gamma}(E^{\prime}_\gamma>Q_D)/n^{\prime}_{\bar{B}}$ (dotted black line), the abundance of photons with energies $E^{\prime}_\gamma$ above the anti-deuteron photodissociation threshold $Q_D$. 
        Initially, nuclear bound states are essentially non-existent, apart from $\antiD$ whose abundance is kept at an exponentially small value due to the high abundance of photodissociating photons, $Y_{\antiD,\gamma}\gg X_{\antiD}$. 
        As the fireball cools down, $Y_{\antiD,\gamma}$ decreases while $X_{\antiD}$ increases. Eventually at $T=T^\prime_{\antiD}\sim 140\,\keV$ [cf.~\eqref{eq:TprimeD}], $Y_{\antiD,\gamma}$ drops sufficiently low that photodissociation becomes inefficient, thus marking the onset of anti-nucleosynthesis: the absence of photodissociation allows $X_{\antiD}$ to rise significantly, which enables nuclear reactions involving $\antiD$ to produce $\antiT$ whose presence, in turn, enables the production of $\antiHe{4}$. 
        The rise in $X_{\antiD}$ also enables $\antiHe{3}$ production but the produced $\antiHe{3}$ quickly converts into $\antiT$. 
        Since the nuclear reaction rates are proportional to the antibaryon number density $n^{\prime}_{\bar{B}}\propto T^{\prime 3}$, most of the anti-nucleus production occurs within the first few $e$-foldings of expansion after $Y_{\antiD,\gamma}$ drops below $X_{\antiD}$. 
        Nuclear species depicted in dashed lines are stable on the timescale of fireball evolution (until it becomes optically thin) but are expected to decay to the species shown by the solid line of the same color during their journey to the solar system. 
    }
\end{figure*}

\begin{figure*}[t]
    \centering
    \includegraphics[width=0.65\textwidth]{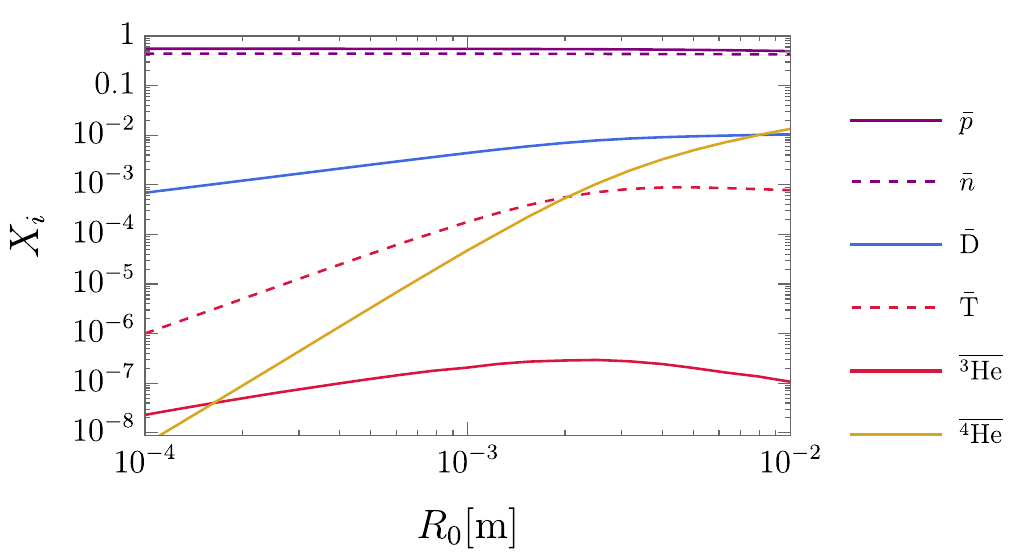}
    \caption{\label{fig:Xvsr0}%
    The abundances of nuclear species $X_i=n^{\prime}_i/n^{\prime}_{\bar{B}}$ (with $i$ as annotated in the legend) released when a fireball with an initial radius $R_0$ becomes optically thin, computed by numerically solving the Boltzmann equations for a simplified nuclear-reaction network, as detailed in \appref{app:reactionNetwork}. 
    Here, we set $T_0=100\,\MeV$ and $\bar{\eta}=10^{-2}$ ($\Gamma=10$).
    While the hierarchy $X_{\bar{p}}\approx X_{\bar{n}}\gtrsim X_{\antiD}\gtrsim X_{\antiT}\gtrsim X_{\antiHe{4}}$ is maintained, the abundances of $\antiD$, $\antiT$, $\antiHe{4}$ released by the fireball scale in reasonable agreement with our analytical predictions, when the appropriate comparisons are made: (a) numerically, $X_{\antiD}\propto (R_0)^{0.8}$, which is only slightly shallower than the analytical $X_{\antiD}\propto R_0$ scaling from \eqref{eq:XD1}; (b) the \emph{relative} $X_{\antiT}\propto X_{\antiD}^3$ scaling holds reasonably well [cf.~the form of $X_{\antiT}^{\text{burn}}$ expressed at \eqref{eq:Xhe32}] leading to $X_{\antiT}\propto (R_0)^{2.4}$ when combined with the numerical result $X_{\antiD}\propto (R_0)^{0.8}$, in reasonably good agreement with these numerics [actually, $X_{\antiT}\propto (X_{\antiD})^{2.8}$ is a slightly better numerical fit for the relative scaling, leading to $X_{\antiT}\propto (R_0)^{2.2}$, which is also a slightly better numerical fit]; and (c) the \emph{relative} $X_{\antiHe{4}}\propto X_{\antiD}^5$ scaling also holds very well [cf.~\eqref{eq:Xhe42}], leading to $X_{\antiHe{4}}\propto (R_0)^{4}$ when combined with the numerical result $X_{\antiD}\propto (R_0)^{0.8}$, again in very good agreement with these numerics. Note however, in connection with (b) and (c), that the na\"ive analytical expectations based on our discussion in the main text would be $X_{\antiHe{3}} \propto R_0^3$ and $X_{\antiHe{4}}\propto R_0^5$, respectively, if one took $X_{\antiD}\propto R_0$ from \eqref{eq:XD1}; see also further discussion in \secref{sec:finalProducts}.
    Besides that, $\antiHe{3}$ is produced promptly in negligible amount, $X_{\antiHe{3}}\lesssim 10^{-6}$ [cf.~\eqref{eq:X3He} and surrounding discussion]. 
    Nuclear species depicted in dashed lines are stable on the timescale of the fireball evolution (until it becomes optically thin) but are expected to subsequently decay to the species shown in with solid lines of the same color on timescales short compared to their propagation time in the Milky Way (e.g., $\antiT$ later decays to $\antiHe{3}$ with a half-life of $\sim 12$\,years in the $\antiT$ rest frame).
    }
\end{figure*}

\subsubsection{Anti-nucleon abundance}
\label{ss:antineutronabundance}

At temperatures $10\,\MeV\lesssim T'\lesssim 200\,\MeV$, nuclear bound states have not formed and all the anti-baryons reside in unbound anti-neutrons and anti-protons. 
Anti-neutrons can, in principle, convert to and from anti-protons through both weak (e.g., $\bar{n}+e^-\leftrightarrow \bar{p}+\nu_e$) and strong (e.g., $\bar{n}+\pi^-\leftrightarrow \bar{p}+\pi^0$) processes~\cite{Kohri:2001jx,Kawasaki:2004qu, Boyarsky:2020dzc}. 
If at least one of these processes is efficient, the relative abundance of these anti-nucleons is initially kept at its chemical equilibrium value, $n^\prime_{\bar{n}}/n^\prime_{\bar{p}}=\text{exp}\left[-(m_n-m_p)/T'\right]$. 
In the early Universe, the matter analogs of both processes were efficient at some point after hadronization. 
Then, strong processes decoupled first as pions rapidly decayed and annihilated away, and hence the pre-BBN freeze-out abundances of neutrons and protons were determined by the later-occurring decoupling of weak interactions. 
By contrast, in the fireball anti-nucleosynthesis scenario we consider, the typically short timescales of the fireball expansion render weak interactions inefficient at all times. 
Consequently, the anti-nucleons freeze out as soon as the pion-mediated strong interconversion processes become inefficient.

After the fireball has thermalized, the following strong-mediated charge exchange reactions (SMCER) are initially in equilibrium~\cite{Pospelov:2010cw} (see discussion about $T'_{\bar{n}\bar{p}}$ below)
\begin{align*}
    &\bar{p}+\pi^+\leftrightarrow \bar{n}+\pi^0\quad (Q=5.9\MeV)\:;\\
    &\bar{n}+\pi^-\leftrightarrow \bar{p}+\pi^0\quad (Q=3.3\MeV)\:.\\
\end{align*}
Additionally,%
\footnote{%
    The reactions $\pi^0 \pi^0 \leftrightarrow \pi^+ \pi^-$, $\pi^0 \rightarrow \gamma \gamma$, $\pi^0 \pi^0 \leftrightarrow \gamma \gamma$, and $\pi^+\pi^- \leftrightarrow \gamma \gamma$ are all in equilibrium.
    } %
$\mu_\gamma=0$, $\mu_{\pi^0}=0$, and $\mu_{\pi^-}=-\mu_{\pi^+}$.
These imply
\begin{align}
    \mu_{\bar{n}}-\mu_{\bar{p}}=\mu_{\pi^+}\:.
\end{align}
The chemical equilibrium anti-neutron--to--anti-proton ratio for $T^\prime\ll m_n-\mu_{\bar{n}}, m_p-\mu_{\bar{p}}$ is thus given by
\begin{align}
    \left(\frac{n_{\bar{n}}^\prime}{n_{\bar{p}}^\prime}\right)_{\rm ch}\approx e^{-\frac{m_n-m_p}{T'}+\frac{\mu_{\pi^+}}{T'}}\:.
\end{align}
where $m_n-m_p\approx 1.3\MeV$.
Note that the chemical potential of the charged pions, $\mu_{\pi^+}$, depends on the physics before the fireball has thermalized via efficient strong and EM interactions. 
Therefore, it is model-dependent. 
For instance, it depends on whether electroweak interactions were ever efficient in this pre-thermalization stage, and on some details of the BSM particle injection process that seeds the fireball.
For simplicity, we neglect the chemical potential of charged pion in our analysis here by assuming%
\footnote{\label{ftnt:Qhadronic}%
    As we discuss \appref{app:antinucleonFO}, this specific assumption is equivalent to assuming that there is a net negative charge in the hadronic sector of the plasma that has a certain very specific value: defining $X_Q \equiv - (n'_Q)^{\text{hadronic}} /n'_{\bar{B}}$ as at \eqref{eq:XQdef}, we would have $X_Q = 0.56$; cf.~\eqref{eq:XpbarXnbarCh}.
    This charge is compensated by opposite charge in the leptonic sector so that the plasma as a whole is net EM-neutral as expected from fireballs seeded by EM-neutral dark states.
    However, our results as stated in the main text are unchanged qualitatively, and change quantitatively by only $\mathcal{O}(1)$ factors, so long as it is approximately true that $X_Q \sim \mathcal{O}(1/2)$ by the time that the SMCER become inefficient.
    We also show in \appref{app:antinucleonFO} that we may even be able to tolerate values as small as $X_Q \sim 10^{-2}$, although that changes some conclusions stated in the main text in a qualitative fashion.}
     %
\begin{align}
    \left|\mu_{\pi^+}\right|\ll m_n-m_p \:.
\end{align}
We discuss the model-dependence of $\mu_{\pi^+}$ and the case when $\mu_{\pi^+}$ is non-negligible in \appref{app:antinucleonFO}.

The pion-mediated strong interactions decouple at a temperature $T'=T^{\prime}_{\bar{n}\bar{p}}$ where the pion abundance becomes sufficiently Boltzmann suppressed that the $\Gamma_{\rm strong}(T')$ found in \eqref{eq:Gammastrong} goes below the fireball expansion rate $1/\tau^{\prime}$ found in \eqref{eq:tauprime}. 
We find that, numerically, $T^{\prime}_{\bar{n}\bar{p}}\approx 6\,\MeV$ invariably in the whole parameter space that is viable for our scenario. 
The freeze-out value of the anti-neutron--to--anti-proton ratio $n^{\prime}_{\bar{n}}/n^{\prime}_{\bar{p}}$ can be approximated by its chemical-equilibrium value at that time
\begin{align}
    \left.\frac{n^{\prime}_{\bar{n}}}{n^{\prime}_{\bar{p}}}\right|_{{\rm ch},\,  T_{\bar{n}\bar{p}}^\prime}\approx e^{-(m_n-m_p)/T_{\bar{n}\bar{p}}^\prime}\approx 0.8 \:.
    \label{eq:nDnpch}
\end{align}
If $\left|\mu_{\pi^+}\right|\gtrsim m_n-m_p$, unlike what we have assumed, then the freeze-out value of $n^{\prime}_{\bar{n}}/n^{\prime}_{\bar{p}}$ and our subsequent results would change; however, as long as $\left|\mu_{\pi^+}\right|\lesssim T^\prime_{\bar{n}\bar{p}}\approx 6\MeV$, these changes are only $\mathcal{O}(1)$ and most of our conclusions remain valid.
See \appref{app:antinucleonFO} for further discussion.

Moreover, while neutron decay is an important phenomenon in the BBN that was realized in the early Universe, in our scenario anti-neutrons do not decay until well after anti-nucleosynthesis finishes.
Assuming that only a small fraction of the $\bar{p}$ and $\bar{n}$ are burned to higher nuclei (true throughout our parameter space of interest), we will thus have, for all times relevant for the anti-nucleosynthesis in the expanding fireball, the following:
\begin{align}
    X_{\bar{n}}(T') &\approx X_{\bar{n}}(T'_{\bar{n}\bar{p}}) \equiv X^{\text{ch}}_{\bar{n}};\\
    X_{\bar{p}}(T') &\approx X_{\bar{p}}(T'_{\bar{n}\bar{p}}) \equiv X^{\text{ch}}_{\bar{p}}\:,
\end{align}
where
\begin{align}
    X^{\text{ch}}_{\bar{n}} &\approx 0.8 X^{\text{ch}}_{\bar{p}}\:, &
    X^{\text{ch}}_{\bar{n}} + X^{\text{ch}}_{\bar{p}} &\approx 1\:,
\end{align}
implying that
\begin{align}
    X^{\text{ch}}_{\bar{n}} &\approx 0.44\:, & 
    X^{\text{ch}}_{\bar{p}} &\approx 0.56\:. \label{eq:XpbarXnbarCh}
\end{align}

\subsubsection{Anti-deuterium production}

Anti-deuterium is produced primarily through the reaction%
\footnote{%
    The anti-deuterium formation releases some amount of energy density to the plasma, given by the total binding energy of the anti-deuterium formed: $B_{\antiD}n^{\prime}_{\antiD}\sim B_{\antiD} \bar{\eta}X_{\antiD}T^{\prime 3}$, where $B_{\antiD}\approx 2.2\,\MeV$. This amounts to a tiny fraction of the radiation energy density $\sim T^{\prime 4}$ in the parameter space of our interest, where $\bar{\eta}\lesssim 10^{-2}$, $X_{\antiD}\lesssim 10^{-2}$, and $T'\gtrsim 100\,\keV$ when the anti-deuterium forms.} %
$\bar{n}+\bar{p}\rightarrow\antiD+\gamma$.
Initially, however, the reverse reaction (photodissociation) is in equilibrium and the high abundance of photons with energies above the anti-deuterium binding energy $B_{\antiD}\approx 2.2\,\MeV$ suppresses the  (quasi-equilibrium) anti-deuterium abundance, which is given by the Saha equation:
\begin{align}
    X_{\antiD,\rm ch}\approx  \bar{\eta}\left(\frac{T^\prime}{m_p}\right)^{3/2}e^{B_\antiD/T^\prime}. \label{eq:XDbarch}
\end{align}
This continues until the abundance of photons with sufficient energy to photodissociate anti-deuterium,
\begin{align}
    Y_{\antiD\gamma}=\frac{n^{\prime}_\gamma(E^{\prime}_\gamma\gtrsim B_\antiD)}{n^{\prime}_{\bar{B}}}\sim \frac{1}{\bar{\eta}}\frac{B_\antiD^2}{T^{\prime 2}}e^{-B_\antiD/T^\prime} \:, 
\end{align}
starts to fall below the anti-deuterium abundance; i.e., $Y_{\antiD\gamma}\sim X_{\antiD}$. 
The temperature at that point can be estimated as 
\begin{align}
    T_{\antiD}^\prime&\approx  \frac{B_\antiD}{4.6-\ln\bar{\eta}+(7/4)\ln\left(4.6-\ln\bar{\eta}\right)}\nonumber\\
    &\approx 140-170\,\keV , \label{eq:TprimeD}
\end{align}
where the displayed range of values $T_{\antiD}^\prime$ corresponds to the range of viable anti-baryon--to--entropy ratios $10^{-3}\lesssim\bar{\eta}\lesssim 10^{-2}$.
As mentioned earlier, in our the parameter space of interest $T_\antiD^\prime$ falls in the spreading phase of the fireball expansion (i.e., it satisfies $T_\antiD^\prime\lesssim T_0/\Gamma^{5/3}$) and, neglecting the mild logarithmic dependence on $\bar{\eta}$, the fireball expansion timescale at decoupling of the photodissociation reactions is
\begin{align}
    \tau'_{\antiD} &\equiv \tau^\prime(T_{\antiD}^\prime)\nonumber\\
    &\approx 5.1\times 10^{-10}\,\text{s}\left(\frac{T_0}{100\,\MeV}\right)\left(\frac{R_0}{\text{mm}}\right)\left(\frac{\Gamma}{10}\right)^{-2/3} \:,
\end{align}
where we have set $T_{\antiD}'=140\keV$ (corresponding to $\bar{\eta}=10^{-3}$). 
After the photodissociation of $\antiD$ decouples at $T_\antiD^\prime$, at which point the fireball expansion timescale is $\tau'_{\antiD}$, anti-deuterium production through $\bar{n}+\bar{p}\rightarrow\antiD+\gamma$ is no longer thwarted, and so the anti-deuterium abundance rises monotonically. 
At around the same time, heavier elements that rely on anti-deuterium burning as an initial step begin to be populated sequentially. 
Since the product of the fireball expansion timescale and the nuclear reaction rates that form any of the light elements scales as $\propto n^{\prime}_{\bar{B}}\left<\sigma v\right>\tau^\prime\propto T^{\prime 2}$ during the spreading phase ($T'\lesssim T_0/\Gamma^{5/3}$), these nuclear reactions are most efficient in populating the light elements in the first fireball-expansion $e$-fold or so after the decoupling of $\antiD$ photodissociation, before the anti-baryon density is significantly diluted by the expansion.

Moreover, as we operate in the regime where anti-deuterium is not efficiently burned to more complex nuclei (see the next sub-subsection), a simple estimate for the final anti-deuterium abundance can be obtained by assuming the anti-deuterium abundance is that generated by neutron--proton fusion reactions operating in a single dynamical expansion timescale at the point of anti-deuterium photodissociation freeze-out. 
We estimate that abundance to be
\begin{align}
    X^{\text{prompt}}_{\antiD}&\approx k_{\bar{2}} n^{\prime}_{\bar{B}}\left<\sigma v\right>_{\bar{n}\bar{p}}\tau'_{\antiD} X_{\bar{n}}^{\text{ch}} X_{\bar{p}}^{\text{ch}}\label{eq:XD} \\
    &\approx 4.0\times 10^{-3} \times \left( \frac{k_{\bar{2}}}{0.60} \right) \nl \times \left(\frac{T_0}{100\,\MeV}\right)^2\left(\frac{R_0}{\text{mm}}\right)\left(\frac{\Gamma}{10}\right)^{-5/3}\label{eq:XD1},
\end{align}
where we took the value of cross-section $\left<\sigma v\right>_{\bar{n}\bar{p}}$ to be that at $T'\sim 140\,\keV$ (the lower end of the range of values for $T'_{\antiD}$); see \appref{app:nuclearCrossSections}.  
In writing the above results, we have used $\bar{\eta}\sim T_0/(\Gamma m_p)$ and we manually inserted an $\mathcal{O}(1)$ prefactor $k_{\bar{2}} \approx 0.60$ in \eqref{eq:XD}, such that the final result is in better agreement with what we obtained by numerically solving the Boltzmann equations at this benchmark point. 
Note also that, in our numerical results, we find the scaling of $X^{\text{prompt}}_{\antiD}$ with $R_0$ is actually closer to $X^{\text{prompt}}_{\antiD}\propto (R_0)^{0.8}$ (cf.~\figref{fig:Xvsr0}); this is important in the context of later results that will raise this result to large powers [cf.~\eqref[s]{eq:Xhe32} and (\ref{eq:Xhe42})].

We are interested in the regime where $X^{\text{prompt}}_{\antiD}\ll 1$: i.e., the anti-deuterium production decouples before its abundance rises to $X_{\antiD}\sim 1$. 

\subsubsection{Anti-deuterium burning}

During and slightly after the production of anti-deuterium, a small fraction of it also burns through the following dominant channels:
\begin{align*}
    \antiD+\antiD&\rightarrow \antiHe{3}+\bar{n} \:,\\
    \antiD+\antiD&\rightarrow \antiT+\bar{p} \:,
\end{align*}
with essentially equal branching fractions (both are strong-mediated nuclear reactions). 
Because we are analyzing production in the regime where anti-deuterium is not efficiently burned to more complex anti-nuclei, we may treat the abundance of anti-deuterium as a fixed source which acts to populate the more complex nuclei over roughly a single dynamical expansion timescale after the anti-deuterium are produced.
As such, the prompt production of $\antiHe{3}$ and $\antiT$ can be estimated as
\begin{align}
    X_{\antiT}^{\text{prompt}}\approx X_{\antiHe{3}}^{\text{prompt}} \approx k_{\bar{3}} n^{\prime}_{\bar{B}}\left<\sigma v\right>_{\antiD\antiD}\tau'_{\antiD} (X^{\text{prompt}}_{\antiD})^2 \:,\label{eq:3Heoverp}
\end{align}
where we have evaluated all the quantities at $T'\sim T'_{\antiD}$, assumed $X_{\antiD}\gg X_{\antiHe{3}}, X_{\antiT}$ around the time of this production, and manually included an $\mathcal{O}(1)$ numerical prefactor $k_{\bar{3}} \approx 0.15$ [cf.~the factor $k_{\bar{2}}$ introduced in \eqref{eq:XD}]. 

Alternative production channels for $\antiHe{3}$ and $\antiT$ are $\antiD+\bar{p}\rightarrow \antiHe{3}+\gamma$ and $\antiD+\bar{n}\rightarrow \antiT+\gamma$, respectively; however, we verified numerically that the former production channel is negligible as long as $X_{\antiD}\gg 10^{-5}$ and the latter is negligible as long as $X_{\antiD}\gg 10^{-4}$, which are always satisfied in the parameter space we consider. 
It is understood that these process are inefficient because they both suffer from photon-emission suppression (i.e., they are electromagnetic-mediated, rather than strong-mediated, nuclear reactions).

The reaction $\antiD+\antiD\rightarrow \antiHe{4}+\gamma$ can also proceed with a branching ratio of $\sim 10^{-7}$ (it is electromagnetically mediated).
Because of this small branching fraction, prompt $\antiHe{4}$ production through this channel, $X^{\text{prompt}}_{\antiHe{4}} \sim n^{\prime}_{\bar{B}} (10^{-7}\left<\sigma v\right>_{\antiD\antiD})\tau_{\antiD}^\prime (X^{\text{prompt}}_{\antiD})^2$, is negligible compared to other, deuterium--tritium-burning channels that we discuss below so long as $X_{\antiT}/X_{\antiD}\gg 10^{-7}$ around the time of production.

\subsubsection{Anti-helium-3 burning}

The strong-mediated reaction $\antiHe{3}+\bar{n}\rightarrow\antiT+\bar{p}$ is also present. 
It is also extremely efficient in part because it is not Coulomb suppressed.
It thus gives the one counterexample to our earlier statement that we can ignore back-reaction on sequentially produced species: because it is exothermic ($Q \approx 0.76\,\text{MeV}$~\cite{Huang_2021a,Wang_2021b}, assuming completely ionized nuclei as appropriate at the relevant temperatures), this reaction burns essentially all the anti-helium-3 that are produced primarily (via $\antiD+\antiD \rightarrow \antiHe{3} + \bar{n}$) to anti-tritium.
As such, the anti-helium-3 abundance is maintained at a very low, quasi-equilibrium level: we numerically found that the residual $X_{\antiHe{3}}$ never exceeds $\sim 10^{-6}$.
At the same time, the $\antiT$ abundance is roughly doubled because our estimate at \eqref{eq:3Heoverp} indicated roughly equal production abundances for the two $A=3$ anti-nuclei before this depletion reaction was accounted for.
Our estimates of the anti-tritium and anti-helium-3 abundances after this burning should therefore be revised to
\begin{align}
    X^{\text{burn}}_{\antiT}&\approx 2 X^{\text{prompt}}_{\antiT} \approx 2 k_{\bar{3}} n^{\prime}_{\bar{B}}\left<\sigma v\right>_{\antiD\antiD}\tau'_{\antiD} (X^{\text{prompt}}_{\antiD})^2\: , \label{eq:XT}\\
    X^{\text{burn}}_{\antiHe{3}}&\lesssim 10^{-6} \label{eq:X3He}\:.
\end{align}

\subsubsection{Anti-tritium burning}

Anti-tritium burns efficiently to anti-helium-4 through the following dominant process:
\begin{align*}
    \antiT+\antiD&\rightarrow\antiHe{4}+\bar{n}\:; 
\end{align*}
assuming that $X_{\antiT}\gtrsim X_{\antiHe{4}}$ throughout the burning, we find that the prompt production is
\begin{align}
    X^{\text{prompt}}_{\antiHe{4}}\approx k_{\bar{4}} n^{\prime}_{\bar{B}}\left<\sigma v\right>_{\antiT\antiD}\tau_{\antiD}^\prime X_{\antiT}^{\text{burn}}X_{\antiD}^{\text{prompt}}\:,
    \label{eq:X4He}
\end{align}
where we again manually introduced an $\mathcal{O}(1)$ prefactor $k_{\bar{4}} \approx 0.20$ to better match our numerical results.

\subsubsection{Final anti-nucleosynthesis products}
\label{sec:finalProducts}

\begin{figure*}[!t]
    \centering
    \includegraphics[width=0.7\textwidth]{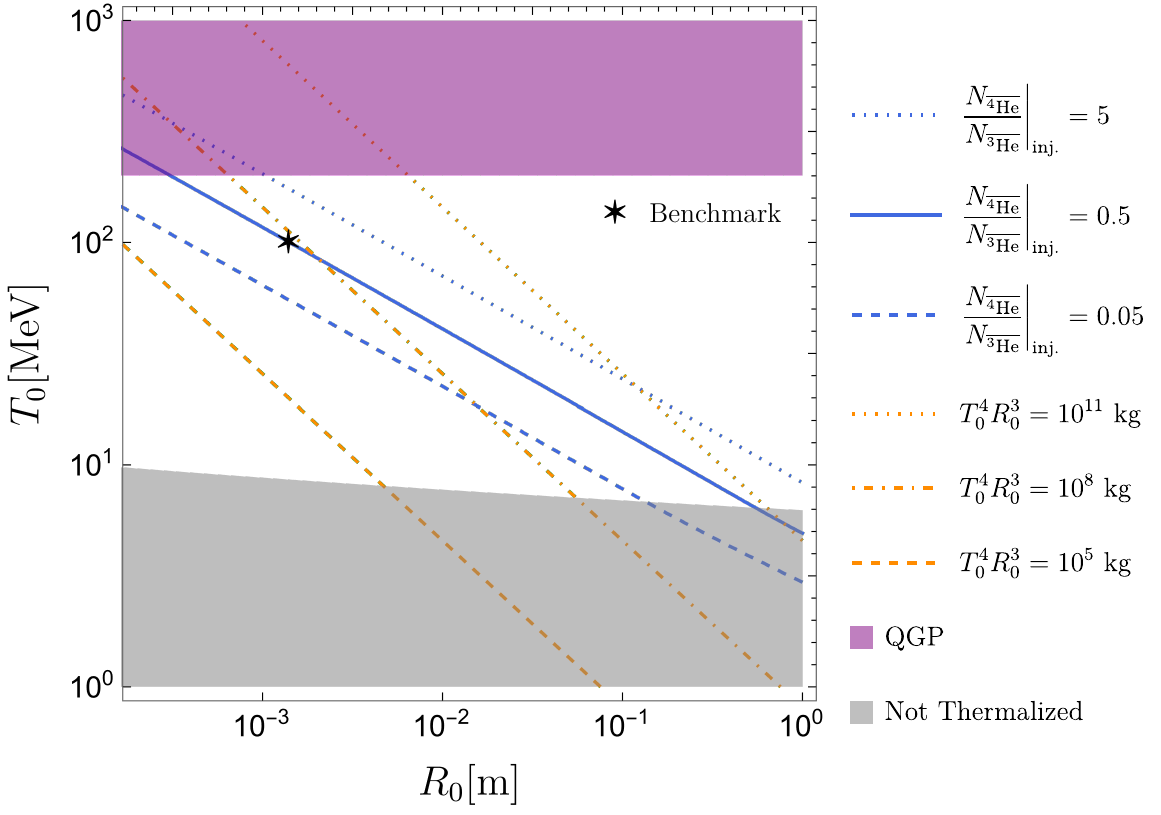}
    \caption{\label{fig:fireballparameterspace}%
    The fireball parameter space. 
    The initial temperature $T_0$ and radius $R_0$ of the fireball are defined at the point right after the fireball plasma has reorganized itself into a shell moving with an average Lorentz factor $\gamma\sim\text{few}$. 
    The various styles of blue lines show contours of constant values of ratio of the \emph{injected} anti-helium-4 and anti-helium-3 abundances, $N_{\antiHe{4}}/N_{\antiHe{3}}|_\text{inj.}$.
    These contours were obtained by numerically solving the Boltzmann equations for a simplified nuclear-reaction network detailed in \appref{app:reactionNetwork}, while fixing the anti-baryon--to--entropy ratio $\bar{\eta}$ such that the terminal Lorentz factor of the shell is $\Gamma\sim T_0/(\bar{\eta}m_p)=10$.
    The orange lines show contours of constant $T_0^4R_0^3$, and provide estimates, up to numerical factors, of the total injection energy in the form of anti-quarks required to create the fireball corresponding to a parameter space point of interest. 
    The black star indicates the benchmark parameter point defined at \eqref{eq:benchmarkParameters}.
    The purple region is the parameter space where the fireball would thermalize at a temperature above the QCD phase transition (making it a quark--gluon plasma [QGP]), a regime we avoid to keep our analysis tractable. 
    In the the gray region, the would-be temperature of the fireball $T_0$ is so low that pions are too Boltzmann suppressed to facilitate the thermalization of the injected anti-quarks; i.e., $\Gamma_{\rm strong}(T_0)\lesssim R_0^{-1}$.
    }
\end{figure*}

\begin{figure*}[!t]
    \centering
    \includegraphics[width=0.7\textwidth]{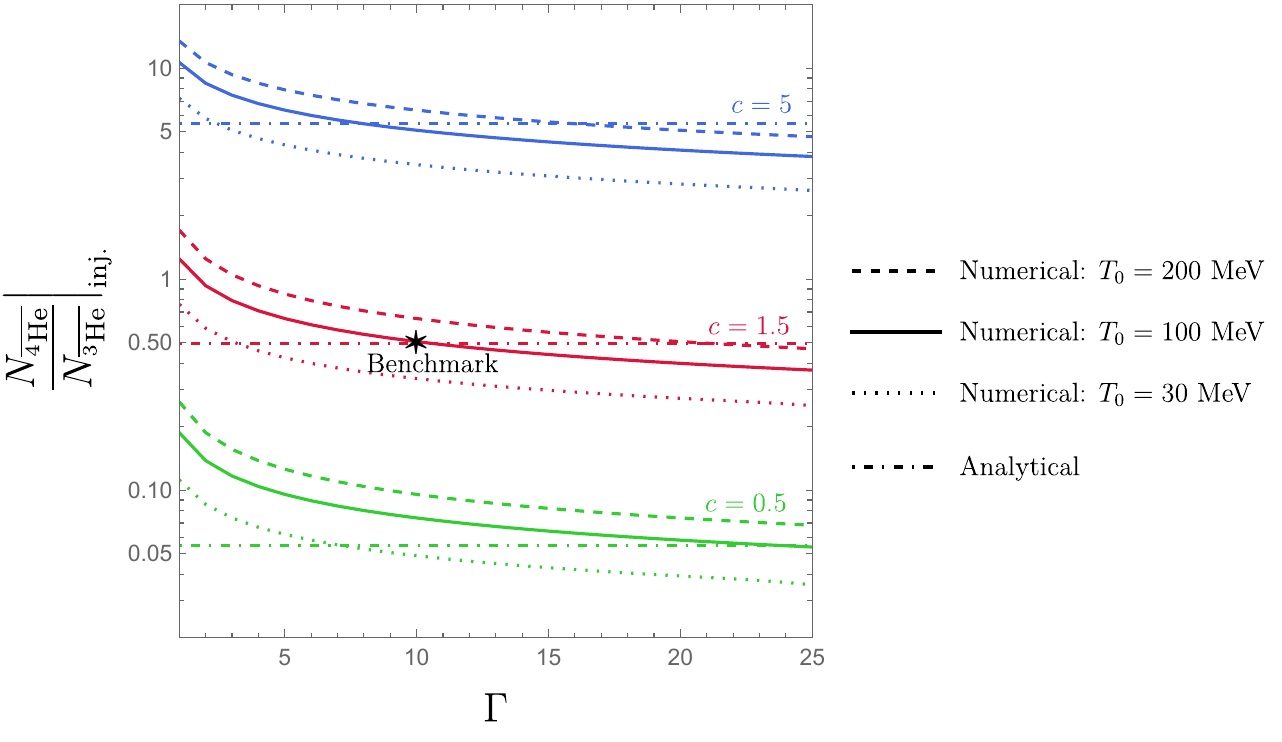}
    \caption{\label{fig:outputspace}%
    Output space. 
    The \emph{injected} anti-helium isotope ratio $N_{\antiHe{4}}/N_{\antiHe{3}}|_\text{inj.}$ and the typical \emph{injected} Lorentz factors $\Gamma$ of the anti-particles are two key observables we seek to explain with the fireball anti-nucleosynthesis scenario we propose in this paper. 
    Publicly available AMS-02 tentative data currently favor \emph{observed} values of  $N_{\antiHe{4}}/N_{\antiHe{3}}|_\text{obs.}\approx 1/2$ and $\Gamma\sim 10$ (this figure does not include propagation effects; see \secref{sec:prop-detection}). 
    Dashed and dotted lines show the injected anti-helium isotope ratio  $N_{\antiHe{4}}/N_{\antiHe{3}}|_\text{inj.}$ obtained by numerically solving the Boltzmann equations for a simplified nuclear-reaction network detailed in \appref{app:reactionNetwork}, while dot-dashed lines show the analytical approximation, \eqref{eq:scalingLaw}. 
    The black star indicates the benchmark parameter point defined at \eqref{eq:benchmarkParameters}.
    As far the numerical results are concerned, this figure is constructed as follows: for each indicated value of $T_0$ (as annotated in the legend), we vary $R_0$ with $\Gamma$ so as to keep $c \equiv (T_0/100\,\MeV)^2(R_0/\text{mm})(\Gamma/10)^{-5/3}$ [cf.~\eqref{eq:cDefn}] fixed to the annotated constant value.
    The analytical prediction on the anti-helium isotope ratio has a precise quadratic dependence
    $N_{\antiHe{4}}/N_{\antiHe{3}}|_\text{inj.}\propto c^2$, and we do not therefore need to make further specific numerical assumptions about the values of $T_0$ and $R_0$; although it is only strictly speaking valid when $c\lesssim 2.1$ [see discussion below \eqref{eq:cDefn}], we also show it here for $c=5$, where it is still reasonably accurate. 
    The numerical computation of $N_{\antiHe{4}}/N_{\antiHe{3}}|_\text{inj.}$ confirms the expected strong dependence on $c$, but it also shows mild sensitivity to $(T_0,R_0,\Gamma)$ variation orthogonal to $c$. 
    }
\end{figure*}

Anti-neutrons have a mean rest-frame lifetime of $\mathcal{O}(15)$ minutes (decaying to an anti-proton), while anti-tritium decays to anti-helium-3 via the beta decay $\antiT\rightarrow \antiHe{3}+e^++\nu_e$ with a rest-frame half-life of 12.3 years.
Even accounting for Lorentz factors $\Gamma \sim 10$, the $\bar{n}$ and $\antiT$ will decay on $\mathcal{O}(100)\,\text{yr}$ timescales (at most) as viewed in the fireball center-of-mass frame. 
After these decays, the remaining light anti-nuclei outputs of our scenario are, at late time, given by
\begin{align}
    X_{\bar{p}}&\approx 1 \:, \label{eq:Xp2}\\
    X_{\antiD}&\approx X^{\text{prompt}}_{\antiD} \qquad \text{[see \eqref{eq:XD}]}\:, \label{eq:XD2} \\
    X_{\antiHe{3}}
        &\approx X_{\antiT}^{\text{burn}} \nonumber\\  
        &\approx \frac{2k_{\bar{3}}}{k_{\bar{2}}}\left(\frac{\left<\sigma v\right>_{\antiD\antiD}}{\left<\sigma v\right>_{\bar{n}\bar{p}}}\right)_{T'_{\antiD}}\frac{(X^{\text{prompt}}_{\antiD})^3}{X^{\text{ch}}_{\bar{n}} X^{\text{ch}}_{\bar{p}}} \label{eq:Xhe3} \\
        &\approx 2.5\times 10^{3}X_{\antiD}^3\: \label{eq:Xhe32},
    \\
    X_{\antiHe{4}}
    &\approx \frac{k_{\bar{4}}}{k_{\bar{2}}} \left(\frac{\left<\sigma v\right>_{\antiT\antiD}}{\left<\sigma v\right>_{\bar{n}\bar{p}}}\right)_{T'_{\antiD}}  \frac{X^{\text{burn}}_{\antiT}(X^{\text{prompt}}_{\antiD})^2}{X^{\text{ch}}_{\bar{n}} X^{\text{ch}}_{\bar{p}} }\label{eq:Xhe4}  \\
    & \approx \frac{2k_{\bar{3}}k_{\bar{4}}}{(k_{\bar{2}})^2}\left(\frac{\left<\sigma v\right>_{\antiT\antiD}\left<\sigma v\right>_{\antiD\antiD}}{(\left<\sigma v\right>_{\bar{n}\bar{p}})^2}\right)_{T'_{\antiD}}  \frac{(X^{\text{prompt}}_{\antiD})^5}{(X^{\text{ch}}_{\bar{n}} X^{\text{ch}}_{\bar{p}})^2 } \label{eq:Xhe41}\\
    &\approx 3.6\times 10^{7} X_{\antiD}^5 \label{eq:Xhe42}
     \:;
\end{align}
while all other products are negligible. 
Note that, in \eqref[s]{eq:Xhe3} and (\ref{eq:Xhe4}), we made use of \eqref[s]{eq:XD} and (\ref{eq:XD2}) to rewrite the factors of $n_{\bar{B}}\tau'_{\antiD}$ in \eqref[s]{eq:3Heoverp} and (\ref{eq:X4He}) in terms of $X^{\text{prompt}}_{\antiD},\ X^{\text{ch}}_{\bar{n}},\ X^{\text{ch}}_{\bar{p}}$, and $\langle \sigma v \rangle_{\bar{n}\bar{p}}|_{T_{\antiD}'}$.
We also inserted the numerical values of the $\mathcal{O}(1)$ coefficients here: $2k_{\bar{3}}/k_{\bar{2}} \approx 0.50$, $k_{\bar{4}}/k_{\bar{2}} \approx 0.33$, and $2k_{\bar{3}}k_{\bar{4}}/(k_{\bar{2}})^2 \approx 0.17$. 
It is important to note that the analytically predicted \emph{relative} scalings of $X_{\antiHe{3}} \propto (X_{\antiD})^{3}$ and $X_{\antiHe{4}} \propto (X_{\antiD})^{5}$ are found to be reasonably accurate in numerical results obtained from solving the Boltzmann equations (cf.~\appref{app:reactionNetwork}), at least for the hierarchy $X_{\bar{p}}\approx X_{\bar{n}}\gtrsim X_{\antiD}\gtrsim X_{\antiT}\gtrsim X_{\antiHe{4}}$ (cf.~\figref{fig:Xvsr0}): numerically, we actually find a scaling somewhat closer to $X_{\antiHe{3}} \propto (X_{\antiD})^{2.8}$, but $X_{\antiHe{4}} \propto (X_{\antiD})^{5}$ is found numerically to be very accurate.
However, the na\"ive absolute scaling with $R_0$ of these results that is implied by combining them with the analytically predicted $X_{\antiD}\propto R_0$ scaling from \eqref{eq:XD1} should be understood with some caution owing to the high powers to which $X_{\antiD}$ is raised and the fact that we find numerically that $X_{\antiD}\propto (R_0)^{0.8}$ is a more accurate scaling result, at least the same hierarchy of the $X_i$ (again, cf.~\figref{fig:Xvsr0}). 
That is, the na\"ive predictions would be $X_{\antiHe{4}}\propto (R_0)^3$ and $X_{\antiHe{4}}\propto (R_0)^5$, whereas we observe numerical scalings in \figref{fig:Xvsr0} more consistent with the $X_{\antiHe{3}}\propto (R_0)^{2.4}$ and $X_{\antiHe{4}}\propto (R_0)^4$ results that would follow from combining the quite accurate \emph{relative} scalings predicted analytically with the more accurate numerical $X_{\antiD}\propto (R_0)^{0.8}$ result [actually, $X_{\antiHe{3}} \approx X^{\text{burn}}_{\antiT}\propto (R_0)^{2.2}$ would be a slightly more accurate result, reflecting the numerically obtained scaling $X_{\antiT} \propto (X_{\antiD})^{2.8}$; cf.~\figref{fig:Xvsr0}].

If we take $T_{\antiD}^\prime=140\,\MeV$, for which $\left<\sigma v\right>_{\bar{n}\bar{p}}\approx 2.0\, \mu\text{b}$, $\left<\sigma v\right>_{\antiD\antiD}\approx 1.9\,\text{mb}$, and $\left<\sigma v\right>_{\antiT\antiD}\approx 16\,\text{mb}$, the anti-helium isotope ratio injected into the interstellar medium for a given $X_{\antiD}$ is given by
\begin{align}
    \left. \frac{N_{\antiHe{4}}}{N_{\antiHe{3}}} \right|_{\text{inj.}} 
    &\approx 1.4\times 10^4 X_{\antiD}^2\nonumber\\
    &\approx 0.22\left[\left(\frac{T_0}{100\,\MeV}\right)^2\left(\frac{R_0}{\text{mm}}\right)\left(\frac{\Gamma}{10}\right)^{-5/3}\right]^2 \:, \label{eq:scalingLaw}
\end{align}
where in the last line we have substituted $X^{\text{prompt}}_{\antiD}$ from \eqref{eq:XD1}.
Hence, to obtain comparable anti-helium-3 and anti-helium-4 abundances, the fireball parameters $(T_0,R_0,\Gamma)$ must be such that the combination of parameters
\begin{align}
    \left(\frac{T_0}{100\,\MeV}\right)^2\left(\frac{R_0}{\text{mm}}\right)\left(\frac{\Gamma}{10}\right)^{-5/3} \equiv c \:,\label{eq:cDefn}
\end{align}
is an $\mathcal{O}(1)$ number; see \figref{fig:fireballparameterspace}. 
According to the simple analytical estimates of this section, the tentative AMS-02--observed anti-helium isotope ratio $N_{\antiHe{4}}/N_{\antiHe{3}}|_\text{inj.}\approx 1/2$ corresponds to $c\approx 1.5$.
Note that this estimate is made here for the \emph{injected} ratio, without regard to the impact of propagation of the anti-nuclei in the Galaxy or the isotope-dependent AMS-02 sensitivity, which we discuss and account for in \secref{sec:prop-detection}.
There, we will show that $X_{\antiD} \approx 5\times 10^{-3}$ is required, corresponding to $c \approx 1.25$.

Note that the sequential production approximation we used to derive the analytical predictions for $X_i$ [\eqrefRange{eq:Xp2}{eq:Xhe42}] is justified if $X_{\antiT}^{\rm burn}\gtrsim X_{\antiHe{4}}^{\rm prompt}$. 
This translates to $\left.N_{\antiHe{4}}/N_{\antiHe{3}}\right|_{\rm inj.}\lesssim 1$, or $c\lesssim 2.1$.

We have also numerically computed the anti-helium isotope ratios $N_{\antiHe{4}}/N_{\antiHe{3}}|_\text{inj.}$ for different fixed values of $c$ using the set of Boltzmann equations described in \appref{app:reactionNetwork}; see \figref{fig:outputspace}. 
While these numerical results show reasonable overall agreement with the $N_{\antiHe{4}}/N_{\antiHe{3}}|_\text{inj.}\propto c^2$ analytical scaling derived at \eqref{eq:scalingLaw}, we find numerically that $N_{\antiHe{4}}/N_{\antiHe{3}}|_\text{inj.}$ still varies mildly with $(\Gamma,T_0,R_0)$ if $c$ is held fixed.
Nevertheless, because we have tuned the numerical constants $k_{\bar{2}}$, $k_{\bar{3}}$, and $k_{\bar{4}}$ in the analytical results to the numerical computations, we find that this mild violation of the $N_{\antiHe{4}}/N_{\antiHe{3}}|_\text{inj.}\propto c^2$ scaling makes the analytical results for that isotope ratio at worst an $\mathcal{O}(2)$ factor discrepant from the numerical results throughout the viable parameter space.
We also find that there is still reasonable agreement of the analytical and numerical results for $c$ as large as $c\sim 5$, notwithstanding the limitation $c\lesssim 2.1$ that we noted previously.

\subsubsection{Non-nuclear outputs}
\label{sec:otherOutputs}

In addition to anti-nuclei, the fireball evolution described above will result in the injection of light SM particles throughout the galaxy. 

When the fireball becomes optically thin, a burst of photons is released, with an average energy $\sim \Gamma T_{\text{thin}}' \sim 10\,\keV$. 
We assume (and this is almost certainly the case) that this dominates the integrated emission from the photosphere throughout the previous expansion.
Since the pions present in the fireball remain in thermal equilibrium until at least $T'_{\bar{n}\bar{p}}\approx 6\,\MeV \ll m_{\pi}$, they are Boltzmann-suppressed prior to decoupling and therefore no significant gamma-ray signal is expected from their decay.

Anti-neutrinos are continually produced through weak interactions, via both inelastic weak interactions and decays.
The dominant scattering production occurs when the fireball first thermalizes and its temperature is the highest. At this point, all the produced anti-neutrinos (which possess an average kinetic energy $\sim T_0$) will escape the fireball since their mean free path is $\ell_\nu \sim 1/(n \sigma) \sim 1/(G_F^2 T_0^5) \sim 100\,\text{mm}\gg R_0$, where $n \sim T_0^3$ is taken to be the (thermal) electron or positron number density since $T_0 \gg m_e$.
Additional neutrinos are produced as the pions within the fireball decay. 
These have an average kinetic energy $\sim \Gamma (m_\pi - m_\mu)$.
There may also be a neutrino contribution immediately on injection of the SM species that seed the fireball, but this is model dependent.

Charged leptons present in the fireball may also be injected into the interstellar medium after the plasma becomes optically thin. The number of such leptons remaining upon annihilation is model dependent; it is set by the initial charge asymmetry in the lepton sector. However, if the fireball as a whole is electrically neutral, this injection should be dominated by positrons with an average energy $\sim \Gamma m_e$, whose number cannot exceed the total number of anti-baryons injected.

The observability of X-ray and lepton bursts is discussed in \secref{sec:bursts}. 
We find that for the benchmark parameters sufficient to explain the AMS-02 candidate anti-helium events, detection of these additional particles is infeasible due to the low count of particles arriving at the Earth and, in some cases, their low energies.

\subsection{Summary}
\label{sec:summary}

We have identified a parameter space (see \figref{fig:fireballparameterspace}) where a sudden and spatially concentrated BSM injection of energetic anti-quarks in our Galaxy triggers (subject to certain properties of the injection) a series of events, dictated purely by Standard Model physics, that lead to relativistic anti-helium anti-nuclei being released with number ratios and Lorentz boosts roughly consistent with AMS-02 observations (we discuss in \secref{sec:prop-detection} how propagation effects modify the observed number-ratios from the injected values we have thus far discussed). 
Here, we summarize this predicted series of events and provide benchmark values for key quantities at various points of the process; we denote these benchmark quantities with a tilde and an appropriate subscript.

Following the anti-quark injection, the anti-quarks rapidly hadronize and thermalize mainly via strong and electromagnetic processes into an optically thick, adiabatically expanding fireball with conserved anti-baryon--to--entropy ratio $\tilde{\bar{\eta}}=10^{-2}$. 
This fireball then undergoes a period of rapid acceleration which turns it into a plasma shell moving with relativistic radial speed, with an average Lorentz factor $\gamma \sim \text{few}$. 
Right at the onset of this phase of its evolution, the temperature, outer radius, and thickness of the shell are in the ballpark of
\begin{align}
    \tilde{T}_0 &=100\,\MeV,          & 
    \tilde{R}_0 &= 1.5\,\text{mm}\ ,   &  
    \Delta \tilde{R}_0&\sim R_0\:. \label{eq:benchmarkParameters}
\end{align}

The subsequent evolution of the plasma shell proceeds in three sequential stages (see \figref{fig:benchexp}):
\begin{enumerate}
    \item \textit{Acceleration:} the shell continues to radially accelerate under its own thermal pressure with its average Lorentz factor $\gamma$ increasing linearly with the shell's outer radius $R$, and keeping its thickness approximately constant, $\Delta R\sim \tilde{R}_0$.
    \item \textit{Coasting:} as $\gamma$ is approaching close to the terminal radial bulk Lorentz boost $\tilde{\Gamma}=10$, the shell enters the second stage of expansion where it simply coasts with an approximately constant Lorentz factor $\gamma\approx \tilde{\Gamma}$, again keeping its thickness approximately constant, $\Delta R\sim \tilde{R}_0$. 
    \item \textit{Spreading:} the expansion timescale becomes long enough that the radial velocity difference between the innermost and outermost layers of the shell causes the shell's thickness to increase significantly over time.
\end{enumerate}

We now describe how the fireball's particle content evolves as it expands and cools down. 
Initially, while anti-nuclei are still absent due to rapid photodissociation of $\antiD$, anti-neutrons and anti-protons are kept in detailed balance by pion-mediated interconversion processes such as $\bar{n}+\pi^-\leftrightarrow \bar{p}+\pi^0$.
This continues until they finally decouple at a comoving temperature $T_{\bar{n}\bar{p}}^\prime\approx 6\,\MeV$, at which temperature their relative abundance freezes out at%
\footnote{
    For the purposes of this summary discussion, we are assuming the appropriate charge asymmetry $X_Q$ on the hadronic sector is achieved at injection (i.e., that $\mu_{\pi^+}=0$ at $T'=T'_{\bar{n}\bar{p}}$); qualitatively similar results are however obtained so long as anti-neutron--to--anti-proton ratio remains $\mathcal{O}(1)$. 
    See discussion in \secref{ss:antineutronabundance} and \appref{app:antinucleonFO}.
}
$n_{\bar{n}}/n_{\bar{p}}\approx 0.8$. 

Rapid photodissociation of any fusion-produced $\antiD$ ceases only deep in the final (spreading) expansion stage, when the comoving temperature, outer radius, and thickness of the plasma shell are about
\begin{align}
    \tilde{T}_{\rm D}^\prime &\approx 140 \,\keV,     &
    \tilde{R}_{\rm D}&\approx 2\,\text{m}\ ,          &
    \Delta \tilde{R}_D\sim 2\,\text{cm}\:.
\end{align}
The following nuclear reactions then proceed to produce light anti-nuclei (see \figref{fig:Xvsx}):
\begin{itemize}
    \item $\antiD$ production through $\bar{n}+\bar{p}\rightarrow\antiD+\gamma$.
    \item $\antiT$ production either (1) directly through $\antiD+\antiD \rightarrow \antiT+\bar{p}$, or (2) indirectly through $\antiD+\antiD\rightarrow \antiHe{3}+\bar{n}$, followed by the highly efficient $\antiHe{3}+\bar{n}\rightarrow\antiT+\bar{p}$. 
    The latter process depletes $\antiHe{3}$ and keeps its abundance low.
    \item $\antiHe{4}$ production through $\antiT+\antiD\rightarrow\antiHe{4}+\bar{n}$.
\end{itemize}
In a way somewhat analogous to how dark matter is produced in freeze-in scenarios, these processes sequentially produce nuclear anti-particles with their final (frozen) numbers satisfying $N_{\bar{p}}\approx N_{\bar{n}}\gtrsim N_{\antiD}\gtrsim N_{\antiT}\gtrsim N_{\antiHe{4}}$, and essentially no other elements (see \figref{fig:Xvsr0}). 

As the shell further expands and decreases in density, the bulk of the plasma eventually becomes transparent to photons when its comoving temperature, outer radius, and thickness are around
\begin{align}
    \tilde{T}_{\rm thin}^\prime& \approx 0.7 \,\keV,        & 
    \tilde{R}_{\rm thin} &\sim 400\,\text{m}\ ,         & 
    \Delta \tilde{R}_{\rm thin}&\sim 4\,\text{m}\:.\label{eq:thinBenchmark}
\end{align} 
At that point, the relativistic anti-nucleosynthetic products and a burst of X-ray photons are released from the plasma shell. 

While traversing the interstellar medium, the $\bar{n}$ decay to $\bar{p}$ and the $\antiT$ decay to $\antiHe{3}$ (these decay timescales are very short compared to the galactic dwell-time). 
Each fireball seeded with a total anti-baryon number $\bar{B}$ therefore contributes to the nuclear anti-particle population in the interstellar medium as follows:
\begin{align}
\begin{split}
    \tilde{N}_{\bar{p}}&\approx \bar{B}\:,\\
    \tilde{N}_{\antiD}&\approx 5.8\times 10^{-3} \bar{B}\:,\\
    \tilde{N}_{\antiHe{3}}&\approx  3.6\times 10^{-4}\bar{B}\:,\\
    \tilde{N}_{\antiHe{4}}&\approx 1.9\times 10^{-4}\bar{B}\:.
\end{split}
\end{align}
Note that these specific numerical results depend on the the benchmark values of the parameters $(T_0,R_0,\bar{\eta})$ that were chosen at \eqref{eq:benchmarkParameters} such that the resulting \emph{injected} ratio of anti-helium isotopes $\tilde{N}_{\antiHe{4}}/\tilde{N}_{\antiHe{3}}|_\text{inj.}$ reproduces the current AMS-02 candidate-event \emph{observed} value of $\approx 1/2$, and the terminal Lorentz boost of the plasma shell is $\Gamma=10$ at injection (see \figref{fig:outputspace}).
These injected values are however somewhat modified by Galactic propagation effects that we discuss and account for in the next section.

\section{Propagation and detection}
\label{sec:prop-detection}

In the previous section, we showed how anti-nucleosynthesis occurring in an expanding thermal fireball state characterized by a certain temperature, radius, anti-baryon content, and net hadronic charge asymmetry could generate, after unstable elements have decayed, both $\antiHe{3}$ and $\antiHe{4}$ in an isotopic ratio broadly consistent with the candidate AMS-02 events.

In this section, we discuss how the properties of the anti-helium (and other species) injected by such fireballs at locations within the Milky Way (MW) are processed by propagation from the source to the AMS-02 detector, as well as the necessary parameters to generate event rates consistent with the candidate AMS-02 observations.

Our analysis is predicated on the following basic assumptions: (1) all seeded fireballs have similar $T_0$, $R_0$, and $\bar{\eta}$ parameters (and hadronic charge asymmetries); and (2) a large enough number of anti-helium producing fireballs have been, and continue to be, seeded at random times up to the present day that we can neglect both spatial and temporal clumpiness in the injection and instead model it as a temporally constant and spatially smooth source.
Additionally, motivated by having dark-matter collisions seed the fireballs (see \secref{sec:DMseedModel}), we assume that (3) the spatial distribution of the injections is $\propto [n_{\text{NFW}}(r)]^2$, where $n_{\textsc{NFW}}(r)$ is a Navarro–Frenk–White (NFW) profile~\cite{NFW},%
\footnote{\label{ftnt:NFWparams}%
    Following \citeR{Cholis:2020twh}, we take the NFW profile to be normalized such that the local average DM density is $\rho_0 = 0.4 \, \text{GeV/cm}^3$~\cite{Catena:2009mf,Salucci:2010qr} at $r\sim 8\,\text{kpc}$~\cite{2019A&A...625L..10G}, and use a scale radius $R_s= 20\,\text{kpc}$~\cite{Cholis:2020twh}.} %
so that the anti-helium injection is occurring dominantly within the Milky Way itself and is peaked toward its center.

\subsection{Cosmic rays} 
\label{sec:Galprop}

The fireball injection model discussed in \secref{sec:FireballAntiNucleo} is such that all cosmic-ray species are initially injected in the vicinity of the fireball with a narrow range of velocities centered around the bulk Lorentz factor%
\footnote{%
    Although many anti-helium nuclei are produced through decay processes (and never thermalize with the fireball) [e.g., the bulk of $\antiHe{3}$ production is from $\antiT$ decay], the associated nuclear decay $Q$ values are sufficiently small that the product nuclei are always non-relativistic in the decay rest frame. As a result, their Lorentz factor in the galactic rest frame at the time of injection does not differ appreciably from $\Gamma$.} %
$\Gamma \sim 10$.

The subsequent motion of these injected cosmic rays to Earth (and hence the AMS-02 detector in low-Earth orbit) is of course diffusive in both position and momentum space~\cite{CRprop}.
In principle, we should thus pass the fireball-injected species to \textsc{galprop}~\cite{Galpropwebsite,Galproppaper} to solve the necessary transport equations and account for various propagation effects; see also \citeR{Poulin:2018wzu}.

Instead of using a modified version of \textsc{galprop} to study the propagation of injected anti-particles, we argue as follows.
\textsc{Galprop} natively solves the transport equation for positively charged nuclei. 
Of course, the opposite sign of the charge for the anti-nuclei does not impact the diffusive nature of the transport~\cite{CRprop,Cholis:2020twh}.
To be sure, there are additional annihilation reactions that can occur for anti-particles interacting with the (dominantly ordinary matter) interstellar medium (ISM), and the inelastic cross-sections scattering with the ISM also differ somewhat for particles vs.~anti-particles.
Nevertheless, at the level of precision at which we work, the annihilation cross-sections can however reasonably be ignored for our purposes, as they constitute a negligible correction to the total inelastic cross-section of the anti-particle species at energies $\gtrsim 10 \,\GeV$~\cite{Moskalenko:2001ya, Antixsec, Profumoantihelium} and can therefore be absorbed into the $\gtrsim 10\%$ uncertainties associated with the propagation~\cite{Cholis:2020twh}.
Ignoring also any other differences in the anti-particle vs.~particle inelastic cross-sections for interaction with the ISM, we employ \textsc{galprop} results for the corresponding positive charged nuclei as \emph{an approximation to} the desired results for the negatively charged anti-nuclei (e.g., we inject primary ${}^4\text{He}$ instead of $\antiHe{4}$, and read off results accordingly, etc.).
This approximation could of course be revisited; however, as we shall see, the results of this approximate treatment indicate a ratio of observed anti-helium fluxes that differs by only an $\mathcal{O}(1)$ numerical factor as compared to the ratio injected by the fireballs; it is therefore unclear whether a modification of propagation code as in \citeR{Poulin:2018wzu} to more correctly treat the anti-particle propagation is justified given other, larger uncertainties in our scenario.%
\footnote{%
    We note that such modification was however important for \citeR{Poulin:2018wzu} to achieve accurate results, as one of the main issues addressed in that work was to refine predictions for the fully propagated anti-particle secondary fluxes produced by the primary \emph{ordinary matter} cosmic-ray spectra.} %

Specifically, we model the injection of anti-cosmic rays of species $i$ by specifying \textsc{galprop} source terms for the corresponding \emph{positively-charged, ordinary-matter species}, which we denote here as ${i^+}$:
\begin{align}
    q_{i^+}(\bm{r},\mathcal{R}) \propto X_i F_i(\mathcal{R}) [n_{\text{NFW}}(r)]^2 \:,
\end{align}
where $X_i$ is taken to be the isotopic abundance for the anti-particle species $i$ from \secref{sec:antinucleosynthesis}, $F_i(\mathcal{R})$ is taken to be a narrow top-hat function centered at the rigidity $\mathcal{R}$ corresponding%
\footnote{%
    While the fireballs inject each species $i$ at a single rigidity, the width of the top-hat (chosen here to be $\sim 10\%$ of the central value for numerical reasons) is inconsequential as long as it is subdominant to the momentum-space diffusion occurring during propagation, which we verify \emph{a posteriori}. 
    In order to specify a fixed $\Gamma$, and hence a different $\mathcal{R} =\mathcal{R}(q,m,\Gamma)$ dependence for each species, \textsc{galprop} had to be run multiple times: in each run, a single species $i^+$ was injected at the required rigidity $\mathcal{R}_i$; the output spectra from each such run were then summed with weights $X_i$.} %
to Lorentz factor $\Gamma$ for species $i$ as injected by the fireball, and $n_{\text{NFW}}(r)$ is the NFW profile. 
We of course then also read off local flux results for the species $i^+$, and impute those to species $i$.
The source terms are normalized such that the imputed total injection rate of anti-baryon number, summed over all $i$ species and integrated over the whole \textsc{galprop} simulation volume, is $\Gamma_{\text{inj.}}$.

Furthermore, we approximate the decay of $\bar{n}$ to $\bar{p}$ and $\antiT$ to $\antiHe{3}$ as occurring instantaneously at the fireball location, and we thus consider only of the anti-particle species $i \in \{\bar{p},\antiD,\antiHe{3},\antiHe{4}\}$ when running \textsc{galprop} via the above procedure, in the ratios specified at%
\footnote{%
   As discussed in \secref{sec:antinucleosynthesis}, while the numerical solutions to the Boltzmann equations are more accurate than the analytical results at \eqrefRange{eq:Xp2}{eq:Xhe42}, the tuning of those analytical results to the numerics via the constants $k_{\bar{2}}$, $k_{\bar{3}}$, and $k_{\bar{4}}$, makes the analytical results sufficiently accurate for our purposes here, notwithstanding the minor violation of the scaling of $N_{\antiHe{4}}/N_{\antiHe{3}}|_{\text{inj.}} \propto c^2$ discussed in \secref{sec:finalProducts}; see also \figref{fig:outputspace}.
   } %
\eqrefRange{eq:Xp2}{eq:Xhe42} as a function of $X_{\antiD}$.

For the \textsc{galprop} diffusion model, we adopt the propagation parameters for ``ISM~Model~I'' in \citeR{Cholis:2019ejx}; within the range of parameters consistent with existing cosmic-ray observations, our results are largely insensitive to the choice of transport model. 
We also neglect the modulation of the cosmic-ray fluxes at Earth due to the heliospheric magnetic field: in the force-field approximation, solar modulation is governed by a single parameter known as the Fisk potential $\phi_F \sim 1\,\text{GV}$ which, at high energies, corrects the flux at most by a factor $\sim e\phi_F/E \ll 1$~\cite{WinklerCRs}.

Note also that the typical mean-free path for an anti-nucleus traveling with $\Gamma \sim 10$ is $\ell_{\text{mfp}} \sim 1\,\text{pc}$, meaning that it stays within the galaxy for a duration $t_{\text{diff}}\sim h_{\text{MW}}^2/\ell_{\text{mfp}} \sim 10^6\,\text{yr}$ where $h_{\text{MW}} \sim 1\,\text{kpc}$ is the thickness of the MW disk. 

\begin{figure}
    \centering
    \includegraphics[width=0.95\columnwidth]{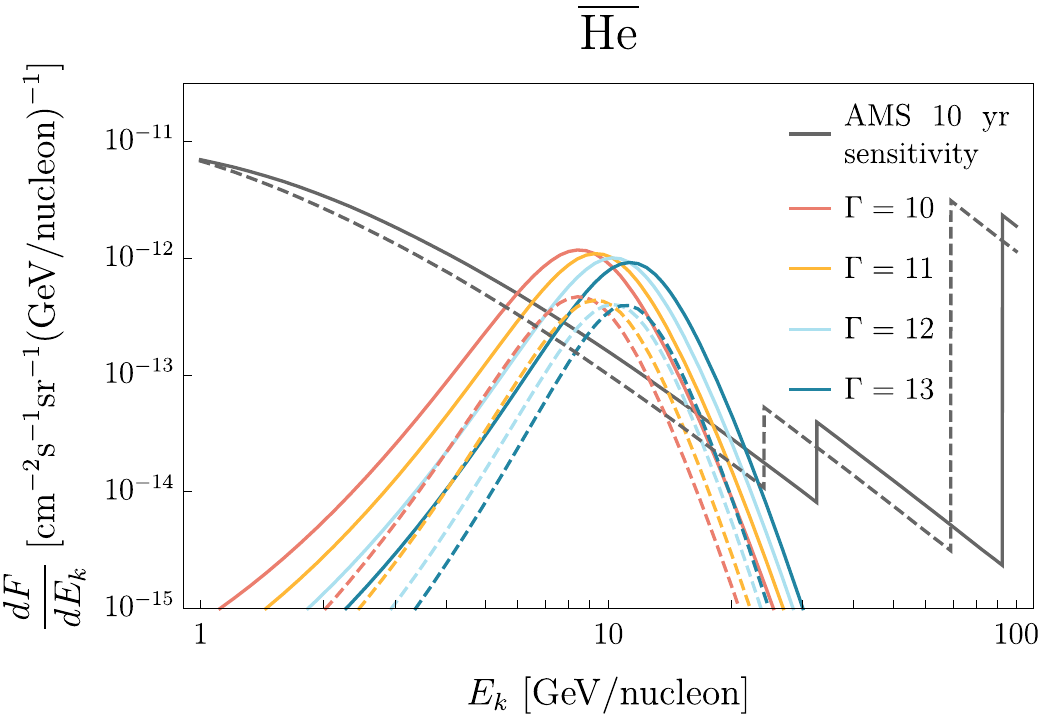}\\[0.5ex]
    \includegraphics[width=0.95\columnwidth]{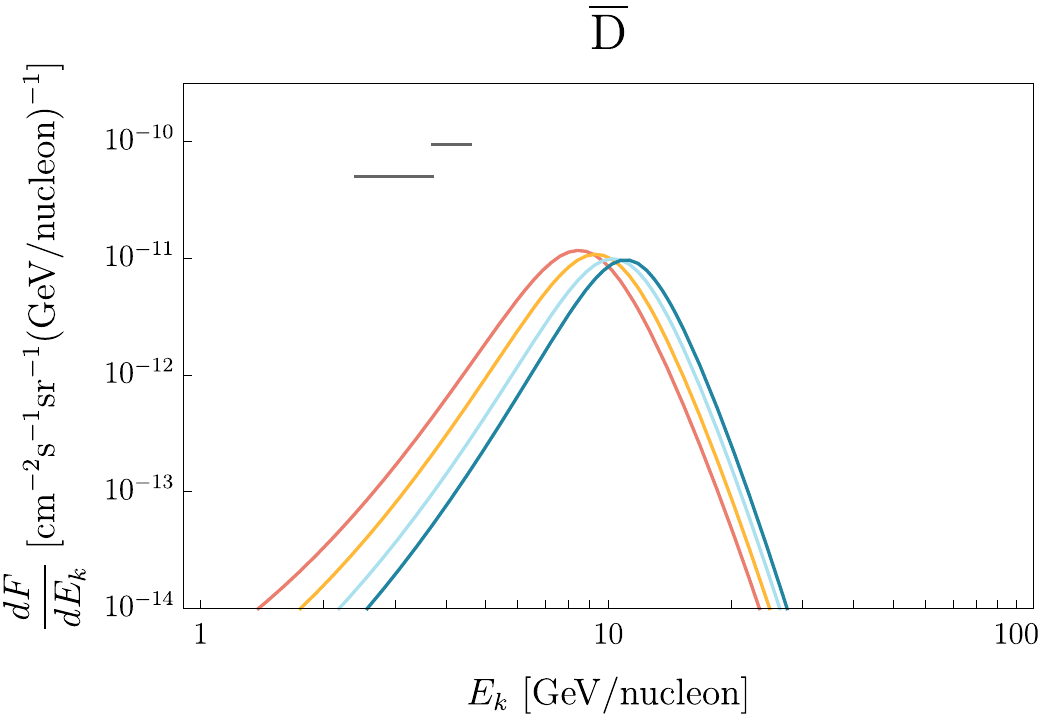}\\[0.5ex]
    \includegraphics[width=0.95\columnwidth]{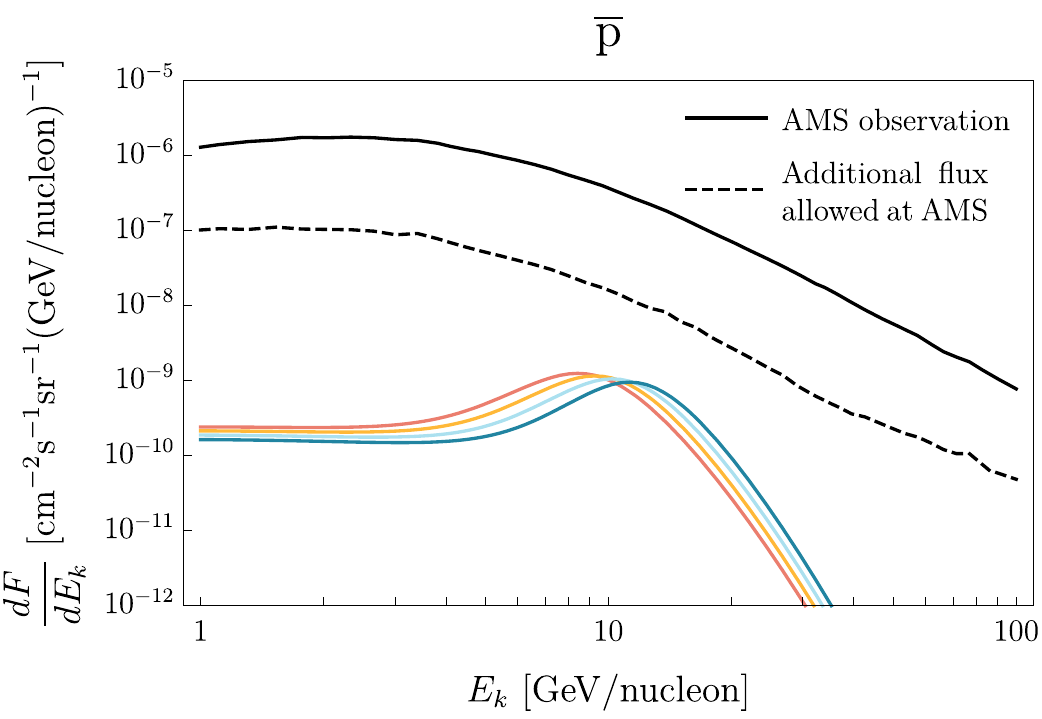}
    \caption{\label{fig:spectra}%
        Predicted spectra of anti-nuclei upon propagation to the Earth as a function of kinetic energy per nucleon $E_k$, for different choices of the initial Lorentz factor $\Gamma$, as annotated by the different colored lines. 
        The best available sensitivity of AMS to anti-helium (top panel) and anti-deuterium (middle panel) events~\cite{Kounine:2010js, AMSantihesens, AMSHeflux, AMSdbarsens, Aramaki:2015pii} are presented for comparison in gray (see text).
        In the top panel, the solid and dashed lines correspond to the isotopes ${}^3\overline{\rm He}$ and ${}^4\overline{\rm He}$, respectively.
        The anti-proton flux observed by AMS is shown (bottom panel) by a solid black line~\cite{AMS:2016oqu}, while the \emph{additional} flux required to exceed the uncertainties on the measured flux is shown as the dashed black line in that same panel; fireball production is a negligible source of galactic $\bar{p}$ at these parameter values. 
        Throughout this figure, we fix $\Gamma_{\rm inj} = 4\times 10^{35} \,\text{anti-nuclei s}^{-1}$, while the isotopic abundances at the fireball are set by $X_{\antiD} = 5\times 10^{-3}$; see \eqref[s]{eq:Xhe3} and~(\ref{eq:Xhe4}). 
       }
\end{figure}

A set of example post-propagation spectra at the position of the Earth is shown in \figref{fig:spectra}, alongside AMS-02 sensitivity curves or observations. 
In each case, the propagated flux of injected anti-particles peaks at kinetic energies corresponding to the Lorentz parameter $\Gamma\sim 10$ [i.e., $\mathcal{O}(10\,\text{GeV/nucleon})$].

\begin{figure}
    \centering
    \includegraphics[width=0.95\columnwidth]{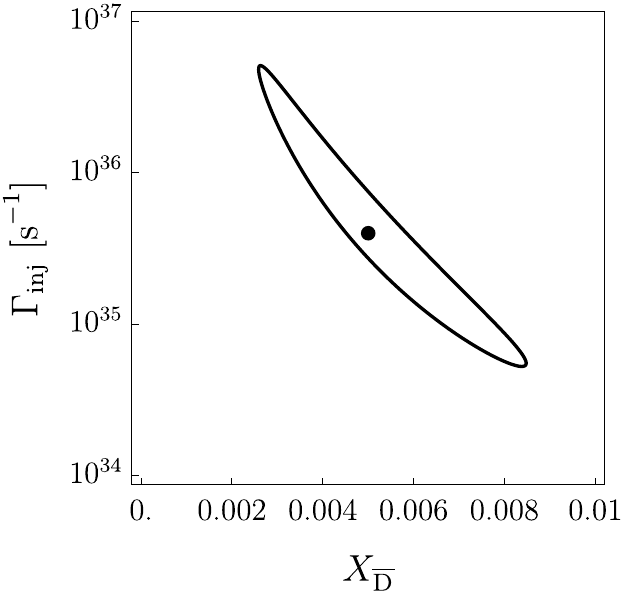}
    \caption{\label{fig:InjRatePlot}%
        The region of parameter space for which AMS-02 would be expected to observe 3 events of ${}^4\overline{\rm He}$ and 6 events of ${}^3\overline{\rm He}$ in $T\sim 10$\,years, with the injection taking place at $\Gamma=10$.
        (The allowed region is approximately identical over the range $8.5 \leq \Gamma \leq 13$.)
        The anti-nucleus injection rate $\Gamma_{\rm inj}$ normalizes the total rate of fireball injection of anti-cosmic rays into the MW, whereas the individual source isotopic ratios $X_i$ are set by the fireball anti-deuterium abundance $X_{\antiD}$: $X_i = X_i(X_{\antiD})$ via \eqrefRange{eq:Xp2}{eq:Xhe42}.
        The predicted number of anti-helium events is allowed to independently vary from the fiducial quantity within the 68\% confidence interval of the joint probability distribution [see \eqref{eq:chisqreq}].
        The black dot shows the fiducial parameter values referred to in \figref{fig:spectra} and \eqref{eq:inj_analytic}.
        }
\end{figure}

The expected number of anti-helium events accumulated at AMS over 10 years (at the 95\% CL) can be computed using the output spectra from \textsc{galprop} and the published AMS sensitivity to the flux ratio between anti-helium and helium~\cite{Kounine:2010js, AMSantihesens}. 
This is done by recasting the anti-helium acceptance of each energy bin in terms of this $\overline{\text{He}}/\text{He}$ sensitivity in conjunction with published helium data~\cite{AMSHeflux}, following the procedure outlined in Appendix B of \citeR{Winkler:2020ltd}; see our \appref{app:AMSantiHeSensitivity} for a brief review.
Assuming the number of events follows Poissonian statistics, and allowing for the joint probability of the predicted numbers of $^3\overline{\text{He}}$ and $^4\overline{\text{He}}$ to deviate from their AMS-02 tentative observed values within the 68\% confidence interval, the $(X_{\antiD},\Gamma_{\text{inj.}})$ parameter space consistent with the AMS-02 candidate anti-helium events is shown in \figref{fig:InjRatePlot}. 
Explicitly, we require~\cite{PDG}
\begin{align}
    (\Delta \chi^2)_{^4\overline{\text{He}}} + (\Delta \chi^2)_{^3\overline{\text{He}}} < 2.3\:.
    \label{eq:chisqreq}
\end{align}
The events for each isotope are assumed to be independent, satisfying
\begin{align}
    (\Delta \chi^2)_i = 2\left[N_i^{\text{th}} - N_i^{\text{obs}} - N_i^{\text{obs}} \ln \left( \frac{N_i^{\text{th}}} {N_i^{\text{obs}}} \right) \right],
    \label{eq:chisqdef}
\end{align}
where $N_i^{\text{th}}$ is the number of events for species $i$ predicted by \textsc{galprop} given a set of model parameters $\{ \Gamma_{\text{inj.}},X_{\antiD} \}$, whereas $N_i^{\text{obs}}$ is the fiducial number of candidate $i$ events reported by AMS-02. 
For the benchmark $X_{\antiD}=5\times 10^{-3}$ shown by the black dot in \figref{fig:InjRatePlot}, the isotope ratio $N_{^4\overline{\text{He}}}/N_{^3\overline{\text{He}}}$ changes from $N_{^4\overline{\text{He}}}/N_{^3\overline{\text{He}}}|_{\text{inj.}} \approx 0.33$ at injection to an observed value of $N_{^4\overline{\text{He}}}/N_{^3\overline{\text{He}}}|_{\text{obs.}}\approx 0.55$--$0.60$ (depending on the choice of $\Gamma$).

Part of this change in the isotope ratio from injection to observation has to do with physical propagation effects. 
For instance, spallation of $^4\overline{\text{He}}$ onto the interstellar medium; however, only $\sim 10$\% of the $\antiHe{3}$ abundance arises from this effect, which moves the isotopic ratio by only an $\mathcal{O}(1)$ factor. 
Moreover, that effect would tend to drive the ratio in the other direction (i.e., it reduces $\antiHe{4}$ and increases $\antiHe{3}$).
We also expect there to be some mild differences in diffusion for different-mass isotopes at the same kinetic energy per nucleon owing to different rigidities.
However, we identify the larger part of the change to arise not from propagation effects at all, but rather from the fact that the AMS-02 anti-helium sensitivity, which we take to be given as the same function of rigidity $\mathcal{R}$ for all anti-helium isotopes (see \appref{app:AMSantiHeSensitivity}), is not flat as a function of $\mathcal{R}$; on the other hand, the post-propagation energy-per-nucleon spectra of $\antiHe{3}$ and $\antiHe{4}$ are almost the same (since they have the same energy-per-nucleon at injection), resulting in rigidity distributions for the two species that peak at different values of $\mathcal{R} = (m/Q)\sqrt{(E/m)^2-1}$ (i.e., when stated in terms of the kinetic energy per nucleon, the AMS-02 sensitivity that we assume  differs for species with different charge-to-mass ratios; see the top panel of \figref{fig:spectra}).
Ultimately, however, this $\mathcal{O}(2)$-factor change could easily be absorbed into a slightly different parameter point if any of the assumptions leading to this effect are found to be inaccurate.

For the case of anti-deuterium, a robust detection relies on the rejection of backgrounds (particularly $\bar{p}$ and $\text{He}$) that are substantially more abundant. 
As described in \citeR{Aramaki:2015pii}, the latest published AMS-02 anti-deuterium sensitivity curve (shown in \figref{fig:spectra}) is based on an earlier superconducting-magnet configuration for AMS-02, rather than the permanent-magnet configuration actually in use, and cuts off at $\approx 5 \,\text{GeV/nucleon}$. 
Nevertheless, taking that sensitivity, only injections taking place with $\Gamma \lesssim 12$ would result in $\gtrsim 0.1$ expected $\antiD$ events in the parameter space of interest, but it would be challenging on the basis of that sensitivity to simultaneously account for the 7 candidate anti-deuterium events reported in \citeR{TingCERNslides2023} in addition to the candidate anti-helium events. 
That said, an updated study of the AMS-02 anti-deuteron sensitivity would be required to accurately determine whether a single choice of parameters $(X_{\antiD}, \Gamma_{\text{inj.}})$ could achieve this.

Finally, anti-proton events arising from the fireball are also expected to be observed at AMS-02; indeed, as seen in \figref{fig:spectra}, this flux dominates those of the other species in the parameter regime of interest. 
However, conventional astrophysical sources are responsible for an anti-proton flux that is greater than that created by the fireballs by a factor of $\sim 10^3$~\cite{AMS:2016oqu}. 
Producing sufficiently many events to exceed the uncertainty on this measurement (tantalizing in light of the anti-proton excess of $\sim 10\%$ at similar energies~\cite{Cholis:2019ejx}) would require a greater anti-nucleus injection rate than is favored by the candidate anti-helium observations; see however \figref{fig:AbundanceswithChemPot} in \appref{app:antinucleonFO} (and related discussion) for an alternative parameter point that may be more interesting from this perspective.
In contrast to the heavier species, secondary production contributes significantly to the $\bar{p}$ flux at low kinetic energy per nucleon, resulting in an approximately flat spectrum that falls off only for $E_k \lesssim 0.7\text{ GeV/nucleon}$ (i.e., below the range shown in the lower panel of \figref{fig:spectra}). 
However, this plateau is observationally unimportant as compared to the peak in the spectrum arising from the primary fireball-injected $\bar{p}$, as the former lies further below the detectable flux level than the latter. 
This feature is absent from the spectra for heavier antinuclei in the ranges plotted owing to the inefficiency of secondary production of heavy nuclei (as discussed in, e.g., \secref{sec:Intro}).

\subsection{Other signatures (indirect detection)}
\label{sec:bursts}

In this section, we examine the detectability of X-ray, anti-neutrino, and positron bursts which arise from the fireball model (discussed in \secref{sec:otherOutputs}).
While fireball injection of anti-nuclei is envisaged as a continuous process compared to the galactic dwell-time for these (diffusively transported) particles, the rapid expansion timescale $\tau'$ and low galactically-integrated injection event rates $\Gamma_{\text{coll.}} \sim 3\,\text{s}^{-1}$ involved (see \secref{sec:BlobCollRates} for this estimate) mean that the non-nuclear outputs from independent fireball injections are temporally well-separated at the Earth and so are best treated independently (with the exception of positrons, whose transport is also diffusive).

At the fluxes required to explain the AMS-02 candidate anti-helium events, there should be $N_{\text{inj.}} \sim \Gamma_{\text{coll.}}T \sim 10^9$ events occurring within the galaxy over $T \sim 10\,\text{yr}$.
Suppose that we made an observation to look for their non-nuclear products that has a duration $T_{\text{obs.}}$ and that observes a fraction $f_{\text{sky}}$ of the whole sky.
In this time, we estimate that the closest observable injection event would occur at a distance $d_{\text{inj.}} \sim 16\,\text{pc} \times (10\,\text{yr}/T_{\text{obs.}})^{1/3}\times f_{\text{sky}}^{-1/3}$ from Earth, if we assume that the injections arise from collisions of the sub-component composite DM states whose benchmark parameters we discuss below in \secref{sec:BlobCollRates}.

Photons are rapidly released in a burst when the fireball arising from the injection becomes optically thin, at size $R_{\text{thin}} \sim 400\,\text{m}$; see \eqref{eq:thinBenchmark} for this benchmark. 
We estimate the energy released in this burst to be $\sim g_* (T_{\text{thin}}')^4 \Gamma^2 R_{\text{thin}}^2\Delta R_{\text{thin}}\sim g_* (T'_{\text{thin}})^4 R_{\text{thin}}^3 \sim 6\times 10^{27}\,\text{erg}$ [using the benchmark at \eqref{eq:thinBenchmark}, and $g_* \sim 2$], which results in the emission of $N_\gamma \sim 6\times 10^{35}$ photons with average energy $\Gamma T'_{\text{thin}} \sim 7\,\keV$.
X-ray telescopes sensitive to such events have an effective area $A_\gamma \sim 1000\,\text{cm}^2$~\cite{INTEGRAL, NICER}, a field of view of at most $\Omega_\gamma \sim 100$ square degrees (i.e., $f_{\text{sky}} \sim 2\times 10^{-3}$), and mission lifetimes of $\mathcal{O}(\text{years})$.
The total expected number of photons arriving at the detector from the closest release that would occur would be $\sim N_\gamma (A_\gamma/d_{\text{inj.}}^2) \sim 0.2 \times f_{\text{sky}}^{2/3} \times (T_{\text{obs.}}/10\,\text{yr})^{2/3}$. 
Even if we conservatively ignore the finite field of view (i.e., set $f_{\text{sky}}\sim 1$) and take $T_{\text{obs.}} \sim 10\,\text{yr}$, this is too faint to observe.
There may additionally be a diffuse background of X-ray photons due to the cumulative injections taking both intra- and extra-galactically over extended periods of time, but we have not estimated this here.

Anti-neutrinos may be injected promptly when the fireball thermalizes, or indirectly through the decay of injected pions.
The fraction of the total energy $g_* T_0^4 R_0^3 \sim 4\times 10^{32}\,\text{erg}$ [see \eqref{eq:benchmarkParameters} for this benchmark; we set $g_* \sim 5.5$] that is carried by the prompt anti-neutrinos is $\sim G_F^2 T_0^5 \tau' \sim 1\%$ (for $\tau' \sim R_0$ as at early times; see \figref{fig:benchexp}), corresponding to $N_{\nu,\,\text{scatter}} \sim 3\times 10^{34}$ particles with an average energy of $T_0 \sim 100\,\MeV$. 
As far as \text{(anti-)neutrinos} from decay are concerned, thermal pions have a number density $n_{\pi} \sim (m_\pi T/(2\pi))^{3/2}\exp(-m_{\pi}/T)$ within the fireball, so the number of neutrinos injected due to their decay is $N_{\nu,\,\text{decay}} \sim n_\pi R^3 (\tau'/\tau_\pi) \lesssim 10^{31}$, where $\tau_\pi \sim 10^{-8}\,\text{s}$ is the charged pion lifetime. 
Their average energy is also $\sim \Gamma (m_\pi - m_\mu) \sim 100\text{ MeV}$.
The decay contribution is thus sub-dominant.
Given a neutrino detector in this energy range, with an area $A_\nu \sim (100\,\text{m})^2$~\cite{Super-K}, roughly $\sim N_\nu A_\nu/(d_{\text{inj.}})^2 \sim 10^3 \times (T_{\text{obs.}}/10\,\text{yr})^{2/3}$ neutrinos would pass through the detector as a result of the nearest injection in the observation time.%
\footnote{%
    We took $f_{\text{sky}}\sim 1$ here to reflect that Earth is quite transparent to neutrinos with energies below a few TeV, so the detector has full-sky coverage; see, e.g., \citeR{Donini:2018tsg}.} %
The probability of a single neutrino interacting in the detector is $p_\nu \sim n_T G_F^2 E_\nu^2 d_T$ where $n_T \sim 3\times 10^{22}/\text{cm}^3$ is the target (e.g., water~\cite{Super-K}) number density, and $d_T$ is the target thickness.
Taking $E_\nu\sim 100\,\MeV$ and $d_T\sim 100\,\text{m}$ as relevant for our assumptions, we find $p_\nu \sim 10^{-13}$, so the expected number of detectable events%
\footnote{%
    By way of comparison, the Kamionkande-II detector detected only 11 neutrino interactions of energies of $\mathcal{O}(10\,\text{MeV})$ from SN1987A~\cite{PhysRevLett.58.1490}, which released $\sim 10^{58}$ neutrinos of all flavors~\cite{PhysRevLett.58.1490} at a distance of $\sim 50\,\text{kpc}$~\cite{LMCdistance}, leading to $\sim (\text{few}) \times 10^{16}$ electron anti-neutrinos passing through the $\mathcal{O}(200\,\text{m}^2)$ detector area (varies slightly depending on the orientation relative to the source)~\cite{PhysRevLett.58.1490}.
    Taking $E_\nu\sim 10\,\MeV$ and $d_T\sim 10\,\text{m}$ gives $p_\nu\sim 10^{-16}$, which is broadly consistent with these numbers, demonstrating the consistency of our estimate in the text.
} %
is $\sim 10^{-10} \times(T_{\text{obs.}}/10\,\text{yr})^{2/3} \ll 1$ for any conceivable observation duration.

Positrons injected by the fireball are potentially observable in two ways: directly as cosmic rays, and indirectly through $511\,\text{keV}$ photons produced when they annihilate with electrons in the ISM. 
However, the injected positron energies are $\sim \Gamma m_e \sim 10\,\text{MeV}$, which falls below the AMS-02 sensitivity threshold (even without accounting for further energy loss during propagation), likely making them unobservable.%
\footnote{%
    Unsurprisingly given that solar-modulation effects severely impact sub-GeV positrons~\cite{2013LRSP...10....3P}, this energy also falls below the lowest-energy positron measurements reported by PAMELA~\cite{PhysRevLett.111.081102}, HEAT~\cite{DuVernois_2001}, CAPRICE94~\cite{Boezio_2000}, AMS-01~\cite{AMS01positrons}, or FERMI~\cite{PhysRevLett.108.011103,Fermi-LAT:2017bpc}.
    Relevant Voyager 1 data for the sum of electrons and positrons down to energies of $\mathcal{O}(10\,\MeV)$ from periods after it crossed the heliopause are available~\cite{doi:10.1126/science.1236408} (see also \citeR{Boudaud:2016mos}); we do not however pursue these constraints further in this work.
    } %
For the 511\,keV emission, the total integrated positron injection rate from fireballs cannot exceed the total integrated anti-baryon injection rate (cf.~the discussion about $X_Q<0$ in \appref{app:antinucleonFO}), which from the numbers shown in \figref{fig:InjRatePlot} is maximally $\Gamma_{\text{inj.}} \sim (\text{few}) \times 10^{36}\,\text{s}^{-1}$; this is much smaller than the integrated positron injection rate of $\sim 10^{43}\,\text{s}^{-1}$ that explains 511\,keV emission in the MW~\cite{Prantzos:2010wi}.
The fireball-injected positrons annihilating therefore likely only contributes a sub-dominant 511\,keV flux.

On the basis of these estimates, these non-nuclear products thus do not appear to be observable; however, a more detailed investigation of these (indirect-detection) signatures of this class of models may be worthwhile in future work.

\section{A dark-matter origin for the fireballs?} 
\label{sec:DMseedModel}

Thus far, we have operated under the assumption that a rapid, localized injection of energetic Standard Model anti-baryons can be achieved in order to seed the fireballs.
A BSM mechanism is required to explain these injections.

In this section, we discuss whether the collisions of large, composite dark-matter states that carry anti-baryon number may be able to provide such a mechanism.
Assuming that a substantial fraction of the mass energy of such colliding DM states can be promptly converted to SM anti-quarks as a result of dynamics triggered by the collision, we demonstrate in \secref{sec:BlobCollRates} that the requisite injection rate of anti-baryon number could be achieved for DM states with certain bulk physical properties (i.e., total mass, number of constituents, and physical size).
In \secref{sec:fireballMapping} we then show that, provided certain benchmarks can be realized in the conversion of the dark-state mass energy to the SM, fireballs with appropriate bulk parameters $(T_0,R_0,\bar{\eta})$ could also be seeded.
As this is suggestive, in \secref{sec:DMmodelIdeas} we then advance some speculations toward specific microphysical DM models in which we suspect this could possibly occur.
We emphasize, however, that we have not settled the question of whether one or more of these models actually do in fact realize the required dynamics.
While we intend to return to this open question in future work, we also encourage other work on this point.

\subsection{Collision rates} 
\label{sec:BlobCollRates}

Suppose that a fraction $0<f_{\textsc{dm}}\leq1$ of the DM energy density is comprised of large, cosmologically stable composite objects.
The existence, formation, and signatures (or lack thereof) of a variety of objects of this type have been subjects of extensive study in the literature~\cite{Witten:1984rs,Krnjaic:2014xza,Wise:2014jva,Wise:2014ola,Hardy:2014mqa,Gresham:2017zqi,Gresham:2017cvl,Gresham:2018anj,Grabowska:2018lnd,Bai:2018dxf,Bai:2018vik,Hong:2020est,Gross:2021qgx,Zhitnitsky:2021iwg,Ebadi:2021cte,Acevedo:2021kly,Acevedo:2020avd,Jacobs:2014yca,Kribs:2016cew,Narain:2006kx,Curtin:2019ngc} (see also \citeR[s]{Asadi:2021yml, Asadi:2021pwo}).
For the purposes of the estimate we give here, we assume%
\footnote{%
    Slightly different assumptions could be made (and indeed are made in the following subsection), but would mostly just lead to parameter-space remappings in the discussion that follows here: for instance, we could take the baryon number per constituent to be something other than $-1$, but still of $\mathcal{O}(-1)$; e.g.,~$-1/3$ per constituent.
    Likewise, the total mass of the composite object could get corrections from binding energy or relativistic motion of its constituents.
} %
that these objects have a mass dominated by $N_{\textsc{dm}}$ constituents, each with mass $m_{\textsc{dm}}$ and baryon number of $-1$, giving them a total mass $M_{\textsc{dm}} \sim N_{\textsc{dm}} m_{\textsc{dm}}$ and total baryon number $B_{\textsc{dm}} = - N_{\textsc{dm}}$; we take their radius%
\footnote{\label{ftnt:radii}%
    Note that the initial fireball radius $R_0$ discussed in \secref{sec:FireballAntiNucleo} can be very different from the dark-blob radius $R_{\textsc{dm}}$.
    } %
to be $R_{\textsc{dm}}$.

Taking a collision cross-section 
\begin{align}
\sigma \sim \pi R_{\textsc{dm}}^2 \Sigma\: , \label{eq:collisionCrossSection}
\end{align} 
where $\Sigma\geq 1$ is a Sommerfeld-like enhancement to the geometrical cross-section, the collision rate of two such blobs integrated over the whole $V_{\textsc{mw}} = 40\times40\times 11\,\text{kpc}^3$ simulation volume used in \textsc{galprop} (see \secref{sec:Galprop}), can be estimated as%
\footnote{%
    The blob collision rate can be significantly modified if the blobs formed binaries in the early Universe~\cite{Diamond:2021dth,Bai:2023lyf,Banks:2023eym}.} %
\begin{align}
    \Gamma_{\text{coll.}} &\sim \int dV \left( \frac{f_{\textsc{dm}} \rho_{\textsc{dm}}(r)}{M_{\textsc{dm}}} \right)^2 (\pi R_{\textsc{dm}}^2\Sigma) v_{\textsc{dm}} \label{eq:GammaColl}\\
    &\equiv V_{\textsc{mw}} \left( \frac{f_{\textsc{dm}} \rho_{0}}{M_{\textsc{dm}}} \right)^2 (\pi R_{\textsc{dm}}^2\Sigma) v_{\textsc{dm}} \times \mathcal{I}_{\textsc{dm}}^{\textsc{mw}} \label{eq:GammaColl2}\\
    &\sim 3 \,\text{s}^{-1} \times \left(\frac{f_{\textsc{dm}}}{ 0.01} \right)^2 \lb( \frac{\Sigma}{1} \rb) \nl\times \lb( \frac{N_{\textsc{dm}}}{ 5\times 10^{36} } \rb)^{-2} \left(\frac{m_{\textsc{dm}}}{ 10\,\GeV}\right)^{-2} \left(\frac{R_{\textsc{dm}}}{1\,\text{m}}\right)^2. \label{eq:GammaColl3}
\end{align}
At \eqref{eq:GammaColl}, $\rho_{\textsc{dm}}(r)$ is the DM energy density at galactocentric radius $r$, as specified in the NFW model~\cite{Navarro:1995iw} (see \secref{sec:Galprop}), and $v_{\textsc{dm}}\sim 10^{-3}$ is the typical velocity of blobs in the galaxy (we take this to be constant over $r$). 
At \eqref{eq:GammaColl2}, we have have defined $\mathcal{I}_{\textsc{dm}}^{\textsc{mw}} \equiv \int (dV/V_{\textsc{mw}})\lb[\rho_{\textsc{dm}}(r)/\rho_0\rb]^2 \approx 0.85.$

In the above numerical estimate, we considered benchmark blobs with radius $R_{\textsc{dm}} \sim 1\,\text{m}$, number of constituents $N_{\textsc{dm}}\sim 5\times 10^{36}$, and constituent mass $10\,\GeV$, giving a total blob mass of $M_{\textsc{dm}}\sim 9\times 10^{10}\,\text{kg}$ (roughly the mass of a typical Main Belt asteroid with a diameter of a few-hundred meters) and an average density $\sim 2\times 10^{7}\,\text{g/cm}^3$ (while much more dense than ordinary matter, this is still less dense than a non-extremal SM white dwarf).%
\footnote{%
    Such an object, while very dense, is still many orders of magnitude larger than its own Schwarzschild radius: $R_S \sim 2M_{\textsc{dm}}/M_{\text{Pl.}}^2 \sim 10^{-16}\,\text{m}$.} %
We have also assumed that they constitute only about $\sim 1$\% of the DM, and have conservatively neglected any Sommerfeld-like enhancement to the collisions rate.

We assume further that, upon such pair-wise collisions, the DM states can be destabilized in such a way that a fraction $0<f_{\textsc{sm}}\leq 1$ of their combined number of constituents is converted to Standard Model anti-quarks, seeding into the SM a total baryon number $B_{\text{inj.}} = - 2f_{\textsc{sm}} N_{\textsc{dm}}$ per collision.
The injection rate of anti-baryon number (cf.~\figref{fig:InjRatePlot}) can then be estimated as 
\begin{align}
    \Gamma_{\text{inj.}} &\sim |B_{\text{inj.}}| \Gamma_{\text{coll.}} \\
    &\sim 4\times 10^{35}\,\text{s}^{-1} \times \lb( \frac{ f_{\textsc{sm}}}{1.5\times 10^{-2}} \rb) \left(\frac{f_{\textsc{dm}}}{0.01}\right)^2 \lb( \frac{\Sigma}{1} \rb) \nl \times \left(\frac{N_{\textsc{dm}}}{5\times 10^{36}}\right)^{-1} 
    \left(\frac{m_{\textsc{dm}}}{10 \,\GeV}\right)^{-2} \left(\frac{R_{\textsc{dm}}}{1\,\text{m}}\right)^2. \label{eq:inj_analytic}
\end{align}

Importantly, for these benchmark bulk DM parameters, we find that this could achieve the injection luminosity required to obtain the correct fiducial number of candidate anti-helium events at AMS-02, $\Gamma_{\text{inj.}} \sim 4\times 10^{35}\,\text{s}^{-1}$ (cf.~\figref{fig:InjRatePlot}), for $f_{\textsc{sm}} f_{\textsc{dm}}^2 \Sigma \sim 10^{-6}$. 
That is, the requisite injection rate could be achieved even with large composite DM that is a sub-component of the total DM density and that is not very efficient in converting constituent anti-baryon number to SM anti-quarks upon collision, all without any Sommerfeld-like enhancement to the collision rate.
Of course, these conclusions change for different DM bulk parameter ranges, per \eqref{eq:inj_analytic}.

Note also that the typical inter-collision time here, $1/\Gamma_{\text{coll.}} \sim 0.4\,\text{s}$, is $\sim 14$ orders of magnitude shorter than the diffusion time $t_{\text{diff}} \sim 10^6\,\text{yr}$ for the charged cosmic rays to escape the MW~\cite{Cholis:2019ejx}, which also justifies \emph{a posteriori} treating the injection as a roughly constant-in-time source term in \textsc{galprop} (see \secref{sec:Galprop}).

\subsection{Mapping to fireball parameters} 
\label{sec:fireballMapping}

While we cannot robustly estimate the fireball parameters $(T_0,R_0,\bar{\eta})$ that would obtain from such collisions without a model, we can show that these dark composite DM states would be able, at the order of magnitude level, to generate approximately the benchmark fireball parameters discussed at \eqref{eq:benchmarkParameters}.

If we assume that $f_{\textsc{sm}}$ is also the fraction of energy injected into the SM anti-quarks, then the energy injected is $E_{\text{inj.}} \sim m_{\textsc{dm}}|B_{\text{inj.}}|$.
If we take $T_0 \lesssim 100\,\MeV$ (verified \emph{a posteriori}), this energy must go to the mass-energy of $|B_{\text{inj.}}|$ anti-nucleons (they dominate the fireball anti-baryonic output) and the remainder to $g_* \sim 5.5$ relativistic species.
That is, $E_{\text{inj.}} \sim m_{\textsc{dm}}|B_{\text{inj.}}| \sim m_p |B_{\text{inj.}}| + g_* ( \pi^2 T_0^4 / 15 ) ( 4\pi R_0^3 / 3 )$.
Therefore, 
\begin{align}
    T_0  &\sim \lb[ \frac{45f_{\textsc{sm}}}{2\pi^3 g_*} (m_{\textsc{dm}}-m_p) \frac{N_{\textsc{dm}}}{R_0^{3}} \rb]^{1/4} \\
    &\sim 62\,\MeV \times \lb( \frac{f_{\textsc{sm}}}{1.5\times 10^{-2}} \rb)^{1/4} \lb( \frac{g_*}{5.5} \rb)^{-1/4} \nl \times \lb( \frac{N_{\textsc{dm}}}{5\times 10^{36} } \rb)^{1/4} \lb( \frac{R_0}{3.6\,\text{mm}} \rb)^{-3/4}  \nl \times \lb( \frac{ m_{\textsc{dm}}-m_p }{10\,\GeV-m_p} \rb)^{1/4}  \:,
\end{align}
where we have assumed that the fireball injection happens in a region of radius%
\footnote{%
    We limit our analysis in this paper to the ``burst'' injection regime in which the timescale for injection $t_*$ is smaller than the region of size $R_*$ over which the injection occurs: $t_*\lesssim R_*$; we have also taken $R_* \sim R_0$ for simplicity, although we could also have $R_* < R_0$ in this regime.
    However, the fireball anti-nucleosynthesis should proceed similarly in the opposite, wind regime ($ t_*\gtrsim R_*$), but with some differences in how the injection properties $(E_{\text{inj.}},B_{\text{inj.}},R_*, t_*)$ are related to the properties and outputs of the resulting thermalized fireball. 
    Parametrically, one can think of the anti-quark injection in the wind regime as a sequence of $N_{\rm burst}\sim t_*/R_*\gtrsim 1$ anti-quark bursts occurring in a region of size $R_*$ continuously one after another, each with an injection energy and anti-baryon number smaller by a factor of $N_{\rm burst}$ relative to their total values $E_{\text{inj.}}$ and $B_{\text{inj.}}$. 
    Consequently, in the wind regime the resulting initial fireball temperature $T_0$ will be lower by a factor of $N_{\rm burst}^{1/4}$ and at the same time the overall anti-nuclei output for the same $T_0$ will be $N_{\rm burst}$ times higher compared to that of the burst regime.
    We refer the reader to \appref{app:BurstWind} for more discussion on this point.
    } %
$R_*$ that is much smaller than $R_{\textsc{dm}}$ (i.e., $R_* \sim R_0\ll R_{\textsc{dm}}$), consistent with that injection occurring as a result of some catastrophic collapse dynamics in the post-collisional evolution (see \secref{sec:DMmodelIdeas}). 

Meanwhile, in the final post-fireball evolution, we have roughly that $E_{\text{inj.}} \sim  \Gamma m_p |B_{\text{inj.}}|$, so $\Gamma \sim m_{\textsc{dm}}/m_p \sim 11 \times ( m_{\textsc{dm}}/10\,\GeV)$, corresponding to $\bar{\eta} \sim 6.1\times 10^{-3}$.
Moreover, the constant $c$ defined at \eqref{eq:cDefn} is $c \sim 1.24$, implying an injected anti-helium isotope ratio of $N_{\antiHe{4}}/N_{\antiHe{3}}|_{\text{inj.}} \sim 0.33$, roughly consistent with an observed isotope ratio of $N_{\antiHe{4}}/N_{\antiHe{3}}|_{\text{obs.}} \sim 0.5$, as discussed in \secref{sec:prop-detection}.

In order to achieve the requisite net negative charge on the hadronic sector of the fireball in order to obtain a value $X_{\bar{p}} \sim 0.5$ around the onset of anti-nucleosynthesis (see the discussion in \secref{ss:antineutronabundance} and \appref{app:antinucleonFO}) while also maintaining net fireball EM neutrality (as required for injections from net-neutral dark objects), it would be sufficient for the injection to take place in such a way that a fraction $X_p$ of the anti-baryons are anti-protons, with a corresponding number $L \sim X_p |B_{\text{inj.}}|$ of positively charged leptons being injected.
On energetics grounds, because $m_e \ll m_p$ and $f_{\textsc{sm}}\ll 1$, this can easily be accomplished with minimal change to the above discussion.
It would however require the initial dark object to carry net-negative lepton number in addition to net-negative baryon number.

While the precise values shown in both this subsection and the previous one should not be taken too literally given the roughness of the estimates, what this nevertheless indicates is that collisions of large, macroscopic dark objects can in principle attain both the requisite injection rates and required fireball conditions to make our anti-helium anti-nucleosynthesis mechanism operate in a phenomenologically viable fashion, provided that the benchmarks and broad model features we have given can be attained in a concrete model.

\subsection{Towards a particle physics model} 
\label{sec:DMmodelIdeas}

The development a detailed microphysical model for the composite dark states, and their evolution in and after collisions is a question we ultimately leave open in this paper.
Nevertheless, in this section, we sketch the outlines of two particle physics models that we believe are promising targets for future investigation, and indicate where the difficulties in understanding their evolution lie.
We intend to return to this point in future work, and also encourage other work on it.

We also note at the outset of this discussion that obtaining a model with a consistent and observationally allowed cosmological evolution in light of the greatly increased density of dark matter in the early Universe%
\footnote{%
    We remind the reader that the \emph{local} DM density in the MW, $\rho_{\textsc{dm}}\approx 0.4\,\text{GeV/cm}^3$~\cite{Catena:2009mf,Salucci:2010qr}, is approximately 5.5 orders of magnitude larger that the present-day average cosmological abundance, $\Omega_{\text{c}}\rho_c \sim 1.3\times 10^{-6}\,\text{GeV/cm}^3$~\cite{2020A&A...641A...6P}, which is in turn roughly 10.5 orders of magnitude down from the abundance at matter-radiation equality (MRE), $(1+z_{\text{eq}})^3 \sim 4\times 10^{10}$.
    The DM density at MRE is thus approximately 5 orders of magnitude larger than locally in the MW today.} %
is an additional model-building constraint that would need to be carefully evaluated in the context of a full model, taking into account also the redshift dependence of the average speed of the collisions.

\subsubsection{Imploding fermion+Yukawa model}

The requisite injection may result from rapid conversions of $N_\chi\sim |B_{\text{inj.}}|/q_B$ fundamental DM particles $\chi$, each carrying baryon number $-q_B$ ($q_B>0)$ [and possibly lepton number], into SM anti-particles via a higher-dimensional effective operator; e.g., something of the form $\bar{\chi}\bar{\chi}\bar{\chi}\bar{q}\bar{q}\bar{q}$ if only anti-baryon number needs to be injected, or something like $\bar{\chi}\bar{\chi}\bar{\chi}\bar{\chi}\bar{q}\bar{q}\bar{q}\bar{\ell}$ if both anti-baryon and anti-lepton number need to be injected primarily (here, $\bar{q}$ and $\bar{\ell}$ represent an anti-quark and anti-lepton, respectively); see discussions in \secref[s]{sec:thermalization} and \ref{ss:antineutronabundance}, and \appref{app:antinucleonFO}.

Simultaneous and spatially concentrated conversions of a large number of $\chi$ particles can be naturally achieved if the $\chi$ particles participating in the conversion existed in the form of blobs; i.e., large composite bound states.
While these blobs, presumably formed in the early Universe, must survive over $\sim 10\,\text{Gyr}$ timescale until the present epoch, they must also rapidly convert to anti-quarks in $\sim 10^{-12}\,\text{s}$ timescale when the time is ripe (within the last $\sim \text{Myr}$ timescale for anti-nuclei to escape the Galaxy). 
To bridge these extremely long and extremely short timescales required for the DM$\rightarrow$SM process, we can imagine the following scenario. 
These blobs may have existed in a metastable state in the sense that they are individually stable, with the rate of the DM$\rightarrow$SM conversion process within each blob satisfying $\Gamma_{\text{DM}\rightarrow\text{SM}}\ll (10\,\text{Gyr})^{-1}$, but the merger of a pair of blobs would trigger a runaway collapse of the merger product accompanied with many orders of magnitude increase in $\Gamma_{\text{DM}\rightarrow\text{SM}}$, eventually causing most of the $\chi$ particles to convert to SM anti-particles when $1/\Gamma_{\text{DM}\rightarrow\text{SM}}\sim 10^{-12}\,\text{s}$.

The stable blob can be kept in equilibrium, for instance, by the balance between the repulsive effect of the Fermi pressure of $\chi$ particles and the compressing effect due to a higher vacuum pressure outside compared to inside the blob.
Such a vacuum pressure difference $P_{\rm vac}^{\rm out}-P_{\rm vac}^{\rm in}>0$ arises naturally in scenarios where the blobs are formed through a cosmological first-order phase transition from a higher-energy false vacuum to a lower-energy true vacuum in the early Universe~\cite{Hong:2020est,Witten:1984rs,Bai:2018dxf, Gross:2021qgx, Asadi:2021pwo,Asadi:2021yml}. 
In those scenarios, the present-epoch setup is that the inside of the blobs remains in the false vacuum while the rest of the Universe is in the true vacuum. 
Correspondingly, the pressure $P_{\rm vac}=-\rho_{\rm vac}$ associated with the vacuum energy $\rho_{\rm vac}$ is more negative inside than outside the blob.

Runaway collapse can be achieved by incorporating additional Yukawa forces between $\chi$ particles with an intermediate-range mediator: long compared to the typical spacing between $\chi$ but short compared to the radius of the blob.
We additionally assume that the $\chi$ are non-relativistic to avoid suppressions in their Yukawa forces (cf.~\citeR[s]{Wise:2014ola,Gresham:2017zqi} and the discussion in the next subsection). 
The total Fermi kinetic energy of the blob, the vacuum energy difference between inside and outside the blob, and the attractive Yukawa potential for an intermediate-range mediator then scale with the blob's number of constituents $N_\chi$ and radius $R_{\rm blob}$ as $\propto N_\chi^{5/3}/R_{\rm blob}^{2}$, $\propto R_{\rm blob}^3$, and $\propto -N_\chi^2/R_{\rm blob}^{3}$, respectively. 
For a given sufficiently small $N_\chi$, the blob's total energy as a function of $R_{\rm blob}$ has a metastable minimum at a radius that is set by the balance between the non-relativistic Fermi pressure and vacuum-pressure difference. 
However, when $N_\chi$ exceeds a certain threshold the metastable minimum ceases to exist and it becomes energetically favorable for $R_{\rm blob}$ to decrease indefinitely, until new effects that deplete the $\chi$ or reverse the tendency to collapse turns on.%
\footnote{\label{ftnt:differentRegime}%
    We note also that this short-range Yukawa regime has different qualitative behavior as compared to the cognate ``saturated'' regime considered in \citeR{Gresham:2017zqi} because the fermions we consider here are still non-relativistic, whereas those considered in \citeR{Gresham:2017zqi} are relativistic by the time the Yukawa becomes short range compared to the size of the blob.
    This modifies both the scaling of the degeneracy pressure with blob radius and the behavior of the Yukawa force, as compared to the saturated case in \citeR{Gresham:2017zqi}.
    } %
Such a runaway collapse can be triggered by the merger of two near-critical blobs.

Aside from the complexity of accurately modeling the envisaged scenario, one can already see competing effects or constraints that make it challenging to present a concrete realization of this scenario. 
First, the requirement to reduce the blob radius by many orders of magnitude during the blob collapse in order to increase $\Gamma_{\text{DM}\rightarrow\text{SM}}$ from $\ll (10\,\text{Gyr})^{-1}$ to $\sim (10^{-12}\,\text{s})^{-1}$ can increase the Fermi momentum of the $\chi$ particles to the point that they become relativistic. 
Once the $\chi$ particles are relativistic, the Yukawa forces become progressively suppressed~\cite{Wise:2014ola,Gresham:2017zqi} and at the same time the Fermi pressure increases faster as the blob radius decreases~\cite{ShapiroTeukolsky:1983}.
Both these effects tend to stop the collapse, potentially preventing $\Gamma_{\text{DM}\rightarrow\text{SM}}$ from reaching $\sim (10^{-12}\,\text{s})^{-1}$. 
Moreover, consistency with collider data (and $\tilde{E}|_{\text{inj.}}/\tilde{B}|_{\text{inj.}}\sim 10\,\GeV$ to obtain $\Gamma \sim 10$) requires raising the cutoff of the higher-dimensional DM$\rightarrow$SM operator to at least above $10\,\TeV$. 
The latter implies a strong suppression on $\Gamma_{\text{DM}\rightarrow\text{SM}}$ unless the blob becomes very dense at the time of conversion.
Preliminary estimates suggest that in order to obtain $1/\Gamma_{\text{DM}\rightarrow\text{SM}}\sim 10^{-12}\,\text{s}$, the energy density of the blob at the time of conversion must be $\gg (100\,\MeV)^4$, which is much higher than the typical energy density of the fireball we seek to create. 
To dilute such a high energy density one may need to introduce another operator that converts the $\chi$ particles to a lighter dark particles $\psi$ much more efficiently than the DM$\rightarrow$SM conversion, such that most of the $\chi$ convert into $\psi$ with a rate $\Gamma_{\chi\rightarrow \psi}\sim (10^{-12}\,\text{s})^{-1}$, and only a small fraction $f_{\rm SM}\sim  10^{-12}\,\text{s} \times \Gamma_{\text{DM}\rightarrow\text{SM}} \ll 1$ of $\chi$ with energy density $\sim (100 \,\MeV)^4$ converts to SM anti-particles during the $\sim 10^{-12}\,\text{s}$ timescale.

\subsubsection{Dark-dwarf model}

The second model that we believe may be interesting to consider is based on the dark states behaving somewhat like white dwarfs in the SM.
The particle content of this model would be a heavy ``dark proton'' $p_d$, a slightly heavier ``dark neutron'' $n_d$, a much lighter ``dark electron'' $e_d$, and a massless ``dark neutrino'' $\nu_d$.
The $p_d$ and $n_d$ carry baryon number $-1$, and both couple with equal strength to a Yukawa force mediated by a scalar $\varphi$.
Additionally, the $p_d$ and $e_d$ are taken to both couple to a dark electromagnetism, with opposite charges.
We also assume the existence of a two higher dimensional 4-fermion operators allowing $e_d + p_d \leftrightarrow n_d + \nu_d$ (and crossed) reactions, and permitting $n_d \rightarrow \bar{q} + \bar{q} + \bar{q}$, where $\bar{q}$ is a SM anti-quark.

With such particle content, it is possible to form dense macroscopic composite objects (``dark dwarfs''~\cite{Gross:2021qgx}) with very large constituent number, consisting of $p_d$ held together by the Yukawa force, and supported against collapse by the (relativistic) degeneracy pressure of the $e_d$.
The mass hierarchies of the particles can be arranged such that, at zero temperature, these objects would initially contain no $n_d$ or $\nu_d$ constituents.

Similar objects, albeit with just a single uncharged constituent fermion supported against collapse due to attractive Yukawa forces by its own degeneracy pressure, have been considered in, e.g., \citeR[s]{Wise:2014jva,Wise:2014ola,Gresham:2017zqi,Grabowska:2018lnd}; see also \citeR{Gross:2021qgx}.
A straightforward extension of these works to the two-component dark dwarf we have in mind indicates that these objects, while not possessing a singular collapse instability (i.e., there is no true Chandrasekhar-like limit),%
\footnote{\label{ftnt:noSingularCollapse}%
    The avoidance of the singular collapse instability arises from the suppression of the coupling of scalar-mediated forces to relativistic fermions, as was pointed out in \citeR{Wise:2014ola} and examined further in \citeR{Gresham:2017zqi}.
    } %
possess of threshold constituent number $N_*$ for which the following is true: for $0<N<N_*$, their radii $R$ are large, say $R \sim R_{\text{below}}$; on the other hand, for $N_* < N < 2N_*$, their radii $R\sim R_{\text{above}}$ are much smaller.%
\footnote{\label{ftnt:WhyNstar}%
    One way to understand this is that the $R(N)$ relationship for $N<N_*$ behaves similarly to the mass--radius relationship for a SM white dwarf (as computed in purely Newtonian gravity, using the full equation of state for degenerate electrons)~\cite{ShapiroTeukolsky:1983}: a slow function of $N$ for $N\lesssim 0.95N_*$, but showing a vary rapid decrease toward $R=0$ for $N\sim N_*$. 
    However, very near $N\sim N_*$, the $p_d$ become relativistic, leading to the softening of the Yukawa coupling to the $p_d$ discussed in \citeR{Wise:2014ola,Gresham:2017zqi}, which avoids a singular crunch but leads to a minimum size $R_* \ll R(0.95N_*)$ near $N_*$ and a slow (power law) increase in $R(N)$ away from $R_*$ above $N_*$.
    } %
Roughly, the hierarchy of these radii is in ratio of the dark electron and proton masses: $R_{\text{above}} \sim (m_{e_d}/m_{p_d}) R_{\text{below}} \ll R_{\text{below}}$.

If the cosmologically stable dark objects have $N_{\textsc{dm}} \sim \epsilon N_*$ with $0.5 \lesssim \epsilon < 1$, then collisions of two such objects could in principle trigger a large change in the bulk properties of the pre- and post-collision states because (assuming no ejection of constituents), the post-collision object would than have $N \sim 2\epsilon N_* > N_*$ constituents.
If the collision were to play out in such a way that both the pre- and post-collision states were at exactly zero (dark) temperature, the idea would heuristically be the following:
before collision, the Fermi energy of the (relativistic) $e_d$ is much smaller than the mass difference of the $n_d$ and $p_d$.
However, when these objects collide and then dynamically relax down to their much smaller post-collision equilibrium size, the $e_d$ density increases greatly, leading to their Fermi energy becoming much larger.
Once the $e_d$ Fermi energy exceeds the $n_d$--$p_d$ mass difference, a large fraction of the $p_d$ can suddenly ``neutronise'' to $n_d$ (we adopt the nomenclature of the cognate process that occurs in the collapse toward a SM neutron star~\cite{ShapiroTeukolsky:1983}).
The $n_d$ in turn are unstable to decay to SM anti-quarks.
As a result, the sudden dynamical ``collapse'' of the incoming dark states to a much smaller and more dense state upon collisions could trigger a catastrophic decay of a large fraction of the mass energy of the initial colliding dark objects to SM anti-quarks.%
\footnote{\label{eq:offShell}%
    Of course, the existence of decay channels for the dark dwarfs through annihilation reactions $e_d+p_d \rightarrow \nu_d + \bar{q} + \bar{q} + \bar{q}$ that proceed through an off-shell $n_d$ must also be considered both for blob stability and for understanding the rapidity of decay to SM quarks during this collapse.
    } %

Unfortunately, the dynamics of this system are actually more complicated than this na\"ive picture.
Initial estimates indicate that for scalar forces with a range larger than the blob sizes (the regime we assume), there is a speed up of the dark states as they collide, leading to the collision taking place in a supersonic regime with respect to the estimated speed of sound in the blobs.
As a result, two things are true: (1) the collision is a violent process that is not amenable to simple analytical treatment, and could lead to the ejection of constituents, or the formation of a large bound object, or (more likely) something more complicated; and (2) as the collision occurs, it is likely that the blob constituents are heated significantly (via, e.g., shock heating), possibly to initial post-collision dark temperatures $T_d$ that are a significant fraction of the dark electron mass $m_{e_d}$.
As a result, the zero-temperature dynamics of the object are likely significantly modified by resulting thermal pressures.
It appears that the resulting state may still lose energy and shrink via surface dark-photon emission, volume relativistic emission of the $\varphi$ particle, and later potentially either volume or surface emission of dark neutrinos.
The resulting state may thus still evolve toward the neutronization regime sufficiently rapidly for this model to be viable.
However, the entirety of this evolution is complicated and requires further study to establish in detail whether it works.

While this model may potentially be able to achieve the requisite destabilization dynamics, obtaining a net-negative EM charge on the anti-hadrons for fireballs seeded by this model (see discussions in \secref[s]{ss:antineutronabundance} and \ref{sec:fireballMapping}, and \appref{app:antinucleonFO}) may be challenging in the exact formulation advanced here.
A modification to this picture, or further model building, may be required.

The scope of work required to investigate this model in full and place the speculative statements in this sub-section on a firm footing is such that we defer its consideration to a future paper.

\section{Discussion and conclusion}
\label{sec:DiscussionConclusion}

In this paper we have studied the anti-nucleosynthesis of elements up to anti-helium-4 in rapidly expanding fireballs of SM plasma that carry net anti-baryon number and that we assume to be seeded within the Milky Way by a BSM process.
For appropriate initial conditions set by the initial radius, temperature, and anti-baryon--to-entropy ratio of the plasma, the evolution of these fireballs is such that their thermal pressure drives the system toward a regime where there is relativistic bulk radial motion of a thin shell of plasma, in which the temperature of plasma falls as the expansion proceeds.
This permits purely SM thermal anti-nucleosynthesis of elements (similar to BBN) to occur in the expanding, cooling thin shell, while the products obtain relativistic boosts with respect to the rest frame of the fireball. 
Eventually, for appropriate parameters, the expansion rate shuts off the anti-nucleosynthesis in a regime where the anti-nucleosynthetic products have not reached their thermodynamic equilibrium values, which allows the abundances of $\antiHe{4}$ and anti-tritium (which later decays to $\antiHe{3}$) produced to not be too dissimilar, and not highly suppressed with respect to anti-deuterium or anti-protons; see \figref[s]{fig:Xvsr0}--\ref{fig:outputspace}.
Once this expanding fireball becomes optically thin, its products cease to be driven by thermal pressures, and are launched at relativistic speeds into interstellar space, all with the same bulk speed (i.e., Lorentz boost).
The anti-tritium subsequently decays to $\antiHe{3}$ on a short timescale, leading to injected amounts of $\antiHe{3}$ and $\antiHe{4}$ that are not too dissimilar.

Assuming that these injections of high-energy anti-nucleons and anti-nuclei are spatially distributed as the square of an NFW profile, and making use of \textsc{galprop} to approximate the galactic transport of these injected products, we showed that the relative fluxes received at Earth are such that one could obtain a ratio of roughly 2:1 for the isotopes $\antiHe{3}$ to $\antiHe{4}$, without being excluded by AMS-02 anti-deuterium or anti-proton constraints; see \figref[s]{fig:spectra} and \ref{fig:InjRatePlot}.
We showed that other products of this scenario (e.g., photons) do not appear to supply additional constraints, but we did not undertake an exhaustive indirect detection study.
We also computed the overall required injection luminosity of anti-baryons into the fireballs we considered in order to obtain $\mathcal{O}(10)$ anti-helium events in the AMS-02 10-year exposure.

Our conclusion that the sizes of the anti-proton (larger) and anti-helium (smaller) fluxes are not too dissimilar so as to violate observational constraints on the anti-proton flux, while also explaining the anti-helium isotope ratio, does depend to some extent what the anti-proton--to--anti-neutron ratio is in the partially thermalized fireball plasma prior to the onset of anti-nucleosynthesis.
Because weak interactions are inefficient in the plasma and weak decay timescales for the charged pions are long compared to the dynamical expansion timescale of the fluid, this value is however dependent on the details of the BSM injection process that seeds the fireball.

In reaching our conclusions as stated above in the main text, we assumed that the charged pions have a small chemical potential (i.e., $|\mu_{\pi^+}| \ll m_n-m_p$) until SMCER reactions such as $\bar{n} + \pi^- \leftrightarrow \bar{p} + \pi^0$ became inefficient, locking in the $\bar{n}$--to--$\bar{p}$ ratio.
This was equivalent to assuming that some combination of the BSM injection and possible SM processes following it were such that they established a certain very specific overall net negative electromagnetic charge (i.e., $X_Q = -(n_{Q}')^{\text{hadron}}/n_{\bar{B}}' >0$; see \appref{app:antinucleonFO}) on the (anti-)hadronic constituents of the fireball no later than the time at which the SMCER begin to be inefficient, with a compensating overall positive charge held by charged leptons in order to allow overall neutrality.
But we also showed in \appref{app:antinucleonFO} that we could relax this assumption on the small pion chemical potential somewhat, which allows $X_Q$ to vary in some range around 0.5 with only $\mathcal{O}(1)$ changes to our results.
However, if $X_Q$ is too small around the time the SMCER reactions become inefficient, a larger anti-proton abundance results for the same anti-helium flux, which can be observationally challenging.
While we argue that SM processes impose $X_Q \gtrsim 5\times 10^{-5}$, we found that $X_Q \gtrsim 10^{-2}$ is likely required for phenomenological viability (see \figref{fig:AbundanceswithChemPot}).
Realizing this thus likely requires that the BSM injection process needs to be able to primarily inject both anti-quarks and anti-leptons of compensating charge.

Finally, we showed that it would be plausible (at least in principle) for the required injection luminosity to be obtained via the collisions of supermassive, composite dark states (possibly a subcomponent of the dark matter), \emph{provided} that these otherwise individually cosmologically stable states can become destabilized in the collision in such a way that activates a decay channel that converts a non-negligible fraction of their mass energy to SM anti-quarks (and possibly some fraction of positively charged leptons), with this taking place both rapidly and in a localized region of space.
We also showed that, assuming certain benchmark bulk behavior in the post-collisional evolution can be realized, this same scenario could seed fireballs with the correct bulk physical values $(T_0,R_0,\bar{\eta})$ to realize anti-nucleosynthesis scenario we advance in a phenomenologically viable fashion.

We did not supply a full particle physics model that could computably realize the required behavior, but we offered two potential model paradigms for these dark objects that we believe may be interesting to investigate further in future work.
The first paradigm is based on a model with a single species of fermions held together by a combination of vacuum pressure differences inside and outside a large agglomeration of such objects, and an intermediate-range Yukawa force, but supported against collapse by their nonrelativistic degeneracy pressure. 
Collisions of pairs of such objects may in certain regimes provoke a catastrophic collapse of the resulting combined state if it exceeds a critical threshold of constituents. 
We can imagine operators that allow some number of these fermions to then mutually annihilate to a collection of SM anti-quarks; modified versions of this behavior could likely realize the opposite-sign lepton injection too.
The dramatic density difference in the pre- and post-collision states of these objects may allow them to be cosmologically stable before collision, but for the annihilation rates to spike high enough during the collapse to permit the requisite injection in a sufficiently rapid timescale. 
The second paradigm is a based on a collision of dark analogs of degenerate white dwarfs, that may undergo rapid collapse after collision, provided the collision produces a merger product above a certain threshold number of constituents.
During this collapse, constituents of the dark dwarfs may be converted to other dark species that are unstable to prompt decay to SM anti-quarks.
In this case, additional model building or a modification of the picture we presented are likely required to achieve the opposite-charge lepton injection.
Neither of these scenarios is however amenable to full analytical control, and important aspects of the evolution of each are thus still unclear.

To be completely clear: it is thus still an open question whether, and how, the requisite BSM injection of SM anti-quarks (and opposite-sign charged leptons) can be achieved via dark-state collisions in order to seed the fireballs whose post-injection SM behavior we have studied.
The most obvious and pressing follow-up is therefore a fuller investigation of these (or other) model paradigms, so as to place this BSM aspect of our scenario on a rigorous footing.
We anticipate returning to this point in future work.

On the observational side, further AMS-02 data will continue to be taken until 2030~\cite{TingCERNslides2023}; combined with the existing data, this may yield further clarity about the status of the AMS-02 candidate anti-helium events.
Additionally, the balloon-borne GAPS experiment~\cite{OSTERIA2020162201} is approved to fly in late-2024~\cite{GAPSNews}, and is anticipated to provide data on lower-energy anti-deuterium and anti-helium fluxes that will be complementary to those from AMS-02.
These data will continue to inform model-building.

The fireball anti-nucleosynthesis scenario we have advanced in this paper provides an interesting and novel alternative SM pathway to anti-helium formation, provided that the BSM seeding issue can be appropriately addressed.
We therefore view this work a partial step toward understanding the origin of the tentatively identified AMS-02 anti-helium events.

\acknowledgments

We thank Anne-Katherine Burns, Junwu Huang, Tim Linden, Vivian Poulin, and Stefano Profumo for useful discussions and correspondence. 
E.H.T.~thanks Xuheng Luo and Ngan H.~Nguyen for useful discussions on another project.
We also thank an anonymous referee for constructive comments that improved the clarity of the presentation of aspects of this work.

This work was supported by the U.S.~Department of Energy~(DOE), Office of Science, National Quantum Information Science Research Centers, Superconducting Quantum Materials and Systems Center~(SQMS) under Contract No.~DE-AC02-07CH11359. 
E.H.T.~acknowledges support by NSF Grant PHY-2310429 and Gordon and Betty Moore Foundation Grant No.~GBMF7946.
D.E.K.~and S.R.~are supported in part by the U.S.~National Science Foundation~(NSF) under Grant No.~PHY-1818899.
S.R.~is also supported by the Simons Investigator Grant No.~827042, and by the~DOE under a QuantISED grant for MAGIS. 
D.E.K.~is also supported by the Simons Investigator Grant No.~144924.
Research at Perimeter Institute is supported by the Government of Canada through the Department of Innovation, Science, and Economic Development, and by the Province of Ontario through the Ministry of Colleges and Universities.
M.A.F.~gratefully acknowledges the hospitality of the Simons Center for Geometry and Physics at Stony Brook University.
M.A.F.~also thanks the Aspen Center for Physics, where parts of this work were undertaken, supported by National Science Foundation (NSF) Grant Nos.~PHY-1607611 and PHY-2210452.

\appendix

\section{Derivation of fireball scaling laws}
\label{app: fireballscalings}

As described in \secref{sec:expansionDynamics}, following a transient and localized injection of energy and anti-baryon number, a thermalized fireball with anti-baryon--to--entropy ratio $\bar{\eta}\ll 1$ is formed.
In our parameter space, the fireball immediately accelerates to semirelativistic bulk velocities corresponding to Lorentz factors $\gamma\sim\text{few}$, at which point its radius and temperature are $R_0$ and $T_0$ respectively, turning into a dense shell with a central underdensity in the process.
At the point where the radius of the fireball is $R_0$, this shell structure is just beginning to form, and the shell thickness can be taken to be $\sim R_0$. 
The subsequent evolution of the shell is dictated by the following relativistic fluid equations which, respectively, encode conservation of anti-baryon number, energy, and momentum in the comoving fluid frame:
\begin{align}
    \partial_t\left[\gamma n_{\bar{B}}^\prime\right]+\frac{1}{r^2}\partial_r\left[r^2\gamma v n_{\bar{B}}^\prime\right]&=0 \:, \label{eq:fluideqB}\\
    \partial_t\left[\gamma^2(\rho'+p')\right]+\frac{1}{r^2}\partial_r\left[r^2\gamma^2v\left(\rho'+p'\right)\right]&=\partial_t p'\:,\label{eq:fluideq0}\\
    \partial_t\left[\gamma^2v(\rho'+p')\right]+\frac{1}{r^2}\partial_r\left[r^2\gamma^2v^2\left(\rho'+p'\right)\right] &=-\partial_rp'  \:,  \label{eq:fluideqi}
\end{align}
where $r$, $t$, $v$ ($\gamma$), $n_{\bar{B}}^\prime$, $\rho'$, and $p'$ are, respectively, the radial position, time, bulk velocity (and its associated Lorentz factor), anti-baryon number density, energy density, and pressure of a fluid element in the fireball.
Primed quantities are defined in the comoving rest frame of the fireball fluid; unprimed quantities are defined in the fireball CoM frame.

In this Appendix we show how, in the ultra-relativistic bulk velocity limit $\gamma\gg 1$, the fireball expansion follows the simple scaling laws used \secref{sec:expansionDynamics}. 
See also \citeR[s]{Piran:1993jm, Bisnovatyi-Kogan:1995cyk} for alternative explicit derivations, and \citeR[s]{Diamond:2023scc,Diamond:2023cto} for related studies.

In the fireball CoM frame, the energy $E_{\rm shell}$ and anti-baryon number $\bar{B}_{\rm shell}$ of a thin radial slice of the shell with radius $r(t)$ and thickness $\delta r(t)$ are given by
\begin{align}
    E_{\rm shell}&\approx 4\pi r^2\delta r \gamma^2(\rho'+p')\:,\\
    \bar{B}_{\rm shell}&\approx 4\pi r^2\delta r \gamma n_{\bar{B}}^\prime \:.
\end{align}
The time derivatives of $E_{\rm shell}$ and $\bar{B}_{\rm shell}$ can be written as
\begin{align}
    \frac{d\bar{B}_{\rm shell}}{dt}&=\frac{\bar{B}_{\rm shell}}{\delta r}\left(\frac{d\delta r}{dt}-\delta r\partial_r v\right)\:,\\
    \frac{dE_{\rm shell}}{dt}&=\frac{E_{\rm shell}}{\delta r}\left(\frac{d\delta r}{dt}-\delta r\partial_r v\right) \nl +\frac{w}{1+w}\delta r\partial_t\left[\frac{1}{\gamma^2}\frac{E_{\rm shell}}{\delta r}\right]\:,
\end{align}
where we have used $d/dt=\partial_t+v\partial_r$ along the radial trajectory of a shell, assumed that the shell evolves adiabatically with a constant equation of state $w=p'/\rho'$, and made use of the fluid equations, \eqrefRange{eq:fluideqB}{eq:fluideqi}. 
For small $\delta r$, we can perform the expansion $d\delta r/dt=\left. dr/dt\right|_{r+\delta r}-\left. dr/dt\right|_{r}=\delta r\partial_r v+(1/2)\delta r^2\partial_r^2v+\ldots$. 
Hence, the content of the round brackets in the above equations is $\mathcal{O}(\delta r^2)$ and thus negligible for sufficiently small $\delta r$.
Moreover, for sufficiently large $\gamma$, the term with the square brackets is also negligible. 
Therefore, in these thin-shell and ultrarelativistic limits we can assume that $E_{\rm shell}$ and $\bar{B}_{\rm shell}$ are approximately constant.

We now focus on a differential radial layer whose initial radius is $\sim R_0$ and neglect for the moment the time evolution of its thickness $\delta r$. 
The change in the thickness of the shell will be important but (as we will show) only at later times. 
In the interest of explaining the AMS-02 relativistic anti-helium candidate events, we are particularly interested in scenarios with $\Gamma\equiv T_0/\bar{\eta} m_p\gg 1$, where each shell is initially radiation dominated (RD) and thus obeys $\rho'+p'\propto T^{\prime 4}$. 
That latter scaling, taken together with $n_{\bar{B}}'\propto T^{\prime 3}$ and the conservation of $E_{\rm shell}$ and $\bar{B}_{\rm shell}$ imply that $\gamma\propto r$ and $T'\propto 1/r$ during RD. 
This `accelerating phase' lasts until $r\sim \Gamma R_0$, whereupon the shell becomes matter dominated (MD). 
Following that, $\rho'+p'\propto  T^{\prime 3}$, which implies that $\gamma\sim \Gamma$ and $T\propto r^{-2/3}$ during MD. 
We refer to this as the `coasting phase'.
To summarize, as long as the shell thickness $\delta r$ is approximately constant, the following scaling laws hold:
\begin{align}
    r&\lesssim \Gamma R_0& :&& \gamma&\propto r\:, & T'&\propto r^{-1}\:;\\
    \Gamma R_0&\lesssim r\lesssim \Gamma^2 R_0 & :  && \gamma&\sim\Gamma=\text{const.}\:,& T'&\propto r^{-2/3}\:.
\end{align}

Due to the differences in the velocities of neighboring radial layers, the constant-thickness approximation employed above breaks down at some point. 
Assuming the velocity gradient is monotonic and approximately uniform over the full thickness of the shell, $\partial_r v\sim \Delta v/ R_0\approx (\Delta \gamma/\gamma^3)/R_0$, and that $\Delta \gamma/\gamma\sim\Delta \gamma_0/\gamma_0\sim 1$, we find $\partial_r v\sim 1/(\gamma^2 R_0)$. 
This implies that the thickness of a differential shell will behave as
\begin{align}
   \delta r(t)\sim \left(1+t\frac{dv}{dr}\right)\delta r_0 \sim \left(1+\frac{1}{\gamma^2}\frac{r}{R_0}\right)\delta r_0 \:, \label{eq:thickness}
\end{align}
where we have used $t\approx  r$; the second term in the $\lb(\cdots\rb)$-bracket encodes the shell-thickness spreading effect.
Since $\gamma\propto r$ during the accelerating phase, the spreading effect in that phase decreases with increasing time/radius, leading to $\delta r \sim \text{const}$. 
On the other hand, in the coasting phase $\gamma \approx \Gamma$, the spreading term grows linearly with $r$.
The spreading term becomes dominant when $r\sim \Gamma^2 R_0$, implying the existence of a second expansion phase during MD where the shell thickness grows as $\delta r\propto r$:  
\begin{align}
    r&\gtrsim \Gamma^2 R_0: & \gamma &\sim \Gamma=\text{const.}, & T'&\propto r^{-1}, & \delta r&\propto r\:.
\end{align}
This `spreading phase' continues until the plasma becomes optically thin to the photons, at which point the assumption of strong coupling of the photon and anti-baryon fluids breaks down. As described in \secref{sec:expansionDynamics}, this occurs when the radius of the fireball is $R_{\rm thin}$, found in \eqref{eq:Rthin}.

\section{Freeze-out ratio of anti-nucleons}
\label{app:antinucleonFO}

The discussion in this Appendix supplements that in \secref{ss:antineutronabundance} and concerns the time period after the fireball has thermalized via strong and EM interactions ($T'\lesssim T_0$) and before the anti-neutron--to--anti-proton ratio has frozen out ($T'\gtrsim T^\prime_{\bar{n}\bar{p}}\approx 6\MeV $) at SMCER decoupling.

While strong and electromagnetic interactions are efficient given the fireball expansion timescales, weak interactions typically operate at timescales that are significantly longer.%
\footnote{%
    For the benchmark fireball with $R_0\sim 1.5\text{ mm}$ and $T_0\sim 100\MeV$, we initially have $\Gamma_{\rm EW}\tau^\prime \sim G_F^2T_0^5R_0\sim 10^{-2}$.} %
This means that the fireball we are considering is only partially thermalized.  
Some assumptions must be made about the chemical potentials of certain particle species, such as leptons and other particles that are in (relative) chemical equilibrium with them, in order to fully characterize the thermal state of the fireball. 
In the main analysis, we made such an assumption by neglecting the $\pi^+$ chemical potential (i.e., setting $|\mu_{\pi^+}|\ll m_n-m_p$) in calculating the chemical-equilibrium anti-neutron--to--anti-proton ratio at \eqref{eq:nDnpch}. 
In this appendix, we depart from  that assumption and instead use charge asymmetry in the lepton (or hadron) sector to parameterize the non-thermalized part of the fireball. 
We then describe the requirements to achieve the assumed negligible $\mu_{\pi^+}$ in terms of the charge asymmetry parameter and comment on the impact of relaxing this assumption on the outputs of fireball anti-nucleosynthesis. 
 
Let the comoving number densities of $\bar{p}$, $\pi^\pm$, and charge residing in the lepton sector, be denoted respectively as $n^\prime_{\bar{p}}$, $n^\prime_{\pi^\pm}$, and $(n^\prime_Q)^{\rm lepton}$. Assuming the fireball is locally charge neutral, we must then have%
\footnote{%
    Charged kaons $K^\pm$ are stable over the timescales of our interest and so should also be present, albeit with much lower abundances compared to $\pi^\pm$ since they are significantly heavier. 
    We expect their influences to be qualitatively similar to that $\pi^\pm$, but that they will not change our conclusions due to their relatively low abundances compared to the pions.} %
\begin{align}
    -n^\prime_{\bar{p}}+\left(n^\prime_{\pi^+}-n^\prime_{\pi^-}\right)+\lb(n_Q^{\prime}\rb)^{\text{lepton}}=0\:. \label{eq:chargeNeutrality}
\end{align}
The inefficiency of weak interactions in the fireball expansion timescales implies that the comoving net charge density in the hadronic sector $(n_Q^{\prime})^{\text{hadron}}=-n^\prime_{\bar{p}}+\left(n^\prime_{\pi^+}-n^\prime_{\pi^-}\right)$ and in the lepton sector $(n_Q^{\prime})^{\text{lepton}}$ are separately approximately conserved. 
Therefore, we can treat%
\footnote{%
    Note that $X_Q$ can in principle be negative, however that implies a negative $\mu_{\pi^+}$, which leads to exponentially suppressed anti-proton abundance $X_{\bar{p}}$ in most cases. 
    We therefore do not consider this case.}
\begin{align}
    X_Q\equiv \frac{-\lb(n_Q^{\prime}\rb)^{\text{hadron}}}{n^\prime_{\bar{B}}}=\frac{+\lb(n_Q^{\prime} \rb)^{\text{lepton}}}{n^\prime_{\bar{B}}}>0 \label{eq:XQdef}
\end{align}
as a fourth parameter, in addition to the three parameters $(T_0, R_0, \bar{\eta})$, characterizing the evolution of the incompletely thermalized fireball.

The anti-neutron--to--anti-proton ratio is a key input for our anti-nucleosynthesis analysis. Its freeze out value is well approximated by its chemical equilibrium value $\left(X_{\bar{n}}/X_{\bar{p}}\right)_{\rm ch}$ at the moment of SMCER decoupling. 
The impact of the charge asymmetry $X_Q$ enters the through dependence of $\left(X_{\bar{n}}\right)_{\rm ch}$ and $\left(X_{\bar{p}}\right)_{\rm ch}$ on it, as dictated by equilibrium thermodynamics. 
To quantify this dependence, we start by assuming
\begin{align}
    &\mu_{\bar{p}}+\mu_{\pi^+}=\mu_{\bar{n}}+\mu_{\pi^0};\\ 
    &\mu_{\pi^-}=-\mu_{\pi^+},\quad \mu_{\pi^0}=0\:;\\
    &T'\ll m_n-\mu_{\bar{n}}, m_{p}-\mu_{\bar{p}}\:;\label{eq:npassumption}\\
    &T'\ll m_\pi\pm \mu_{\pi^+}\label{eq:npiassumption}\:;
\end{align}
where the first and second lines are justified by efficient (see discussion below) strong and EM interactions such as $\bar{p}+\pi^+\leftrightarrow \bar{n}+\pi^0$, $\pi^+\pi^- \leftrightarrow \gamma\gamma$, and $\pi^+\pi^-\leftrightarrow \pi^0\pi^0$ being in equilibrium; the third and fourth lines are justified as long as the charge densities carried by $\bar{p}$ and $\pi^\pm$ are $\ll T^{\prime 3}$.
Hence, we can approximate the chemical-equilibrium abundances $\left(n^\prime_{\bar{p}}\right)_{\rm ch}$ and $\left(n^\prime_{\pi^+}-n^\prime_{\pi^-}\right)_{\rm ch}$ as~\cite{Mukhanov:2005sc,Kolb:1990vq}
\begin{align}
    \left(n^\prime_{\bar{p}}\right)_{\rm ch}&\approx n^\prime_{\bar{B}} \lb[ \exp\lb(-\frac{m_n-m_p}{T'}+\frac{\mu_{\pi^+}}{T'}\rb)+1 \rb]^{-1}\:,\label{eq:npapprox}\\
    \left(n^\prime_{\pi^+}-n^\prime_{\pi^-}\right)_{\rm ch}&\approx 2\left(\frac{m_\pi T'}{2\pi}\right)^{3/2}\
    \exp\lb(-\frac{m_\pi}{T'}\rb)\text{sinh}\left(\frac{\mu_{\pi^+}}{T'}\right)\:.\label{eq:npiapprox}
\end{align}
The charge neutrality relation \eqref{eq:chargeNeutrality} thus reduces to 
\begin{align}
    X_Q &=\lb[ \exp\lb(-\frac{m_n-m_p}{T'}+\frac{\mu_{\pi^+}}{T'}\rb)+1\rb]^{-1}\nonumber \\[1ex]& \quad
    -\dfrac{2\left(\dfrac{m_\pi T'}{2\pi}\right)^{3/2}\exp\lb(-\dfrac{m_\pi}{T'}\rb)\text{sinh}\left(\dfrac{\mu_{\pi^+}}{T'}\right)}{\bar{\eta}\left(\dfrac{\zeta(3)}{\pi^2}\right)g_* T^{\prime 3}}\:. \label{eq:XQ}
\end{align}
For simplicity, we will set $g_*=5.5$. 
\eqref{eq:XQ} can be solved to give $\mu_{\pi^+} = \mu_{\pi^+}(X_Q,T')$ . 
For $0\leq X_Q\leq 1$ and $10^{-3}\leq\bar{\eta}\leq 10^{-2}$, we found that $(m_\pi-|\mu_{\pi^+}|)/T'$ varies monotonically from $\sim 1$ to $\sim 30$ as $T'$ is varied from $100\MeV$ to $5\MeV$, providing an \emph{a posteriori} justification toward the lower range of $T'$ for the assumption at \eqref{eq:npiassumption} that was used to justify the approximation at \eqref{eq:npiapprox}. 

\begin{figure*}[!t]
    \centering
    \includegraphics[width=0.49\textwidth]{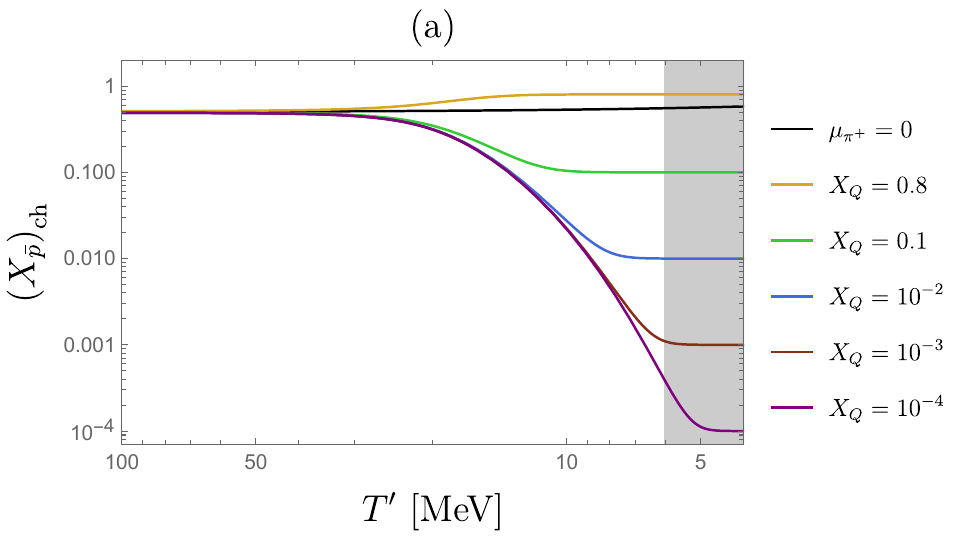}
    \includegraphics[width=0.49\textwidth]{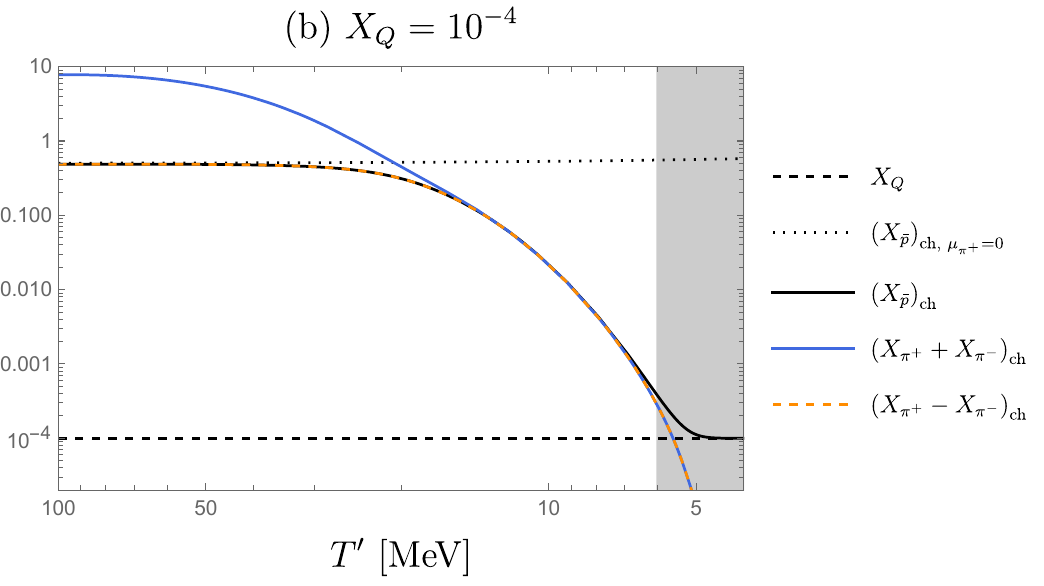}\\
    \includegraphics[width=0.49\textwidth]{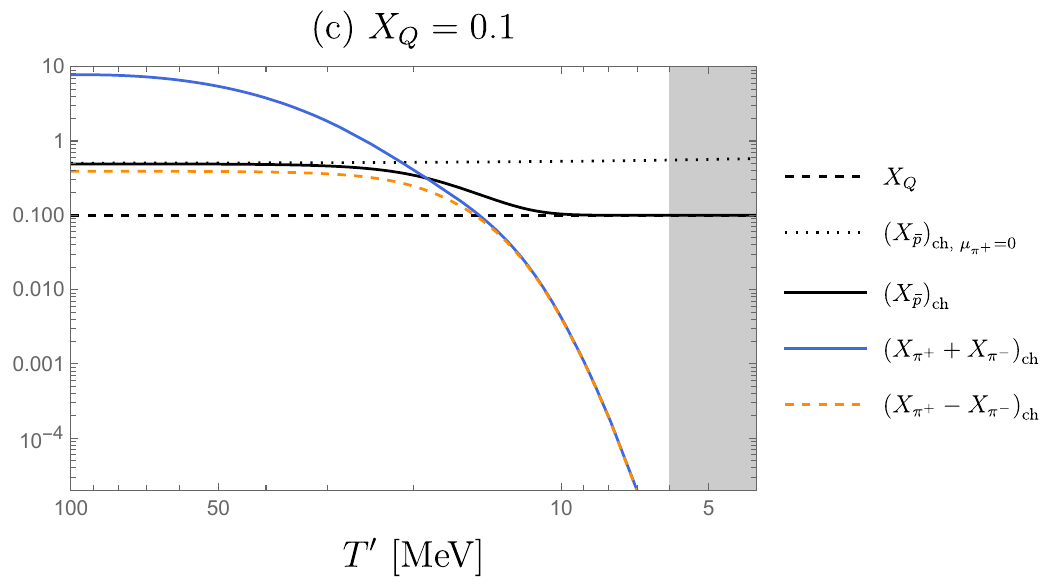}
    \includegraphics[width=0.49\textwidth]{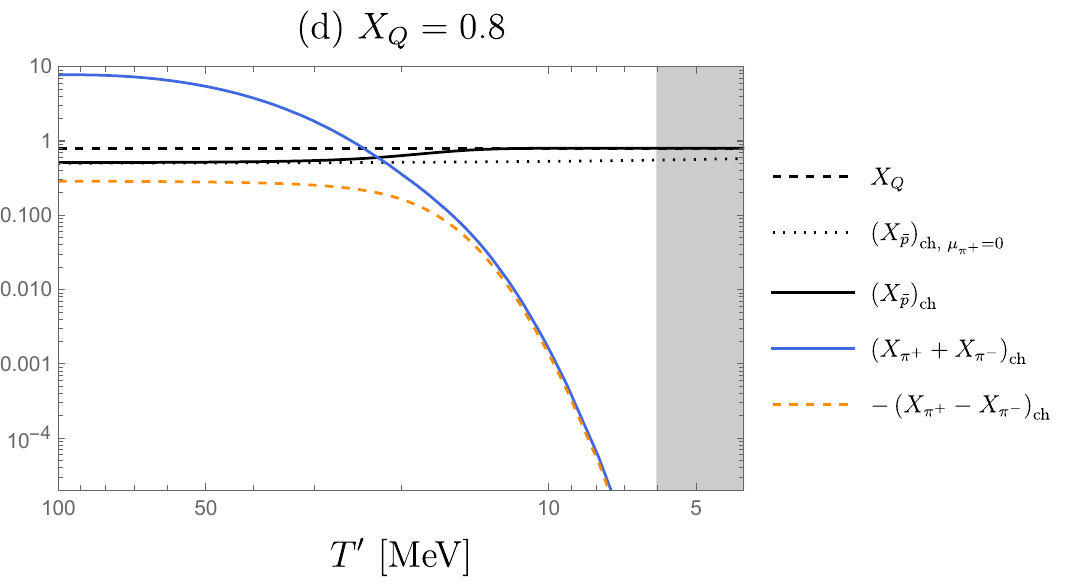}
    \caption{\label{fig:piplusmin}%
    Chemical-equilibrium abundances for different charged-pion chemical-potential $\mu_{\pi^+}$ assumptions. 
    The benchmark parameters $T_0=100\MeV$, $\bar{\eta}=10^{-2}$, $R_0=1.5\text{ mm}$ are assumed in these plots. 
    \textsc{Panel}~(a): The chemical-equilibrium abundance of anti-protons $\left(X_{\bar{p}}\right)_{\rm ch}$ as a function of the comoving temperature $T'$ for different values of $X_Q$ [defined at \eqref{eq:XQdef}]; the $\left(X_{\bar{p}}\right)_{\rm ch}$ for $\mu_{\pi^+}=0$ is also shown for comparison. 
    \textsc{Panels}~(b)--(d): The chemical-equilibrium abundances of anti-protons $\left(X_{\bar{p}}\right)_{\rm ch}$, symmetric population of charged pions $\left(X_{\pi^+}+X_{\pi^-}\right)_{\rm ch}$, and anti-symmetric population of charged pions $\pm \left(X_{\pi^+}-X_{\pi^-}\right)_{\rm ch}$ as a function of the comoving temperature $T'$ for the selected parameter values $X_Q=10^{-4}$, $X_Q=0.1$, and $X_Q=0.8$, respectively.
    For comparison, we also show $X_Q$ and $\left(X_{\bar{p}}\right)_{\rm ch}$ for $\mu_{\pi^+}=0$ (as assumed in \secref{ss:antineutronabundance}). 
    The gray-shaded regions are where we expect $X_{\bar{p}}$ and $X_{\pi^\pm}$ to be given by their freeze-out values, instead of the chemical-equilibrium values, $\left(X_{\bar{p}}\right)_{\rm ch}$ and $\left(X_{\pi^\pm}\right)_{\rm ch}$, that are shown in (or inferable from) these plots.}
\end{figure*}

Knowing  $\mu_{\pi^+}(X_Q,T')$ allows us to compute the chemical-equilibrium abundances of anti-protons and charged pions; see \figref{fig:piplusmin}. 
At high temperatures, chemical potentials are insignificant and consequently particles are thermally populated in a democratic way, yielding $\left(X_{\bar{p}}\right)_{\rm ch}\approx \left(X_{\bar{n}}\right)_{\rm ch}\approx 0.5$ and (to keep charge neutrality) $\left(X_{\pi^+}-X_{\pi^-}\right)_{\rm ch}\approx 0.5$. 
On the other hand, at low temperatures, thermodynamics favors storing charge asymmetry in $\bar{p}$ rather than $\pi^\pm$. Very approximately, this can be understood heuristically in the following way. 
We have assumed a net anti-baryon number for the plasma.
Because baryon number is conserved in the plasma, a fixed number of anti-nucleons (forming the net anti-baryon asymmetry) therefore automatically exist in the plasma at any temperature, either as $\bar{n}$ or $\bar{p}$; specifically, the mass of these particles need not be extracted from the thermal bath in order for them to be in existence.
We have also assumed a non-zero negative charge asymmetry on the hadronic sector, and charge is also conserved in the plasma.
An anti-proton $\bar{p}$ kills two birds with one proverbial stone: it can carry both baryon number and charge.
By contrast, pions do not carry baryon number and therefore are not automatically in existence in the plasma in any abundance: they must be created thermally; i.e., their rest-mass energy must be extracted from the thermal bath.
Effectively, it therefore costs much more energy to create a massive $\pi^-$ thermally from the bath in order to store negative charge on it, than it does to simply store the charge on $\bar{p}$. 
As a result, unless $X_Q\lesssim 10^{-4}$, most of the charge asymmetry $X_Q$ typically ends up in anti-protons, such that $X_{\bar{p}}\approx X_Q$ well before the SMCER interactions decouple.

\begin{figure*}[!t]
    \centering
    \includegraphics[width=0.66\textwidth]{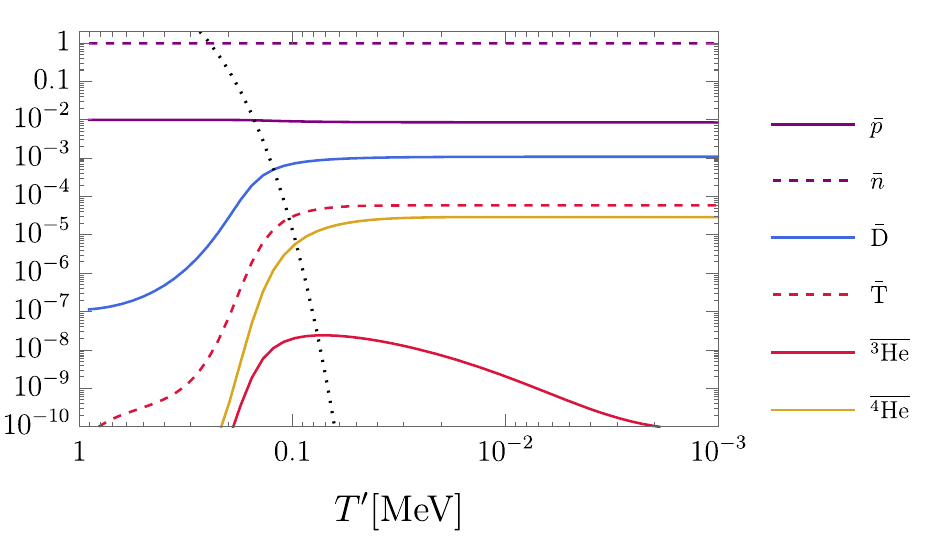}
    \caption{\label{fig:AbundanceswithChemPot}%
    The abundances of nuclear species  $X_i=n^{\prime}_i/n^{\prime}_{\bar{B}}$ (solid and dashed colored lines, as identified in the legend) as a function of the comoving fireball temperature $T'$, computed by numerically solving the Boltzmann equations for a simplified nuclear-reaction network detailed in \appref{app:reactionNetwork}, for a fireball with $T_0=100\,\MeV$, $R_0=1.5\,\text{cm}$, $\bar{\eta}=10^{-2}$ ($\Gamma=10$). 
    Note that this plot is similar to \figref{fig:Xvsx}; however, instead of neglecting the charged pion chemical potential, here we set $X_Q=10^{-2}$ such that $\mu_{\pi^+}=28.8\MeV$, $X_{\bar{p}}=0.01$, and $X_{\bar{n}}=0.99$, all evaluated at $T'\approx 6\MeV$. 
    Note that $R_0$ here is an order of magnitude larger than in \figref{fig:Xvsx}.}
\end{figure*}

Starting at a sufficiently high $T_0$ ($\gtrsim 10\MeV$), the actual comoving number densities of $\bar{n}$, $\bar{p}$, and $\pi^\pm$ closely track their chemical-equilibrium values (obtained from equilibrium thermodynamics as we have computed above) until they eventually freeze out once $\pi^\pm$ become too Boltzmann suppressed to keep the SMCER interactions efficient. 
We found that the decoupling temperature%
\footnote{%
    We computed $T'_{\bar{n}\bar{p}}$ using the criterion $\Gamma^\prime_{\rm strong}\tau'\sim 1$, where we took $\Gamma^\prime_{\rm strong}\sim \text{min}\left[\left(n^\prime_{\pi^+}\right)_{\rm ch},\left(n^\prime_{\pi^-}\right)_{\rm ch}\right]\left<\sigma v\right>_{\rm strong}$ which generalizes \eqref{eq:Gammastrong}. 
    Furthermore, we compare the rate $\Gamma^\prime_{\gamma\gamma\rightarrow \pi^+\pi^-}\sim n^\prime_\gamma(E^\prime_\gamma\gtrsim m_\pi)\times (\alpha_{\rm EM}^2/m_\pi^2)\sim \alpha_{\rm EM}^2T'e^{-m_\pi/T'}$ of the pair production process $\gamma\gamma\rightarrow\pi^+\pi-$ and find that $\Gamma^\prime_{\gamma\gamma\rightarrow \pi^+\pi^-}/\Gamma^\prime_{\rm strong}\sim 1.3\times (6\MeV/T')^{1/2}$. 
    Therefore, the strong and electromagnetic interactions freeze out at similar temperatures.} %
$T'_{\bar{n}\bar{p}}$ varies slightly in the range $6\MeV\lesssim T'_{\bar{n}\bar{p}}\lesssim 8\MeV$ when $X_Q$ is varied in the range $0\leq X_Q\leq 1$. 
Importantly, for $X_Q \gtrsim 10^{-3}$, we find that $X_{\bar{p}}(T')\rightarrow X_Q$ already for $T'>T'_{\bar{n}\bar{p}}$; see \figref{fig:piplusmin}.
This means that we can indeed rely on our equilibrium thermodynamics arguments above to establish that the charge asymmetry $X_Q$ is maintained on the anti-protons in this regime.

In general, we find that having $X_Q\sim 1-X_Q\sim \mathcal{O}(1)$ results in $X_{\bar{n}}/X_{\bar{p}}\sim \mathcal{O}(1)$ at SMCER decoupling; see \figref{fig:piplusmin}.
Because there are only $\mathcal{O}(1)$ quantitative changes to the anti-nucleosynthetic abundance results in the main text for more general values for $X_{\bar{n}}/X_{\bar{p}} \sim \mathcal{O}(1)$ than the value of $X_{\bar{n}}/X_{\bar{p}} \approx 0.8$ that we used in \secref{ss:antineutronabundance} (under the more restrictive assumption of $\mu_{\pi^+}=0$), this implies that the same qualitative conclusions that we reached in the main text will continue to hold for any values of $X_Q\sim 1-X_Q \sim \mathcal{O}(1)$. 
This is turn is significant because, in order to achieve $\mu_{\pi^+}=0$ exactly at $T'_{\bar{n}\bar{p}}$ (as we assumed in \secref{ss:antineutronabundance}) we would need to choose $X_Q$ such that it almost exactly matches the freeze-out anti-proton abundance in the absence of $\mu_{\pi^+}$: i.e., $X_Q \approx \left(X_{\bar{p}}\right)_{\text{ch}}|_{\mu_{\pi^+}=0}\approx 0.56$.
Were that exact value a crucial input to the anti-nucleosynthesis analysis, this would constitute a fine tuning because $X_Q$ is supposed to be set by pre-thermalization physics which has no connection to the post-thermalization physics (although see the discussion in the last paragraph of this Appendix below).
Our analysis in this Appendix thus establishes that our main-text analysis is \emph{not} finely tuned.

Next, we discuss how the fireball outputs are affected if $\mu_{\pi^+}\gg m_n-m_p$. 
In this case, we generically have $X_{\bar{p}}\ll 1$ at SMCER decoupling. 
However, even then, a $X_{\antiHe{4}}/X_{\antiHe{3}}\sim 1/2$ anti-helium isotope ratio can still be obtained without overproducing anti-protons; i.e., $X_{\bar{n}}+X_{\bar{p}}\lesssim 10^5 X_{\antiHe{4}}$ when the fireball becomes optically thin, in order to avoid violating the $1\sigma$ uncertainties on the AMS-02 anti-proton flux~\cite{AMS:2016oqu}; see \figref{fig:spectra}.
To demonstrate this, we show in \figref{fig:AbundanceswithChemPot} the evolution of nuclear abundances in this $\mu_{\pi^+}\gg m_n-m_p$ case with $X_Q=10^{-2}$ and other parameters chosen to explain the AMS-02 candidate anti-helium events. 
The freeze-out abundances shown in \figref{fig:AbundanceswithChemPot} are marginally consistent with the anti-proton flux observed at AMS-02. 
To explain the anti-helium isotope ratio with $X_Q\lesssim 10^{-2}$ would however lead to an overproduction of anti-protons. Therefore, compatibility with the AMS-02 data requires $X_Q\gtrsim 10^{-2}$. 
Note, however, that the marginal set parameters we picked for \figref{fig:AbundanceswithChemPot} to complement $X_Q=10^{-2}$ could be interesting from the perspective of the $\sim 10\%$ anti-proton excess~\cite{Cholis:2019ejx}.

We have so far treated the charge asymmetry $X_Q$ as a free parameter. 
In actuality, it depends \emph{both} on the properties of the initial particle injection, which are model-dependent but could naturally give $X_Q\sim 1$, \emph{and} post-injection SM processes. 
In case the initial injection does not produce a considerable charge asymmetry, some degree of $X_Q$ can arise spontaneously from non-equilibrium Standard Model processes occurring before or after thermalization. 
Electroweak process such as $\bar{n}+e^-\leftrightarrow \bar{p}+\nu_e$ and its variants can transfer charges between the hadron and lepton sectors; however, they have a preferred charge-transfer direction only at sufficiently low temperatures, $T'\lesssim m_n-m_p$, at which point electroweak interaction rates are extremely suppressed (given the dynamical timescales involved in the fireball expansion).
A higher contribution to $X_Q$ arises from the decay of (a small fraction of) charged pions. 
As shown in \figref{fig:piplusmin}(b), even if we start with $X_Q$ being virtually zero, a charge asymmetry of $X_{\pi^+}-X_{\pi^-}\approx 0.5$ is automatically present in $\pi^\pm$ at high temperatures (SMCER reactions are efficient, so if $X_Q\ll 1$, we have $X_{\bar{p}} \sim X_{\pi^+}-X_{\pi^-}$). 
Within the initial fireball expansion timescale (during which the bulk plasma motion is non-relativistic), a fraction $10^{-4} \lesssim R_0/\tau_{\pi^\pm}\lesssim 10^{-1}$ (for $1\,\text{mm} \lesssim R_0\lesssim 1\,\text{m}$) of the charged pions decay to charged muons [$\pi^\pm\rightarrow \mu^\pm+\nu_\mu(\bar{\nu}_\mu)$ with $\tau_{\pi^\pm} \sim 2.6\times 10^{-8}\,\text{s}$~\cite{PDG}], thereby generating $X_Q\sim(X_{\pi^+}-X_{\pi^-})(R_0/\tau_{\pi^\pm})$ in the range $5\times 10^{-5} \lesssim X_Q\lesssim 5\times 10^{-2}$.
Unfortunately, at least for the benchmark $\sim 1\,\text{mm}$ parameters we used in the main text (or the benchmark $R_0 \sim 1\,\text{cm}$ used in \figref{fig:AbundanceswithChemPot}), this spontaneously generated $X_Q$ is too small to be phenomenologically viable in light of the AMS anti-proton results (see discussion above).

\section{Simplified nuclear reaction network}
\label{app:reactionNetwork}

Anti-neutrons and anti-protons in the fireball plasma are initially held in chemical equilibrium by charged-pion--mediated anti-neutron--anti-proton interconversions (i.e., SMCER) at $10\lesssim T'\lesssim 200\,\MeV$. 
As found in \secref{ss:antineutronabundance}, such interconversion process decouples relatively early, at $T_{\bar{n}\bar{p}}^\prime\approx 6\,\MeV$, when there are virtually no nuclear bound states present. 
On the other hand, photodissociation stalls the synthesis of anti-deuterium (and the whole nuclear chain) until the temperature is considerably below the anti-deuterium binding energy $B_D\approx 2.2\,\MeV$, at $T_{\rm D}^\prime\sim 140-170\,\keV$, by which point the anti-neutron--to--anti-proton ratio has completely frozen out. 
Hence, we can treat the anti-neutron--anti-proton decoupling separately from the whole nuclear reaction chain. 
In our numerical procedure, we thus first solve for the freeze-out ratio of $n^\prime_{\bar{n}}/n^\prime_{\bar{p}}$ ignoring nuclear reactions, finding $n^\prime_{\bar{n}}/n^\prime_{\bar{p}}\approx 0.8$ per \eqref{eq:nDnpch} [assuming $\mu_{\pi^+}=0$; similar results could be derived for more general $X_Q$ as discussed in \appref{app:antinucleonFO}], and use this as an input when solving the Boltzmann equations governing the subsequent nuclear reactions.

We now describe our simplified nuclear reaction network governing the evolution of relative abundances of elements, $X_i=n^\prime_i/n^\prime_{\bar{B}}$. 
We consider only elements with $A\leq 4$, namely $\bar{p}$, $\bar{n}$, $\antiD$, $\antiT$, $\antiHe{3}$, and $\antiHe{4}$. 
Heavier species with $A>4$ such as $\overline{\text{Li}}$ and $\overline{\text{Be}}$ isotopes are virtually absent and have negligible impacts on the $A\leq 4$ elements that we consider. 
Conservation of baryon number ensures $\sum_i A_i X_i=1$, with $A_i$ being the atomic mass number of the nuclear species $i$. 
This can be used to solve for the abundance of one element, which we take to be $X_{\bar{p}}$, given the other abundances
\begin{align}
    X_{\bar{p}}=1-\sum_{i\neq \bar{p}}A_i X_i \:.
\end{align}
We set as the initial conditions
\begin{align}
    X_{\bar{p}}&=0.56\:, & X_{\bar{n}}&=0.44\:, & X_{i\notin\{ \bar{n},\bar{p}\}} &=0\:,
\end{align}
per \eqref{eq:nDnpch} at an initial temperature%
\footnote{%
    This value of the initial temperature $T'=1\,\MeV$ is arbitrarily chosen. As long as it lies in the range $200\,\keV\lesssim T'\lesssim 6\,\MeV$, the exact value of the temperature at which the initial conditions are set is not important.} %
$T'=1\,\MeV$, and evolve the abundances of elements other than $\bar{p}$ with the following simplified nuclear reaction network~\cite{Mukhanov:2003xs}:
\begin{widetext}
\begin{align}
    -\frac{dX_{\bar{n}}}{d\ln T'} & \approx 
    \frac{\epsilon_{\rm nuc}(T')}{\text{mb}}\left(\begin{array}{l}%
        -\left<\sigma v\right>_{\bar{p}\bar{n}} X_{\bar{p}} X_{\bar{n}}+\left<\sigma v\right>_{\antiD\gamma}X_{\antiD}Y_{\antiD\gamma}+ \left<\sigma v\right>_{\antiD\antiD} X_{\antiD}^2+ \left<\sigma v\right>_{\antiT\antiD} X_{\antiT}X_{\antiD}-\left<\sigma v\right>_{\antiD\bar{n}} X_{\antiD}X_{\bar{n}} \\[2ex]   
        -\left<\sigma v\right>_{\antiHe{3}\bar{n}} X_{\antiHe{3}} X_{\bar{n}}+\left<\sigma v\right>_{\antiHe{3}\gamma}X_{\antiHe{3}}Y_{\antiHe{3}\gamma}
    \end{array}\right) \label{eq:Bolt1}\:;\\[2ex]
    -\frac{dX_{\antiD}}{d\ln T'}&\approx 
    \frac{\epsilon_{\rm nuc}(T')}{\text{mb}}\left(\begin{array}{l}%
        \left<\sigma v\right>_{\bar{p}\bar{n}} X_{\bar{p}} X_{\bar{n}}-\left<\sigma v\right>_{\antiD\gamma}X_{\antiD}Y_{\antiD\gamma}-\left<\sigma v\right>_{\antiD\antiD} X_{\antiD}^2-\left<\sigma v\right>_{\antiD\bar{p}}X_{\antiD}X_{\bar{p}}-\left<\sigma v\right>_{\antiD\bar{n}} X_{\antiD}X_{\bar{n}}\\[2ex]%
        -\left<\sigma v\right>_{\antiT\antiD} X_{\antiT}X_{\antiD}-\left<\sigma v\right>_{\antiHe{3} \antiD}X_{\antiHe{3}}X_{\antiD}
    \end{array}\right) \label{eq:Bolt2}\:; \\[2ex]
    -\frac{dX_{\antiT}}{d\ln T'}& \approx
    \frac{\epsilon_{\rm nuc}(T')}{\text{mb}}\Big(%
        \left<\sigma v\right>_{\antiD\antiD} X_{\antiD}^2+\left<\sigma v\right>_{\antiD\bar{n}}X_{\antiD}X_{\bar{n}}+\left<\sigma v\right>_{\antiHe{3}\bar{n}} X_{\antiHe{3}} X_{\bar{n}}-\left<\sigma v\right>_{\antiT\antiD} X_{\antiT}X_{\antiD}%
    \Big) \label{eq:Bolt3}\:; \\[2ex]
    -\frac{dX_{\antiHe{3}}}{d\ln T'}& \approx 
    \frac{\epsilon_{\rm nuc}(T')}{\text{mb}}\left(\begin{array}{l}%
        \left<\sigma v\right>_{\antiD\antiD} X_{\antiD}^2+\left<\sigma v\right>_{\antiD\bar{p}} X_{\antiD}X_{\bar{p}}-\left<\sigma v\right>_{\antiHe{3}\gamma}X_{\antiHe{3}}Y_{\antiHe{3}\gamma}-\left<\sigma v\right>_{\antiHe{3}\bar{n}} X_{\antiHe{3}} X_{\bar{n}} \\[2ex]%
        -\left<\sigma v\right>_{\antiHe{3}\antiD} X_{\antiHe{3}} X_{\antiD}
    \end{array}\right)\:; \label{eq:Bolt4}\\[2ex]
    -\frac{dX_{\antiHe{4}}}{d\ln T'}&\approx 
    \frac{\epsilon_{\rm nuc}(T')}{\text{mb}}\Big(%
        \left<\sigma v\right>_{\antiT\antiD} X_{\antiT}X_{\antiD}+\left<\sigma v\right>_{\antiHe{3}\antiD} X_{\antiHe{3}} X_{\antiD}
    \Big)\:, \label{eq:Bolt5}
\end{align}
\end{widetext}
where we have defined the dimensionless quantity
\begin{align}
    \epsilon_{\rm nuc}(T')=\bar{\eta} n_\gamma^\prime\times \text{mb} \times \tau'(T')  \:, \label{eq:epsilonnuc} 
\end{align}
which quantifies the efficiency of nuclear reactions on the timescale of fireball expansion. 
Apart from anti-deuterium photodissociation, we include in the above set of equations only thresholdless (forward) reactions. 
The reverse processes to these reactions are endothermic, and are suppressed by factors of the form $e^{-Q/T'}$ (where $Q$ is the threshold energy of the reverse reaction; i.e., the energy liberated in the forward reaction) in their thermally averaged cross section as well as the smallness (verified \emph{a posteriori}) of the $X_i$ factors relative to those of the forward reactions. 
It is nevertheless important to take into account anti-deuterium photodissociation due to its uniquely low threshold energy and the potentially large photon abundance. 
The abundance of photons energetic enough to photodissociate the nuclear species $i$ is
\begin{align}
    Y_{i\gamma}=\frac{n^\prime_{\gamma}(E^\prime_\gamma > Q_{i})}{n^\prime_{\bar{B}}}\sim \frac{Q_i^2}{\bar{\eta} T^{\prime 2}} e^{-Q_i/T'} \:.
\end{align}
where $Q_i$ is the threshold energy of that process. 
The photodissociation cross-sections as functions of energy typically rise sharply at $E_{\gamma}\approx Q_{i}$ and fall of smoothly at higher $E_{\gamma}$. 
We approximate the thermal averaged photodissociation cross-sections $\left<\sigma v\right>_{i\gamma}$ with their near-threshold values. 
We consider only the photodissociation processes $\antiD+\gamma\rightarrow \bar{p}+\bar{n}$, with $Q_{\antiD}=2.2\,\MeV$ and $\left<\sigma v\right>_{\antiD\gamma}\approx 2.5\,\text{mb}$~\cite{Cyburt:2002uv}. The other ones have significantly higher threshold energies $Q_i$ and are consequently much less efficient due to the exponentially suppressed $E_\gamma\gtrsim Q_{i}$ photon abundance (as considered at the temperature $T'\sim 100\,\keV$ when anti-nucleosynthesis is efficient).

\section{Nuclear cross sections}
\label{app:nuclearCrossSections}

\begin{figure*}
    \centering
    \includegraphics[width=0.6\textwidth]{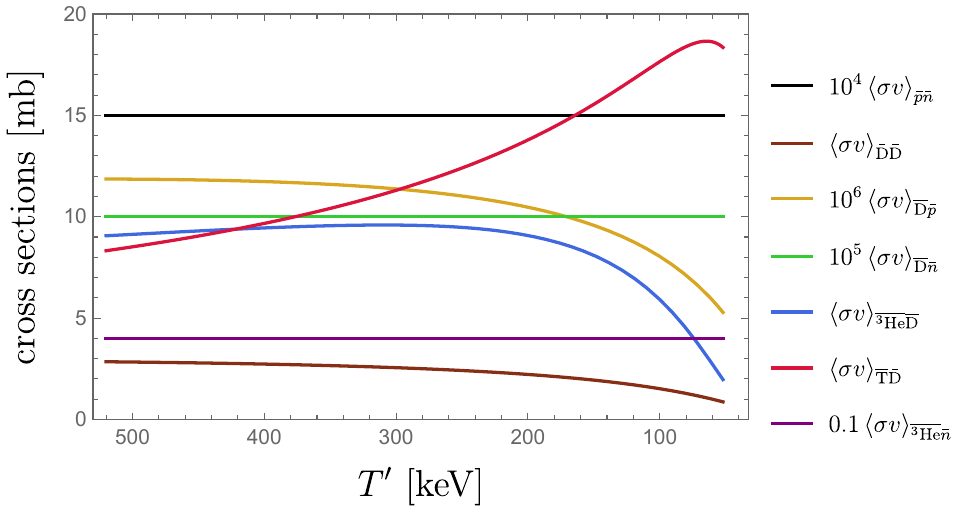}
    \caption{\label{fig:crosssections}%
    The assumed values of thermal-average cross sections for (anti-)nuclear reactions as a function of temperature $T'$.  
    } 
\end{figure*}

We assume that the nuclear cross-sections for purely anti-matter processes are equal to those of the analogous reactions involving matter. 
For completeness, we list here the thermal averaged nuclear cross sections from~\cite{Wagoner:1972jh, Serpico:2004gx,Gorbunov:2011zz,Esmailzadeh:1990hf} and express them in terms of the temperature ratio $T_9\equiv T'/(10^9\,\text{K})=T'/(86\,\keV)$:
\begin{itemize}
    \item $\bar{n}+\bar{p}\rightarrow\antiD+\gamma$ ($Q=2.22\,\MeV$)
    \begin{align*}
        \left.\left<\sigma v\right>_{\bar{n}\bar{p}}\right|_{T'\sim 100\,\keV}&\approx  2\times 10^{-3}\,\text{mb} \:.
    \end{align*}
    \item $\antiD+\antiD\rightarrow \antiHe{3}+\bar{n}$ ($Q=3.26\,\MeV$) and $\antiD+\antiD\rightarrow \antiT+\bar{p}$ ($Q=4.03\,\MeV$). For each,
    \begin{align*}
        \left<\sigma v\right>_{\rm \antiD\antiD}\approx 100\,\text{mb}\times T_9^{-2/3}e^{-4.3 T_9^{-1/3}}\:.
    \end{align*}
    \item $\antiD+\bar{p}\rightarrow \antiHe{3}+\gamma$ ($Q=5.49\,\MeV$)
    \begin{align*}
        \left<\sigma v\right>_{ \antiD\bar{p}}\approx 3\times 10^{-4}\,\text{mb}\times  T_9^{-2/3}e^{-3.7 T_9^{-1/3}}   \:.    
    \end{align*}
    \item $\antiD+\bar{n}\rightarrow \antiT+\gamma$ ($Q=6.24\,\MeV$)
    \begin{align*}
        \left.\left<\sigma v\right>_{ \antiD\bar{n}}\right|_{T'\sim 100\,\keV}\approx 1\times 10^{-4}\, \text{mb}   \:.    
    \end{align*}
    \item $\antiT+\antiD\rightarrow\antiHe{4}+\bar{n}$ ($Q=17.59\,\MeV$)
    \begin{align*}
        \left<\sigma v\right>_{ \antiT\antiD}\approx 30\,\text{mb}\times T_9^{-2/3}e^{-0.5 T_9^{-1}} \:.
    \end{align*}  
    \item $\antiHe{3}+\antiD\rightarrow\antiHe{4}+\bar{p}$ ($Q=18.35\,\MeV$)
    \begin{align*}
        \left<\sigma v\right>_{ \antiHe{3}\antiD}\approx 30\,\text{mb}\times T_9^{-1/2}e^{-1.8 T_9^{-1}} \:.
    \end{align*}
    \item $\antiHe{3}+\bar{n}\rightarrow\antiT+\bar{p}$ ($Q=0.76\,\MeV$)
    \begin{align*}
        \left.\left<\sigma v\right>_{ \antiHe{3}\bar{n}}\right|_{T'\sim 100\,\keV}\approx  40\,\text{mb}   \:.    
    \end{align*}
\end{itemize}
We show these cross sections in \figref{fig:crosssections}, rescaled as appropriate to fit them all on the same plot. 
Most important to our analysis are cross-sections at temperatures $T'\sim 100$--$200\,\keV$ where anti-nucleosynthesis mainly occurs. 
We neglect the mild temperature dependences of the cross-sections for processes that do not suffer from Coulomb-barrier suppression.

\section{Changes in the number of degrees of freedom}
\label{app:dof}

Here we account for the effect of changes in the number of degrees of freedom $g_*$ on the fireball expansion dynamics (see \secref{sec:expansionDynamics}), neglecting the small differences between the $g_*$ for entropy and energy density~\cite{Husdal:2016haj}; i.e., we assume $g_{*S}= g_{*\rho}=g_*$.
We first write the energy and entropy conservation equations, respectively, for each of the shells (assumed relativistic, with $v\sim 1$) in more general forms:
\begin{align}
    \gamma^2 (\rho'+p') r^2\delta r&\approx \text{constant}\:, \label{eq:EconsGen} \\
    \gamma s' r^2\delta r&\approx\text{constant}\:.\label{eq:SconsGen}
\end{align}
In cases of our interest, the fireball can be approximated as consisting of only radiation (photons and relativistic massive species) and (anti-)matter (dominated by anti-baryons), allowing us to write 
\begin{align}
    \rho'+p'\approx \begin{cases}
        s'T', &T'\gg \bar{\eta} m_{p}\\
        \bar{\eta}s' m_{ p}, &T'\ll \bar{\eta}m_{ p}
    \end{cases} \:,
\end{align}
where $s'T'$ and $\bar{\eta}s' m_{p}$ are the contributions from radiation and (anti-)matter, respectively, and we used $\bar{\eta} \equiv n_{\bar{B}}/s'$.
Eliminating $r$ from the energy and entropy conservation equations, and using $\Gamma\equiv T_0/(\bar{\eta} m_{p})$, we find
\begin{align}
    \gamma\left(T'+\frac{T_0}{\Gamma}\right)\sim \text{constant}\:. \label{eq:gammaScaling}
\end{align}
Substituting $s' \propto g_{*}(T') \times (T')^3$, \eqref{eq:gammaScaling}, and \eqref{eq:thickness} into \eqref{eq:SconsGen}, we obtain an equation that determines the evolution of $T'$ with $r$:
\begin{align}
    &g_*(T') \left(\frac{T^{'3}}{T'+T_0/\Gamma}\right)r^2 \nl \times \left[R_0+\left(\frac{T'+T_0/\Gamma}{T_0}\right)^2r\right] \sim \text{constant}\:,\\
\end{align}
which shows how the temperature-radius relation $T'(r)$ is modified in the presence of changes in $g_*(T')$ compared to the $T_{g_*=2}^\prime(r)$ for constant $g_*=2$ assumed in the main text. 
During RD ($T'\gtrsim T_0/\Gamma$), we thus have
\begin{align}
        T' &\sim \lb[ \frac{g_*(T')}{g_*(T_0)}\rb]^{-1/2}T_{g_*=2}^\prime & \text{[RD]}\:;\\
        \rho^{\prime} &\sim g_*(T') \times T^{\prime 4}\propto \lb[\frac{g_*(T')}{g_*(T_0)}\rb]^{-1} \rho_{g_*=2}^\prime & \text{[RD]}\:,
\end{align}
while during MD ($T'\gtrsim T_0/\Gamma$) we have 
\begin{align}
    T' &\propto \lb[\frac{g_*(T')}{g_*(T_0/\Gamma)}\rb]^{-1/3}T_{g_*=2}^\prime & \text{[MD]}\:;\\[1ex]
    \rho^{\prime}&\propto g_*(T') \times T^{\prime 3}\propto [g_*(T')]^{0} & \text{[MD]}\: .
\end{align}
Notice that the impact of a change in $g_*$ is weak if it occurs during matter domination. 

In our scenario, the effective number of relativistic degrees of freedom of the fireball $g_*$ is given by
\begin{align}
    g_*(T')=\begin{cases}
        12\ , & 100\,\MeV\lesssim T'\lesssim 200\,\MeV\:; \\
        5.5\ , &\phantom{00}1\,\MeV\lesssim T'\lesssim 100\,\MeV\:; \\
        2\ , &\phantom{\, 100\,\MeV\lesssim}T'\lesssim 1 \,\MeV\:;
    \end{cases}
\end{align}
reflecting, respectively, the contributions from the sets of particles $\{\pi^{0,\pm},e^\pm,\mu^\pm,\gamma\}$, $\{e^\pm,\gamma\}$, and $\{\gamma\}$ in each temperature range.
That is, we have two possible ``jumps'' in $g_*$, happening at $T'\sim 100\,\MeV$ and $T'\sim 1\,\MeV$ (the transitions of course are smooth, but abrupt). 
Since we require $10\,\MeV\lesssim T_0\lesssim 100\,\MeV$ and $\Gamma\sim 10$, only the jump at $T'\sim 100\,\MeV$ has a chance to spoil the accuracy of the constant-$g_*$ approximation adopted in the main analysis.
This jump at $T'\sim 100\,\MeV$ amounts to an error in the analytical estimate for $T'(r)$ by a factor of $\sqrt{12/5.5}\approx 1.5$ and that for $\rho'(r)$ by a factor of $12/5.5\approx 2.2$. 
These are within the level precision we are aiming for, given that our analysis does not fully capture the effects of the inhomogeneity and initial evolution of the fireball, which would also introduce $\mathcal{O}(1)$ uncertainties in our estimates for $T'(r)$. 
Note that the Lorentz factor $\gamma$ remains to be related to the temperature $T'$ as given in \eqref{eq:gammaScaling}. 

Although $T'$ appears at various places in the Boltzmann \eqrefRange{eq:Bolt1}{eq:Bolt5} [it plays the role of time, appears in the cross sections, and controls $Y_{i\gamma}$], its relation to the shell radius $r$, namely $T'(r)$ [which is affected by changes in $g_*(T')$], only enters  via $\epsilon_{\rm nuc}(T')$, as defined in \eqref{eq:epsilonnuc}, through the comoving expansion timescale $\tau'(T')$. 
The latter is given by \eqref{eq:tauprime} in the absence of changes in $g_*(T')$ and would change by a mild $\mathcal{O}(1)$ factor when the effects of changes in $g_*(T')$ are included. 
This amounts to a slight shift in the value of $T_0$ in relation to $R_0$ that yields a given $\tau'$. 
Since $\tau'$ controls the output nuclear abundances, the lines of constant anti-helium isotope ratio in, e.g., \figref{fig:fireballparameterspace}, would be displaced, though only slightly.

\section{Prompt (`burst') or slow (`wind') injection}
\label{app:BurstWind}

The duration of the anti-quark injection $t_*$ can be shorter $t_*\lesssim R_*$ (burst regime) or longer $t_*\gtrsim R_*$ (wind regime%
\footnote{%
    See \citeR{Chang:2022gcs} for a discussion of thermalization and hydrodynamics in the wind regime in an analogous setup (but completely different context).}%
) than the spatial size of the injection $R_*$.

Following the injection, the anti-quarks will have the following number density profile if they do not interact with one another and simply free stream away
\begin{align}
    n_{B,0}(r)&\sim \zeta_*\times \frac{E_*}{m_{\textsc{dm}} R_{*}^3}\times  \text{min}\left(1,\frac{R_{*}^2}{r^2}\right) \:,
\end{align}
where $r$ is the radius from the center of a spherical injection region of radius $R_*$, $E_*$ is the total energy of the injected SM particles, and the factor
\begin{align}
    \zeta_*\sim \text{min}\left(1,\frac{R_{\rm *}}{t_{\rm *}}\right) \lesssim 1
\end{align}
accounts for the burst vs.~wind branching of cases: in the wind case the density of the injected anti-quarks within $R_*$ is diluted by a factor of $R_*/t_*$.

Regardless of the relative sizes of $t_*$ and $R_*$, the probability that an anti-quark undergoes a process of any sort before arriving at a radius $r$, $P(r)\sim n_{B,0}(r)\sigma v_{\rm rel}\times r\propto r^{-1}$, decreases with $r$ for $r\gg R_{*}$, where we have assumed%
\footnote{%
    When geometrical effects are taken into account, one would find that $v_{\rm rel}$ goes down with $r$ as the motion of the injection particles become increasingly radial~\cite{MacGibbon:2007yq}, which means the probability $P(r)$ would actually reduce even faster with $r$, further strengthening the argument we are making.
    } %
that the relative velocity is relativistic $v_{\rm rel}\sim 1$ and the cross section $\sigma$ is independent of $r$. 
It follows that in order for these particles to thermalize, the thermalization rate needs to be efficient inside the injection region; i.e., $\Gamma_{\rm th}\gtrsim R_*^{-1}$ at $r\lesssim R_*$. 

Once the fireball can achieve thermalization and has attained a semirelativistic bulk radial velocity $v\sim 1$, its subsequent evolution is completely described by the fireball temperature $T_0$ and radius $R_0$ when it first thermalizes (as well as $X_Q$), and these are given by
\begin{align}
    T_0&\sim \left(\frac{\zeta_*E_*}{R_*^3}\right)^{1/4} \:,\label{eq:T0wind}\\
    R_0&\sim R_* \:.\label{eq:R0wind}
\end{align}
Note that the anti-baryon--to--entropy ratio $\bar{\eta}$ does not depend on the relative size between $t_*$ and $R_*$.

As explained in \appref{app: fireballscalings}, the plasma outflow can be treated as a series independent differential radial slices which separately go through nearly the same thermal and hydrodynamical evolution. 
In each of these slices, anti-nucleosynthesis proceeds as described in \secref{sec:FireballAntiNucleo}. 
The only difference is that $T_0$ and $R_0$ are now more generally given by \eqref[s]{eq:T0wind} and (\ref{eq:R0wind}). 
Since the abundances of nuclear species $\left.X_i\right|_{\rm inj.}$ released by each radial slice are completely determined by the set of parameters $(T_0,R_0,\bar{\eta},X_Q)$ which are essentially the same for all the shells, the numbers of anti-nuclei or anti-nucleons $N_i|_{\rm inj.}$ released into the ISM are still given by $N_i|_{\rm inj.}=B \times X_i|_{\rm inj.}$ regardless of whether the injection is in the burst or wind regime.

See also \citeR[s]{Fiorillo:2023cas,Fiorillo:2023ytr} for related studies.

\section{AMS-02 anti-helium sensitivity}
\label{app:AMSantiHeSensitivity}

We review here the procedure referred to in the main text of \secref{sec:Galprop} that we used to obtain an estimate of the AMS-02 anti-helium sensitivity as a function of rigidity $\mathcal{R}$.
This procedure was specified in Appendix B of \citeR{Winkler:2020ltd}; our exposition here merely reproduces the argument in that reference (adding some additional detail per \citeR{LindenCorrespondence}) and is given only to make our presentation self-contained.

For a particle species $i$ with rigidity-dependent flux at AMS-02 given by $\Phi_i(\mathcal{R})$, the number of events $N_i$ in the rigidity range $\mathcal{R}_a \leq \mathcal{R} \leq \mathcal{R}_b$ observed at AMS-02 in a time $T$ can be written as~\cite{Winkler:2020ltd}
\begin{align}
    N_i(\mathcal{R}_a,\mathcal{R}_b;T) \equiv \int_{\mathcal{R}_a}^{\mathcal{R}_b} \Phi_i(\mathcal{R}) \zeta_i(\mathcal{R},T) d\mathcal{R}\:, \label{eq:zetaDefn}
\end{align}
where the species-dependent ``acceptance'' $\zeta_i(\mathcal{R},T)$ is defined here [as in \citeR{Winkler:2020ltd}, where it was called $\eta_i(\mathcal{R})$] to fold in all relevant effects such as integration time $T$, detector effective area, trigger efficiency, etc. [i.e., it subsumes the factors $\mathcal{A}_i$, $\epsilon_i$, and $T_i$ in Eq.~(1) of \citeR{AMSHeflux} (or the factors of $\mathcal{A}^A_i$, $\epsilon^A_i$, and the isotope-dependent integration time implicit in the rate $\Gamma^A_i$ in Eq.~(1) of \citeR{PhysRevLett.123.181102})].

\newcommand{\AntiHeFluxSM}{Ref.~\cite[Suppl.]{AMSHeflux}}
Following \citeR{Winkler:2020ltd}, we consider first ordinary helium and look to the Supplemental Material of \citeR{AMSHeflux} (hereinafter, ``\AntiHeFluxSM'')~(see also \citeR{AMS:2021nhj}), which gives values for $\Phi_{\text{He}}(\mathcal{R})$ and its statistical uncertainty $u[\Phi_{\text{He}}(\mathcal{R})]_{\text{stat.}}$ over narrow rigidity bins in the range $\mathcal{R} \in [1.92,3\times 10^3]\,\text{GV}$.
These data can be used to extract $\zeta_{\text{He}}(\mathcal{R})$ in the following manner~\cite{Winkler:2020ltd,LindenCorrespondence}.
Suppose that one of the aforementioned rigidity bins is centered at $\mathcal{R}$, and has width $\Delta \mathcal{R}$; it then follows from \eqref{eq:zetaDefn} that we can write the number of events in that rigidity bin as 
\begin{align}
    \Delta N_{\text{He}}(\mathcal{R},T) \approx \Phi_{\text{He}}(\mathcal{R}) \zeta_{\text{He}}(\mathcal{R},T) \Delta \mathcal{R}\:; \label{eq:singleBin}
\end{align}
correspondingly, the Poisson statistical uncertainty on that number is 
\begin{align}
    u[\Delta N_{\text{He}}(\mathcal{R},T)]_{\text{stat.}} &\approx \sqrt{\Delta N_{\text{He}}(\mathcal{R},T)}\\
    & = \sqrt{\Phi_{\text{He}}(\mathcal{R}) \zeta_{\text{He}}(\mathcal{R},T) \Delta \mathcal{R}}\:. \label{eq:uNpoisson}
\end{align}
But we also have from \eqref{eq:singleBin} that the statistical uncertainty on the flux is
\begin{align}
    u[\Phi_{\text{He}}(\mathcal{R},T)]_{\text{stat.}}\approx \frac{ u[\Delta N_{\text{He}}(\mathcal{R},T)]_{\text{stat.}} }{  \zeta_{\text{He}}(\mathcal{R},T) \Delta \mathcal{R} }\:. \label{eq:uPhiProp}
\end{align}
Therefore, from \eqref[s]{eq:uNpoisson} and (\ref{eq:uPhiProp}), we have
\begin{align}
    u[\Phi_{\text{He}}(\mathcal{R},T)]_{\text{stat.}}&\approx \sqrt{\frac{ \Phi_{\text{He}}(\mathcal{R})}{ \zeta_{\text{He}}(\mathcal{R},T) \Delta \mathcal{R}} }\:; \\
    \Rightarrow \zeta_{\text{He}}(\mathcal{R},T) &\approx \frac{ \Phi_{\text{He}}(\mathcal{R})}{ (u[\Phi_{\text{He}}(\mathcal{R},T)]_{\text{stat.}})^2 \Delta \mathcal{R}}\:.
\end{align}
All the quantities on the RHS of the last line are known for $T = \tau_{0}\sim 30\,\text{months}$ from \AntiHeFluxSM, allowing us to extract $\zeta_{\text{He}}(\mathcal{R},\tau_0)$ as
\begin{align}
   \zeta_{\text{He}}(\mathcal{R},\tau_0) &\approx \frac{ \Phi_{\text{He}}(\mathcal{R})}{ (u[\Phi_{\text{He}}(\mathcal{R},\tau_0)]_{\text{stat.}})^2 \Delta \mathcal{R}}\:. \label{eq:singleBinZeta}
\end{align}
We also then have that
\begin{align}
    \Delta N_{\text{He}}(\mathcal{R},\tau_0) \approx \lb[ \frac{ \Phi_{\text{He}}(\mathcal{R})}{ u[\Phi_{\text{He}}(\mathcal{R},\tau_0)]_{\text{stat.}} } \rb]^2 \:. \label{eq:singleBinNumber}
\end{align}

It remains to translate this to an anti-helium sensitivity.
We follow the argument given in \citeR{Winkler:2020ltd} to do this.
In \citeR{Kounine:2010js}, a projection for the 95\% confidence upper limit on the ratio of the anti-helium flux to the measured helium flux, $r_{95}$, assuming no observed events in an integration time $T$ in the range $(\mathcal{R}_1 = 1\,\text{GV}) \leq \mathcal{R} \leq (\mathcal{R}_2 = 50\,\text{GV})$ is given as 
\begin{align}
    r_{95}(T) \approx 5\times 10^{-10} \times \frac{18\,\text{yr}}{T}.
\end{align}
Because the Poisson distribution function $f_{\text{P}}(\lambda;n) \equiv \lambda^n \exp[-\lambda]/n!$ has $f_{\text{P}}(2.996,0) = 0.05$, it follows that the number of events in that rigidity range corresponding to the flux of anti-helium $\Phi_{\antiHe{}}^{95}(\mathcal{R},T_{95})$ that saturates that 95\% CL upper limit in an integration time $T_{95}$ would be
\begin{align}
    N^{95}_{\antiHe{}}\approx 2.996 \approx 3\:.
\end{align}
We then assume that the acceptances for anti-helium and helium are proportional to each other, with a rigidity-independent proportionality constant $\kappa$:
\begin{align}
    \zeta_{\antiHe{}}(\mathcal{R},T) \approx \kappa \zeta_{\text{He}}(\mathcal{R},T)\:. \label{eq:acceptProp}
\end{align}
We thus have that
\begin{align}
    N^{95}_{\antiHe{}} &= \int_{\mathcal{R}_1}^{\mathcal{R}_2} \Phi^{95}_{\antiHe{}}(\mathcal{R},T_{95}) \zeta_{\antiHe{}}(\mathcal{R},T_{95}) d\mathcal{R}\\
    &= \kappa r_{95}(T_{95}) \int_{\mathcal{R}_1}^{\mathcal{R}_2} \Phi_{\text{He}}(\mathcal{R}) \zeta_{\text{He}}(\mathcal{R};T_{95})d\mathcal{R} \\
    &= \kappa r_{95}(T_{95}) \times N_{\text{He}}\lb(\mathcal{R}_1,\mathcal{R}_2;T_{95}\rb) \\
    &= \kappa r_{95}(T) \times N_{\text{He}}\lb(\mathcal{R}_1,\mathcal{R}_2;T\rb) \:, \label{eq:lastLine}
\end{align}
where we used at the last line that $r_{95}(T) \propto 1/T$ and that $N_{\text{He}}(T) \propto \zeta_{\text{He}}(T) \propto T$ for steady-state helium fluxes.
Note that \eqref{eq:lastLine} no longer makes reference to the timescale $T_{95}$.
Solving \eqref{eq:lastLine} for $\kappa$ and substituting into \eqref{eq:acceptProp}, we have 
\begin{align}
    \zeta_{\antiHe{}}(\mathcal{R},T) &\approx \frac{N^{95}_{\antiHe{}}\times \zeta_{\text{He}}(\mathcal{R},T)}{ r_{95}(T) \times N_{\text{He}}\lb(\mathcal{R}_1,\mathcal{R}_2;T\rb) }\\
    &= \frac{N^{95}_{\antiHe{}}\times \zeta_{\text{He}}(\mathcal{R},\tau_0)}{r_{95}(T) \times N_{\text{He}}\lb(\mathcal{R}_1,\mathcal{R}_2;\tau_{0}\rb) }\:, \label{eq:zetaConstr}
\end{align}
again using at the last line that $N_{\text{He}}(T) \propto \zeta_{\text{He}}(T) \propto T$ for steady-state helium fluxes.
Note that $\zeta_{\antiHe{}}(T) \propto T$ as expected because $r_{95}(T)\propto 1/T$.

Although we should take $\mathcal{R}_1=1\,\text{GV}$ from the above argument, we shift this to $\mathcal{R}_1 \rightarrow 1.92\,\text{GV}$, the lower limit of available data from \AntiHeFluxSM; likewise, we shift $\mathcal{R}_2 \rightarrow 52.5\,\text{GV}$, the nearest upper bin edge to $\mathcal{R}\sim 50\,\text{GV}$ in \AntiHeFluxSM.

We can then (a) use \eqref{eq:singleBinNumber} to construct the number of helium events in each rigidity bin given in \AntiHeFluxSM\ and sum them up to find $N_{\text{He}}\lb(\mathcal{R}_1,\mathcal{R}_2;\tau_{0}\rb)$; (b) use \eqref{eq:singleBinZeta} to construct $\zeta_{\text{He}}(\mathcal{R},\tau_0)$, again using the data in \AntiHeFluxSM; and (c) construct $\zeta_{\antiHe{}}(\mathcal{R},T)$ using \eqref{eq:zetaConstr}.
Once we have the acceptance for anti-helium, we can compute event numbers using \eqref{eq:zetaDefn}.

Note that we have assumed here throughout that the same acceptance \emph{as a function of rigidity} applies for all the species with $Z = \pm 2$ [i.e., all the \mbox{(anti-)helium} isotopes].
The analysis in Ref.~\cite[\S4.7]{Sonnabend:2023kzy} supports that assumption at the level of a few tens of percent at low rigidity, and better than 15\% at high rigidity ($\mathcal{R} \gtrsim 10\,\text{GV}$).
We have also verified that the reconstruction procedure for the helium acceptance based on \eqref{eq:singleBinZeta} and the data in \AntiHeFluxSM\ reproduces the acceptance that can be constructed from Fig.~4.29 in Ref.~\cite[\S4.7]{Sonnabend:2023kzy} (after accounting for differing integration times) to within at worst $\sim 50$\%, and usually within 30\% or better (with agreement generally becoming better for higher rigidity) in most of the rigidity bins from \AntiHeFluxSM.
This is acceptably accurate for our purposes.

We also tried to apply this acceptance reconstruction technique to the separated ${}^3\text{He}$ and ${}^4\text{He}$ isotope data presented for a much smaller range of low rigidities ($2 \,\text{GV}\lesssim \mathcal{R} \lesssim 15 \,\text{GV}$) in the Supplemental Material of \citeR{PhysRevLett.123.181102} (see also \citeR{AMS:2021nhj}). 
However, a na\"ive application of \eqref{eq:singleBinZeta} to those data for each isotope separately gives results for acceptances for ${}^3\text{He}$ that are generally a factor of $\mathcal{O}(10)$ smaller than those for ${}^4\text{He}$.
However, we believe that to be a spurious result for two reasons, and therefore disregard it: (1) the raw number of events selected for analysis in \citeR{PhysRevLett.123.181102} were $10^8$ events for ${}^4\text{He}$ and $1.8\times 10^7$ events for ${}^3\text{He}$, while the (somewhat rigidity-dependent) flux ratio for ${}^3\text{He}$ to ${}^4\text{He}$ was also reported to be around 10--15\%; because the event ratio is similar to the flux ratio, it seems impossible for the isotope acceptances to differ by as much as a factor of 10; and (2) the result conflicts with the effective acceptance curves%
\footnote{%
    The effective acceptance in \citeR{Sonnabend:2023kzy}, $A_{\text{eff}}(\mathcal{R})$, is defined as $A_{\text{eff}}(\mathcal{R}) = \zeta(\mathcal{R},T) / T$; it folds in both effective area and trigger-efficiency effects as defined in \citeR{PhysRevLett.123.181102}, but not integration time.
} %
shown in \citeR{Sonnabend:2023kzy}, which that reference used to closely reproduce official (combined) AMS-02 helium flux results of \citeR[s]{AMS:2021nhj,AMSHeflux} to within $\sim10\%$.
We suspect that the acceptance reconstruction procedure of \citeR{Winkler:2020ltd} that we have reviewed here is simply inaccurate as applied to the isotope-separated ${}^3\text{He}$ data in the Supplemental Material of \citeR{PhysRevLett.123.181102} because those data are a small, $\mathcal{O}{(20\%)}$, sub-component of the total helium flux data that must be unfolded by AMS-02 to obtain the individual isotope results.
Moreover, as applied to the ${}^4\text{He}$ data in the Supplemental Material of \citeR{PhysRevLett.123.181102}, it gives results with much larger variations from rigidity bin to rigidity bin than those obtained from applying it to the earlier (combined) helium data in \AntiHeFluxSM\ (with integration-time differences accounted for).
Because of these issues with these reconstructions and because they also only cover a lower rigidity range than where we need the results, we do not use $\zeta$ as reconstructed from the isotope-separated data in \citeR{PhysRevLett.123.181102}.

A comment is also in order on the assumption of steady-state helium fluxes. 
AMS-02 has reported data on the time variation of helium fluxes in \citeR[s]{PhysRevLett.123.181102,PhysRevLett.128.231102,AMS:2021nhj}.
Between the years 2011 and 2019, the fluxes at very low rigidities, $R\in[1.71,1.92]\,\text{GV}$ have increased by a factor of $\sim 2$~\cite{PhysRevLett.128.231102}, while those for rigidities $\mathcal{R}\gtrsim 5\,\text{GV}$ have varied by $\lesssim 10\%$~\cite{PhysRevLett.128.231102}; see also \citeR{PhysRevLett.123.181102} for alternative presentation showing changes of similar magnitude between 2011 and 2017.
The variations of the fluxes over the timescale $\tau_0$, which covers the years 2011--3, are much smaller: they change by only a few tens of percent around their average values, even at low rigidity~\cite{PhysRevLett.128.231102}.
In the high-rigidity regime of most interest to the AMS-02 anti-helium candidate events, the assumption of steady-state helium flux in the above derivation is thus well justified.
And while there may be a mild violation of the scaling of $N_{\text{He}}\lb(\mathcal{R}_1,\mathcal{R}_2;T\rb) \propto T$ arising from the changing flux at the low end of the rigidity range, we estimate that this effect has only $\mathcal{O}(1)$-factor overall impact on our analysis.

\bibliography{references}
\bibliographystyle{JHEP}
\end{document}